\documentclass[a4paper,twocolumn,11pt,accepted=2022-12-12]{quantumarticle}
\pdfoutput=1
\usepackage[utf8]{inputenc}
\usepackage[english]{babel}
\usepackage[T1]{fontenc}
\usepackage{amsmath}
\usepackage{hyperref}

\usepackage{tikz}
\usepackage{lipsum}

\usepackage{epsfig,amsmath,amssymb,color,dsfont,upgreek,physics}
\usepackage{mathrsfs}
\usepackage{mathtools}
\begin{document}
\title{Achieving the quantum field theory limit in far-from-equilibrium quantum link models}
\author{Jad C.~Halimeh}
\email{jad.halimeh@physik.lmu.de}
\affiliation{INO-CNR BEC Center and Department of Physics, University of Trento, Via Sommarive 14, I-38123 Trento, Italy}
\orcid{0000-0002-0659-7990}
\author{Maarten Van Damme}
\affiliation{Department of Physics and Astronomy, University of Ghent, Krijgslaan 281, 9000 Gent, Belgium}
\author{Torsten V.~Zache}
\affiliation{Center for Quantum Physics, University of Innsbruck, 6020 Innsbruck, Austria}
\affiliation{Institute for Quantum Optics and Quantum Information of the Austrian Academy of Sciences, 6020 Innsbruck, Austria}
\author{Debasish Banerjee}
\affiliation{Theory Division, Saha Institute of Nuclear Physics, HBNI, 1/AF Bidhan Nagar, Kolkata 700064, India}
\author{Philipp Hauke} 
\affiliation{INO-CNR BEC Center and Department of Physics, University of Trento, Via Sommarive 14, I-38123 Trento, Italy}

\begin{abstract}
Realizations of gauge theories in setups of quantum synthetic matter open up the possibility of probing salient exotic phenomena in condensed matter and high-energy physics, along with potential applications in quantum information and science technologies. In light of the impressive ongoing efforts to achieve such realizations, a fundamental question regarding quantum link model regularizations of lattice gauge theories is how faithfully they capture the quantum field theory limit of gauge theories. Recent work [Zache, Van Damme, Halimeh, Hauke, and Banerjee, \href{https://journals.aps.org/prd/abstract/10.1103/PhysRevD.106.L091502}{Phys.~Rev.~D \textbf{106}, L091502 (2022)}] has shown through analytic derivations, exact diagonalization, and infinite matrix product state calculations that the low-energy physics of $1+1$D $\mathrm{U}(1)$ quantum link models approaches the quantum field theory limit already at small link spin length $S$. Here, we show that the approach to this limit also lends itself to the far-from-equilibrium quench dynamics of lattice gauge theories, as demonstrated by our numerical simulations of the Loschmidt return rate and the chiral condensate in infinite matrix product states, which work directly in the thermodynamic limit. Similar to our findings in equilibrium that show a distinct behavior between half-integer and integer link spin lengths, we find that criticality emerging in the Loschmidt return rate is fundamentally different between half-integer and integer spin quantum link models in the regime of strong electric-field coupling. Our results further affirm that state-of-the-art finite-size ultracold-atom and NISQ-device implementations of quantum link lattice gauge theories have the real potential to simulate their quantum field theory limit even in the far-from-equilibrium regime.
\end{abstract}
\maketitle
\tableofcontents
\section{Introduction}
In today's impressive level of control and precision in quantum synthetic matter (QSM) \cite{Bloch2008,Lewenstein_book,Blatt_review,Hauke2012}, the quest for realizing complex quantum many-body systems on state-of-the-art quantum simulators has been gaining a lot of traction. Exotic phenomena that had been confined to the theoretical realm just a few years ago, have recently been observed experimentally such as dynamical phase transitions \cite{Jurcevic2017,Zhang2017dpt,Flaeschner2018}, prethermalization \cite{Gring2012,Langen2015,Neyenhuis2017}, many-body localization \cite{Schreiber2015,Choi2016,Smith2016}, many-body dephasing \cite{Kaplan2020}, topological spin liquids \cite{semeghini2021probing}, the ground state of the toric code \cite{Satzinger2021}, and discrete time crystalline order \cite{Mi2021}.

This advancement in QSM has also recently facilitated a flurry of experimental realizations of lattice gauge theories (LGTs) \cite{Martinez2016,Klco2018,Kokail2019,Klco2020,Lu2019,Goerg2019,Schweizer2019,Mil2020,Yang2020,zhou2021thermalization}, which are particularly attractive to investigate in such setups for several reasons \cite{Wiese_review,Zohar2015,Dalmonte2016,MariCarmen2019,Alexeev_review,aidelsburger2021cold,zohar2021quantum,klco2021standard}. From a physics point of view, gauge theories are fundamental descriptions of interactions between elementary particles as mediated by gauge bosons, and they give rise to many fascinating salient features \cite{Weinberg_book,Gattringer_book,Zee_book}. The latter arise due to local constraints that are the principal property of gauge theories and that must be satisfied at every point in space and time. This \textit{gauge invariance} is manifest as, e.g., Gauss's law in quantum electrodynamics (QED), which is what leads to a massless photon and a long-ranged Coulomb law. Furthermore, intriguing salient features from a condensed matter perspective also arise in LGTs, such as quantum many-body scars \cite{Bernien2017,Surace2020,Banerjee2021,Aramthottil2022,Desaules2022a,Desaules2022b} and disorder-free localization \cite{Smith2017,Brenes2018,smith2017absence,Metavitsiadis2017,Smith2018,Russomanno2020,Papaefstathiou2020,karpov2021disorder,hart2021logarithmic,Zhu2021}, which are newly discovered paradigms of ergodicity breaking even in certain cases when the underlying model is itself ergodic. From a technological point of view, the reliability and stability of gauge invariance is crucial in modern QSM realizations of LGTs involving both matter and gauge degrees of freedom, where gauge-breaking errors are unavoidable due to the plethora of local constraints to be controlled. This has led to various theoretical works on schemes to stabilize gauge invariance in such implementations \cite{Zohar2011,Zohar2012,Banerjee2012,Zohar2013,Hauke2013,Stannigel2014,Kuehn2014,Kuno2017,Yang2016,Kuno2017,Negretti2017,Dutta2017,Barros2019,Halimeh2020a,Lamm2020,Halimeh2020e,kasper2021nonabelian,vandamme2021reliability,halimeh2021gauge,halimeh2021stabilizing,vandamme2021suppressing,halimeh2021stabilizingDFL,halimeh2021enhancing}. As such, not only do QSM realizations of LGTs permit the possibility of addressing questions in condensed matter and high-energy physics on easily accessible low-energy table-top devices, they also further push forward the technological level of precision and control in QSM platforms.

Beyond the reliable implementation of gauge invariance, there is the issue of whether quantum link model (QLM) \cite{Wiese_review,Chandrasekharan1997} regularizations of LGTs do indeed retrieve the physics of gauge theories in the quantum field theory limit. In QLMs, the Hilbert space of the $\mathrm{U}(1)$ Abelian gauge fields is regularized by representing them with quantum spin-$S$ operators of finite-dimensional Hilbert spaces. This approximation is very well-suited for current QSM experiments, which are limited in terms of both total Hilbert space and volume. Even though there has recently been works showing that QLM regularizations of lattice gauge theories achieve the quantum field theory limit at relatively small link spin length and system size in equilibrium \cite{Buyens2017,zache2021achieving}, exactly how much QLMs need to scale in the local Hilbert space of the gauge link (parametrized by the link spin length $S$), the lattice spacing $a$, and volume in order to achieve this limit in the far-from-equilibrium regime has remained an open question. Given the recent advancement in large-scale QSM implementations of $\mathrm{U}(1)$ quantum link models \cite{Yang2020} including experiments on their quench dynamics \cite{zhou2021thermalization}, it is crucial to investigate this question in order to understand how faithfully such experiments can model true high-energy physics phenomena. 

In this work, we consider the paradigmatic spin-$S$ $\mathrm{U}(1)$ QLM in $1+1$D to address this question. We show through exact diagonalization and infinite matrix product state calculations that quench dynamics of the return rate and chiral condensate in this model rapidly approach both the Wilson--Kogut--Susskind ($S\to\infty$) and continuum ($a\to0$) limits in various parameter regimes already at small values of the link spin length $S$. In fact, we find that the Wilson--Kogut--Susskind (WKS) limit is already achievable at relatively small $S\lesssim4$ up to all accessible evolution times in all considered regimes. We also show that the approach to the thermodynamic limit can be quite fast in certain gauge superselection sectors and parameter regimes. An experimentally relevant conclusion from our work is that in the weak electric-field coupling regime, the WKS limit leads to quench dynamics showing great quantitative agreement across different choices of the target gauge superselection sector and independently of whether $S$ is half-integer or integer. Our results indicate that current QSM implementations of LGTs are not so far away from probing the quantum field theory limit of gauge theories as initially feared.

The rest of our paper is organized as follows: In Sec.~\ref{sec:model}, we introduce the $1+1$D spin-$S$ $\mathrm{U}(1)$ QLM and the quench protocol we employ. In Secs.~\ref{sec:Kogut} and~\ref{sec:continuum}, we present our main numerical results on the approach of the quench dynamics to the WKS and continuum limits, respectively. We conclude in Sec.~\ref{sec:conc}. Our main results are supplemented with Appendix~\ref{app:PH_trafo}, in which we review the particle-hole transformation employed in our model Hamiltonian, and with Appendix~\ref{app:GS}, where we provide and discuss the ground-state phase diagram of the spin-$S$ $\mathrm{U}(1)$ QLM.

\section{Model and quench protocol}\label{sec:model}
We consider the spin-$S$ $\mathrm{U}(1)$ QLM given by the Hamiltonian \cite{Hauke2013,Yang2016,Kasper2017}
\begin{align}\nonumber
	\hat{H}=&-\frac{J}{2a\sqrt{S(S+1)}}\sum_{j=1}^{L-1}\big(\hat{\sigma}^-_j\hat{s}^+_{j,j+1}\hat{\sigma}^-_{j+1}+\text{H.c.}\big)\\\label{eq:H}
	&+\frac{\mu}{2}\sum_{j=1}^L\hat{\sigma}^z_j+\frac{g^2a}{2}\sum_{j=1}^{L-1}\big(\hat{s}^z_{j,j+1}\big)^2,
\end{align}
where the matter field at site $j$ is represented by Pauli matrices in the form of ladder operators $\hat{\sigma}_j^\pm$ and occupation operator $\hat{n}_j=(\hat{\sigma}_j^z+\mathds{1})/2$, and the gauge and electric fields on the link connecting matter sites $j$ and $j+1$ are represented by the spin-$S$ operators $\hat{s}^\pm_{j,j+1}$ and $\hat{s}^z_{j,j+1}$, respectively. The total number of matter sites is denoted by $L$, and the lattice spacing is $a$. The coupling strength of the electric field is denoted by $g$, whereas the mass of the matter field is indicated by $\mu$. We will set the energy scale $J=1$ throughout the paper. The local generator of the $\mathrm{U}(1)$ gauge symmetry of Hamiltonian~\eqref{eq:H} is
\begin{align}\label{eq:Gj}
\hat{G}_j=(-1)^j\big(\hat{n}_j+\hat{s}^z_{j-1,j}+\hat{s}^z_{j,j+1}\big),
\end{align}
where gauge invariance is encoded in the commutation relation $\big[\hat{H},\hat{G}_j\big]=0,\,\forall j$. The eigenvalues $q_j$ of $\hat{G}_j$, known as \textit{background charges}, are integers and range from $-2(-1)^jS$ to $(-1)^j(2S+1)$. A set of these eigenvalues over volume defines a \textit{gauge superselection sector} $\mathbf{q}=(q_1,q_2,\ldots,q_L)$. Furthermore, we note that here we have employed a particle-hole transformation to get the forms of Eqs.~\eqref{eq:H} and~\eqref{eq:Gj}, the details of which can be found in Appendix~\ref{app:PH_trafo}. We also provide and discuss the phase diagram of the spin-$S$ $\mathrm{U}(1)$ QLM in Appendix~\ref{app:GS}. In the limits of $S\to\infty$ and $a\to0$, Hamiltonian~\eqref{eq:H} describes quantum electrodynamics in $1+1$D, known as the Schwinger model. We are interested in investigating the approach of quench dynamics under Eq.~\eqref{eq:H} to both these limits, particularly since state-of-the-art large-scale cold-atom experimental realizations \cite{Yang2020,zhou2021thermalization} are mostly feasible at finite-$S$ and finite-$a$ representations.

Within the chosen physical sector $\mathbf{q}^\text{tar}=(q_1^\text{tar},q_2^\text{tar},\ldots,q_L^\text{tar})$, we prepare our system in the ground state $\ket{\psi_0}$ of Eq.~\eqref{eq:H} at an initial value $\mu_\text{i}=J$ of the mass, and then quench to a final value of $\mu_\text{f}=-J$. This type of quench is motivated by interpreting a change of the sign of the fermion mass $\mu \leftrightarrow -\mu$ as a change of the topological angle $\theta$ by $\pi$. The topological $\theta$ angle is responsible for a term $\propto(\theta-\pi)\sum_j(-1)^j\hat{s}^z_{j,j+1}$ that can explicitly break the global $\mathbb{Z}_2$ symmetry of the spin-$S$ $\mathrm{U}(1)$ QLM~\eqref{eq:H}, which can in turn give rise to intriguing phenomena in this model such as confinement \cite{Zohar2012,Surace2020}. In Ref.~\cite{Zache2019}, quenches of this $\theta$-angle were studied in the massive Schwinger model as a toy model for the relaxation of the analogous term $\propto\theta_\text{QCD}$ in $(3+1)-$D quantum chromodynamics (QCD). In one elegant explanation~\cite{PhysRevLett.38.1440}, the experimental fact that $\theta_\text{QCD} \approx 0$ is explained in terms of a dynamical field, the ``axion''; quenching $\theta$ is thus interpreted as an instantaneous change of this axion field. In contrast to \cite{Zache2019}, where the case $\theta = \pi \rightarrow 0$ (or equivalently $-\mu<0 \rightarrow \mu>0$) was studied, we focus in this article on the reverse quench. However, and as we explain in Appendix~\ref{app:GS}, performing our quench as such in the superselection sector $q_j=0,\,\forall j$, is equivalent to the quench $\theta = \pi \rightarrow 0$ in the superselection sector $q_j=(-1)^j$, and vice versa, because these two sectors are related by a particle-hole transformation.

In the wake of this quench, we are interested in the ensuing dynamics of the Loschmidt return rate \cite{Heyl2013,Heyl_review} and chiral condensate
\begin{subequations}
\begin{align}\label{eq:RR}
r(t)&=-\lim_{L\to\infty}\frac{1}{L}\log\big\lvert\bra{\psi_0}\ket{\psi(t)}\big\rvert^2,\\\label{eq:n}
n(t)&=\frac{1}{L}\sum_{j=1}^L\bra{\psi(t)}\hat{n}_j\ket{\psi(t)},
\end{align}
\end{subequations}
respectively, with $\ket{\psi(t)}=e^{-i\hat{H}t}\ket{\psi_0}$, where $\hat{H}$ is Eq.~\eqref{eq:H} at $\mu=\mu_\text{f}$. Note that here we are using both these quantities to assess convergence to the quantum field theory limit of the underlying $\mathrm{U}(1)$ gauge theory. Even though we calculate the dynamics of the Loschmidt return rate~\eqref{eq:RR}, our work is not intended as an exhaustive study of dynamical quantum phase transitions in gauge theories \cite{Huang2019,Zache2019}.

In the following, we present our numerical results obtained in part from the infinite matrix product state (iMPS) technique based on the time-dependent variational principle \cite{Haegeman2011,Haegeman2016,Vanderstraeten2019}, which works directly in the thermodynamic limit, and also from exact diagonalization (ED) codes that we have built for the purposes of benchmarking and finite-size analysis. Both our ED \cite{LaGaDyn} and iMPS \cite{MPSKit} codes utilize the gauge symmetry of Hamiltonian~\eqref{eq:H}, working directly in the relevant target superselection sector and therefore allowing us to reach larger sizes in ED and longer evolution times in iMPS. In the latter, we find that convergence of our results is achieved at a maximal bond dimension of $D_\text{max}=350$ and a time-step of $\delta t=0.001/J$ for our most stringent calculations.

\begin{figure*}[t!]
	\centering
	\includegraphics[width=.48\textwidth]{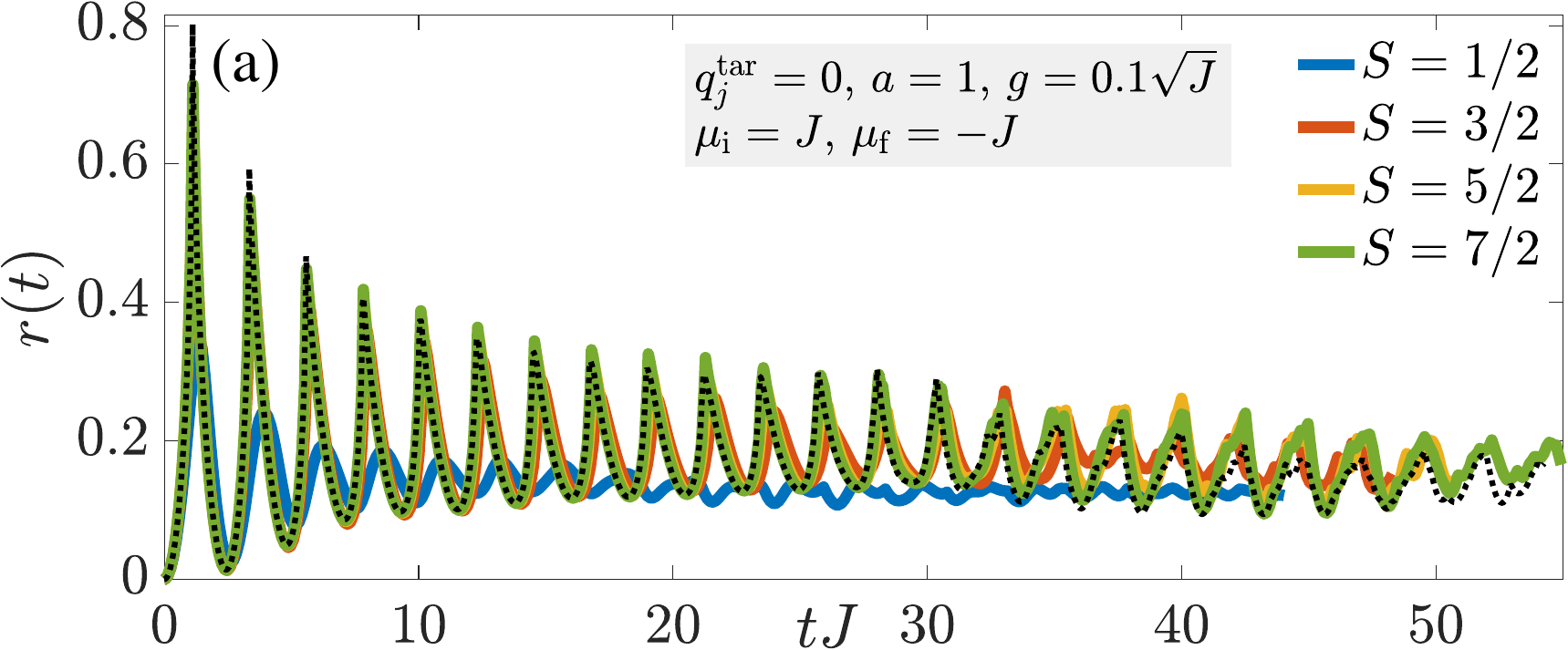}\quad\includegraphics[width=.48\textwidth]{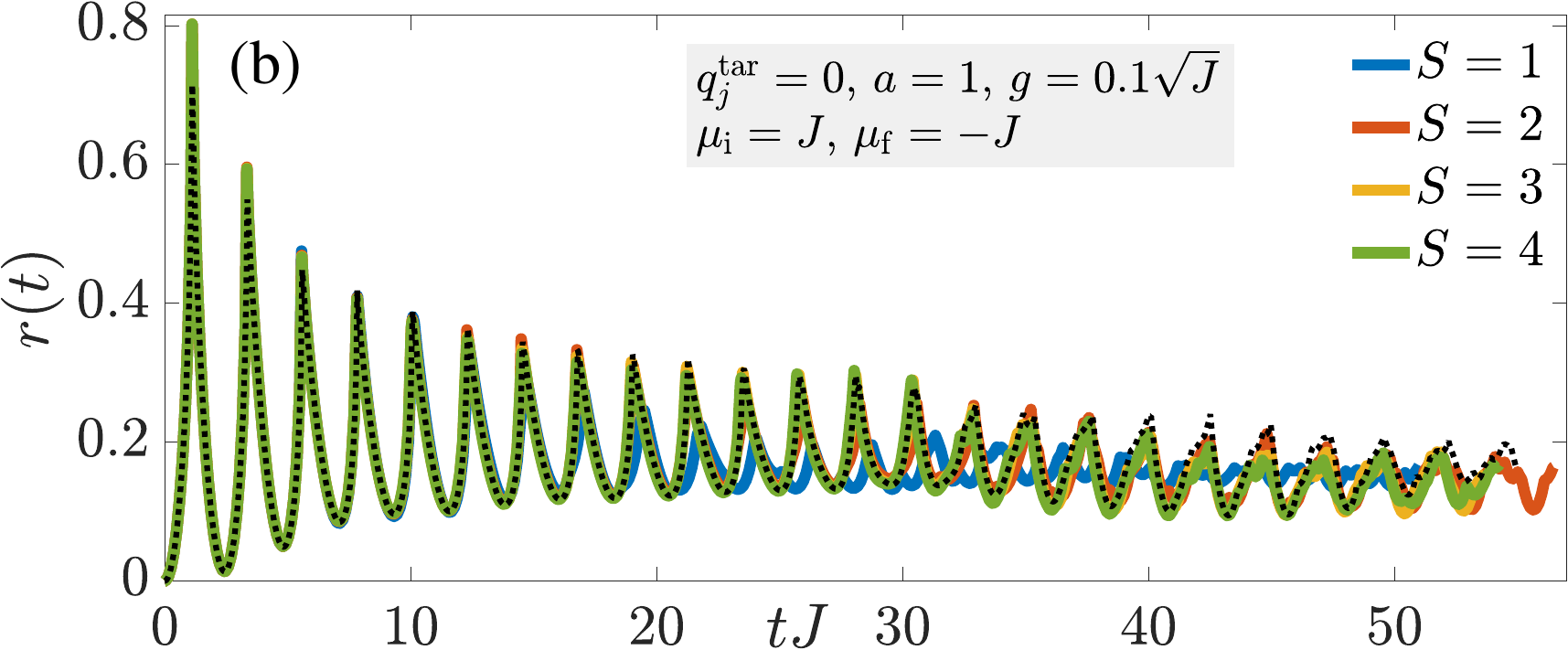}\\
	\vspace{1.1mm}
	\includegraphics[width=.48\textwidth]{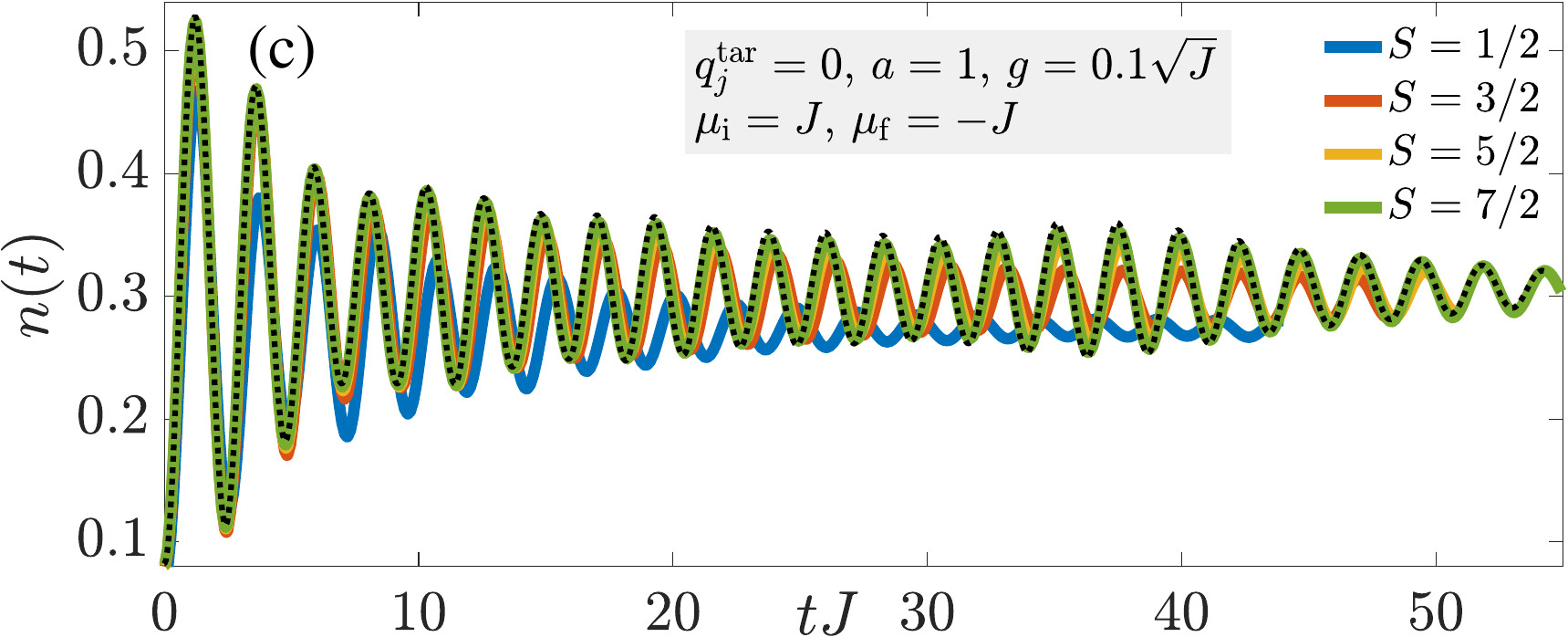}\quad\includegraphics[width=.48\textwidth]{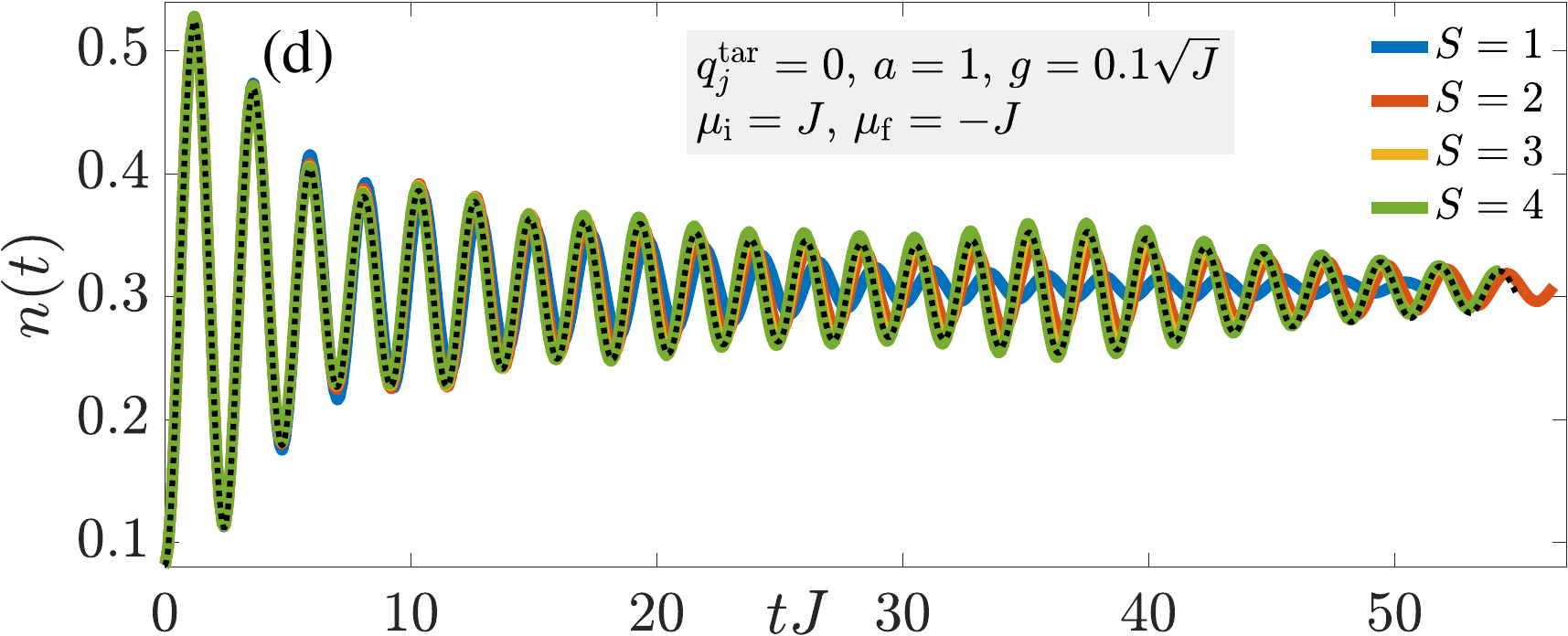}\\
	\vspace{1.1mm}
	\includegraphics[width=.48\textwidth]{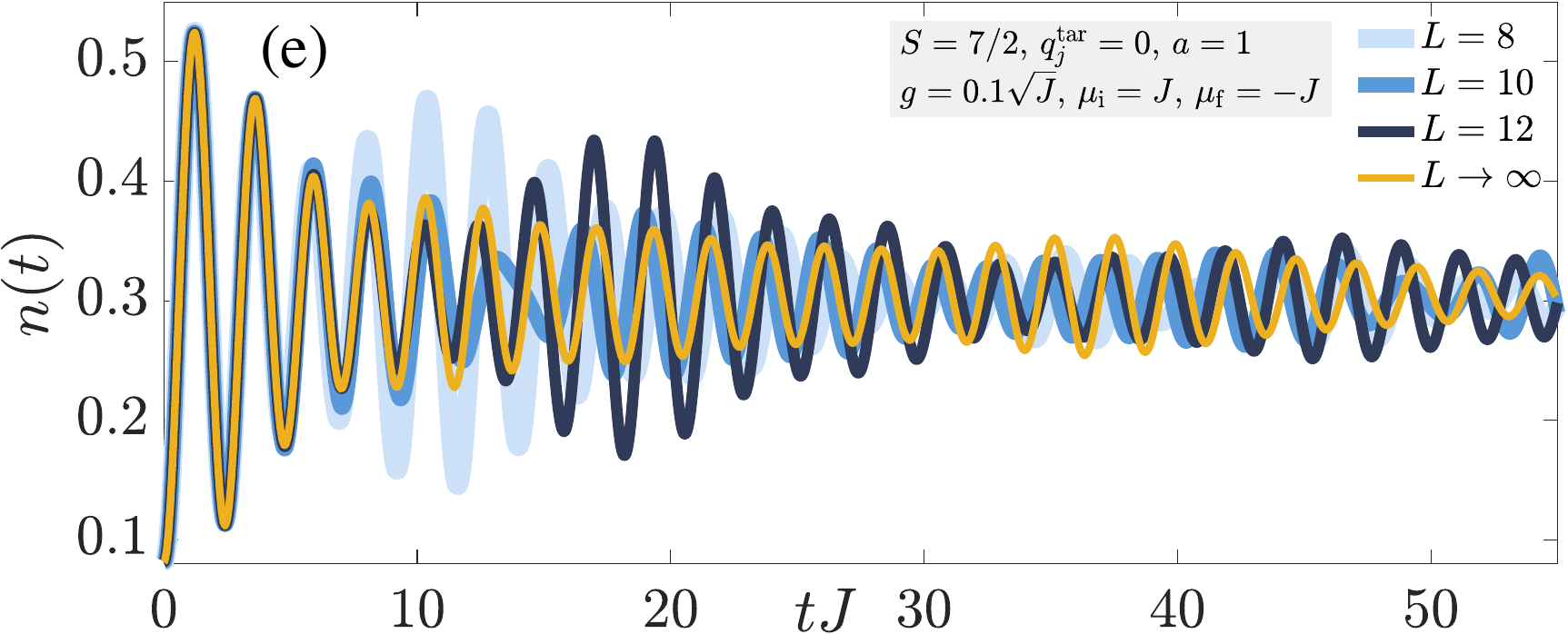}\quad\includegraphics[width=.48\textwidth]{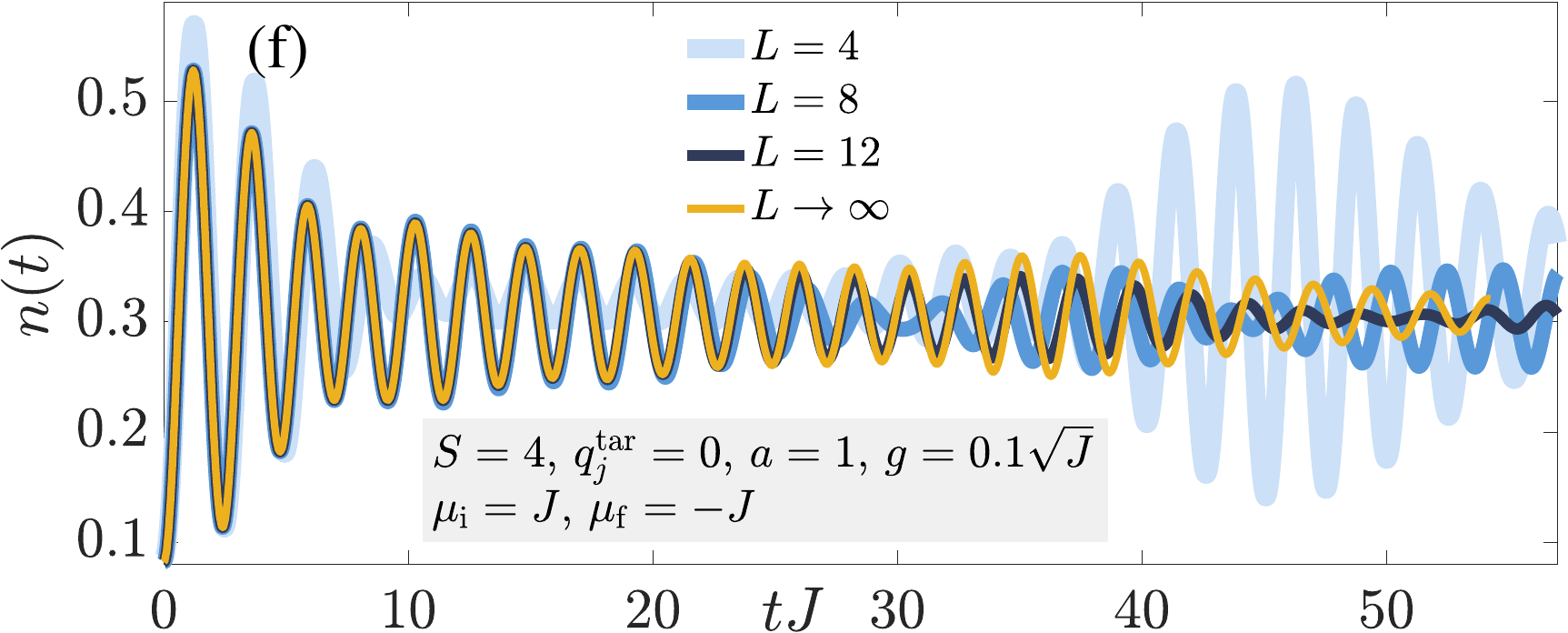}
	\caption{(Color online). Quench dynamics in the spin-$S$ $\mathrm{U}(1)$ quantum link model~\eqref{eq:H} in the weak-coupling regime with $g=0.1\sqrt{J}$ and $a=1$, starting in the ground state of Hamiltonian~\eqref{eq:H} at $\mu=J$ within the superselection sector $q_j^\text{tar}=0,\,\forall j$, and quenching the mass to $\mu=-J$. The results are obtained from the infinite matrix product state technique. Time evolution of (a,b) the return rate~\eqref{eq:RR} and (c,d) the chiral condensate~\eqref{eq:n} both show fast convergence for (a,c) half-integer and (b,d) integer link spin length $S$, although the case of integer $S$ shows overall faster convergence than the case of half-integer $S$. Both the converged return rate and the chiral condensate show good quantitative agreement for half-integer and integer $S$, as evidenced by the dotted black lines for $S=4$ in (a,c) [taken respectively from (b,d)] and for $S=7/2$ in (b,d) [taken respectively from (a,c)]. The approach to the thermodynamic limit in the quench dynamics of the chiral condensate in the weak-coupling regime for link spin lengths (e) $S=7/2$ and (f) $S=4$. The finite-size results are obtained from exact diagonalization, while those in the thermodynamic limit are calculated using the infinite matrix product state technique. In this target sector, we see much faster convergence to the thermodynamic limit for integer than for half-integer $S$.}
	\label{fig:Sscaling_q0_g0.1_a1} 
\end{figure*}

\begin{figure*}[t!]
	\centering
	\includegraphics[width=.48\textwidth]{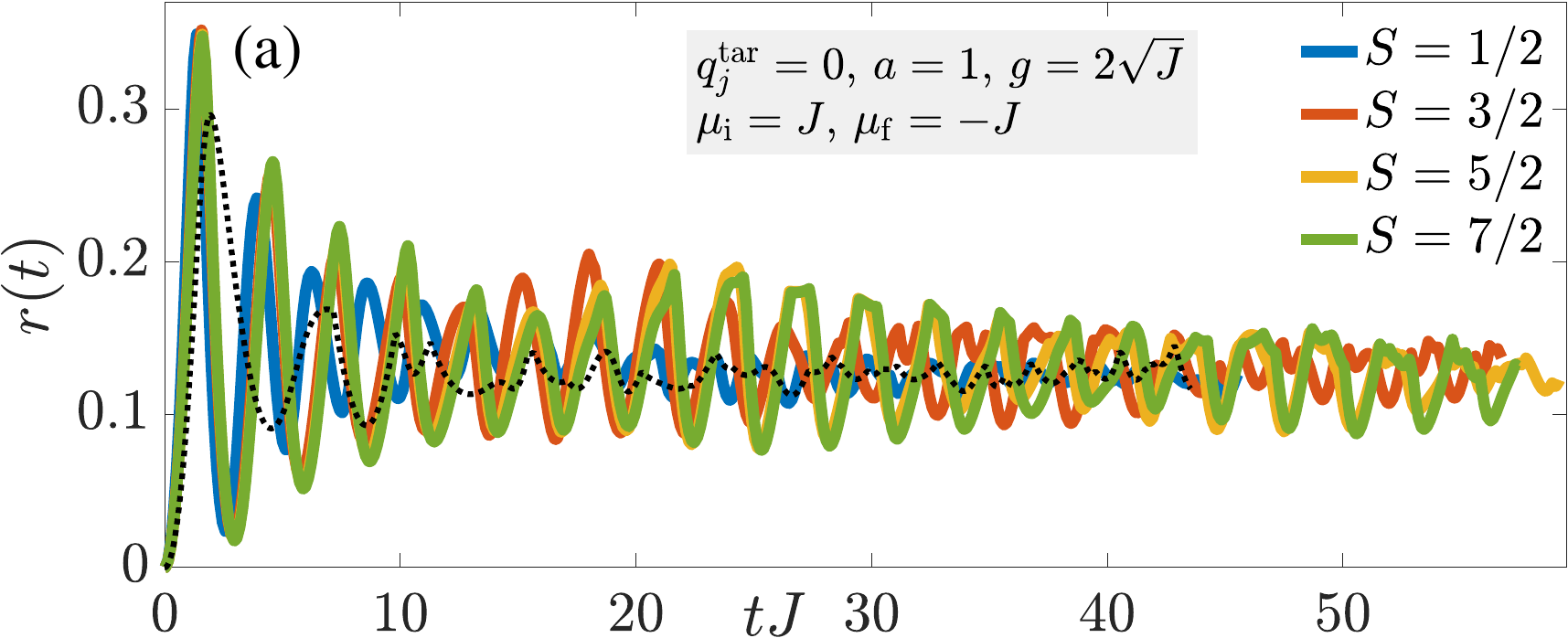}\quad\includegraphics[width=.48\textwidth]{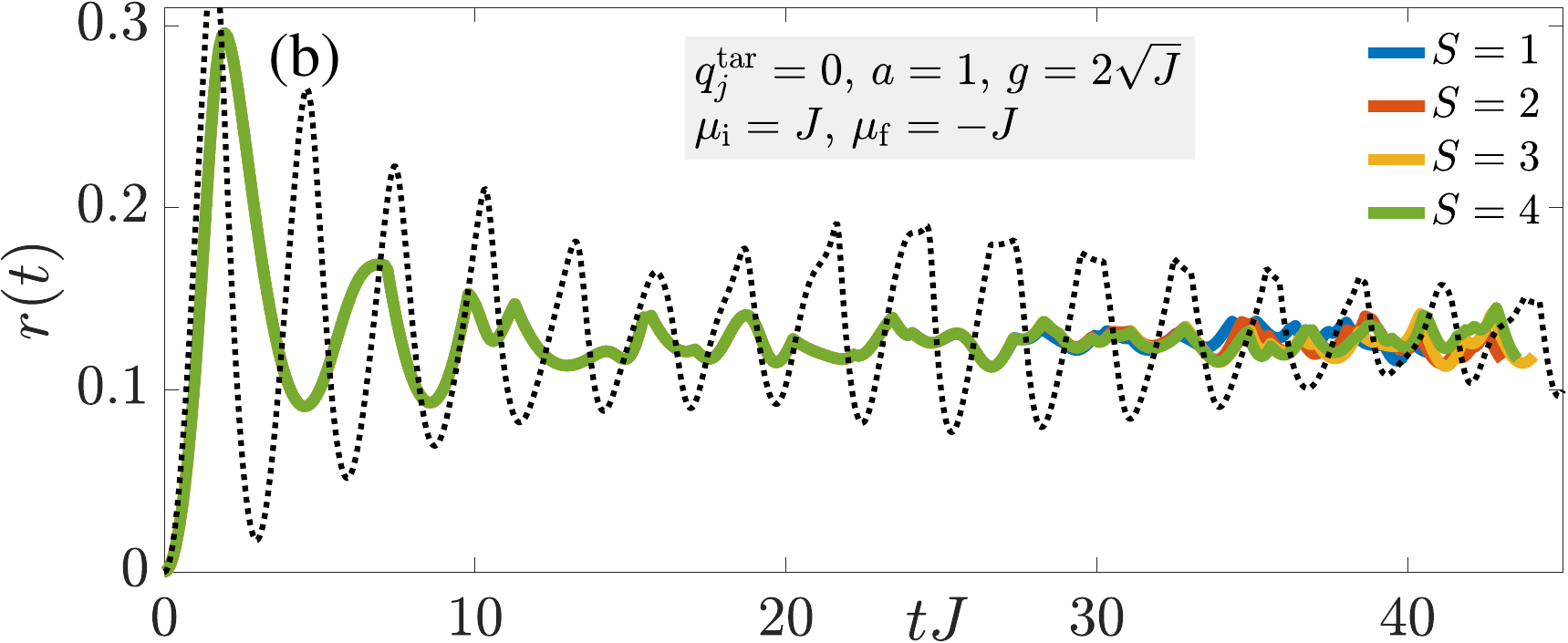}\\
	\vspace{1.1mm}
	\includegraphics[width=.48\textwidth]{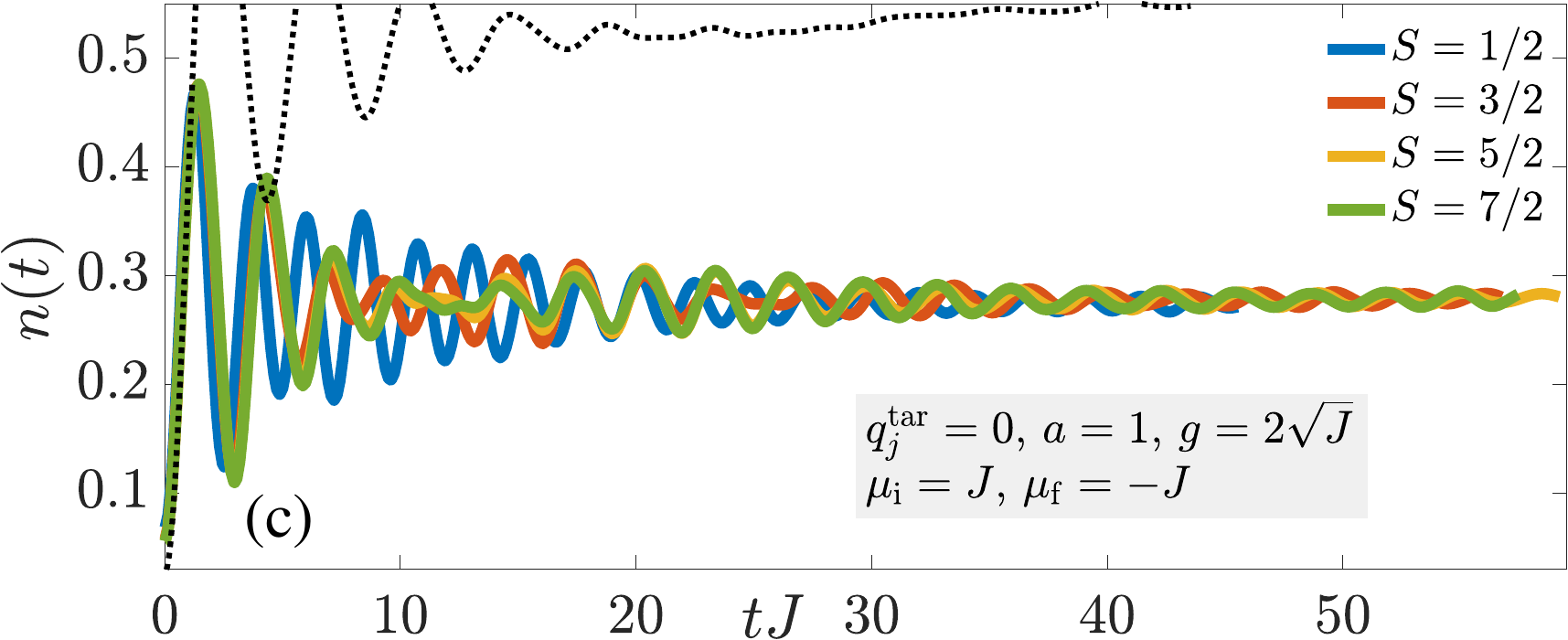}\quad\includegraphics[width=.48\textwidth]{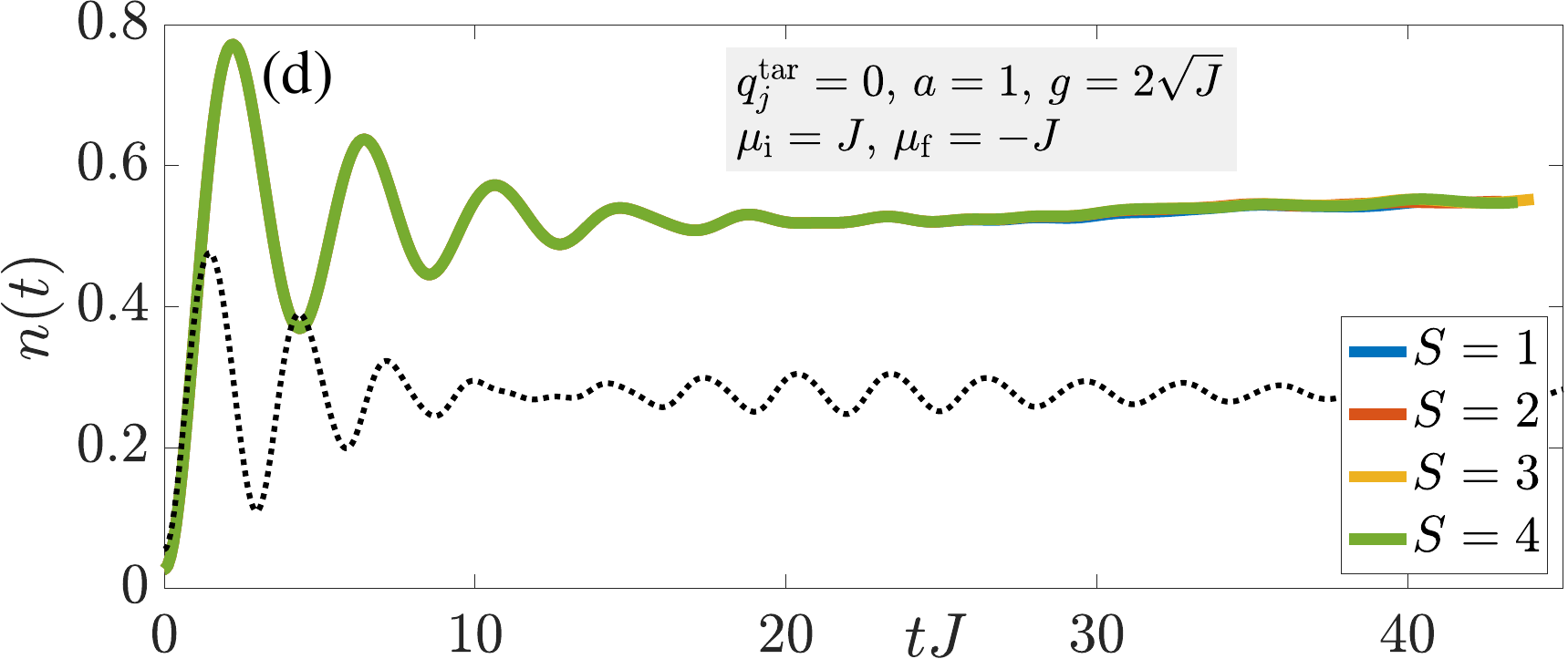}\\
	\vspace{1.1mm}
	\includegraphics[width=.48\textwidth]{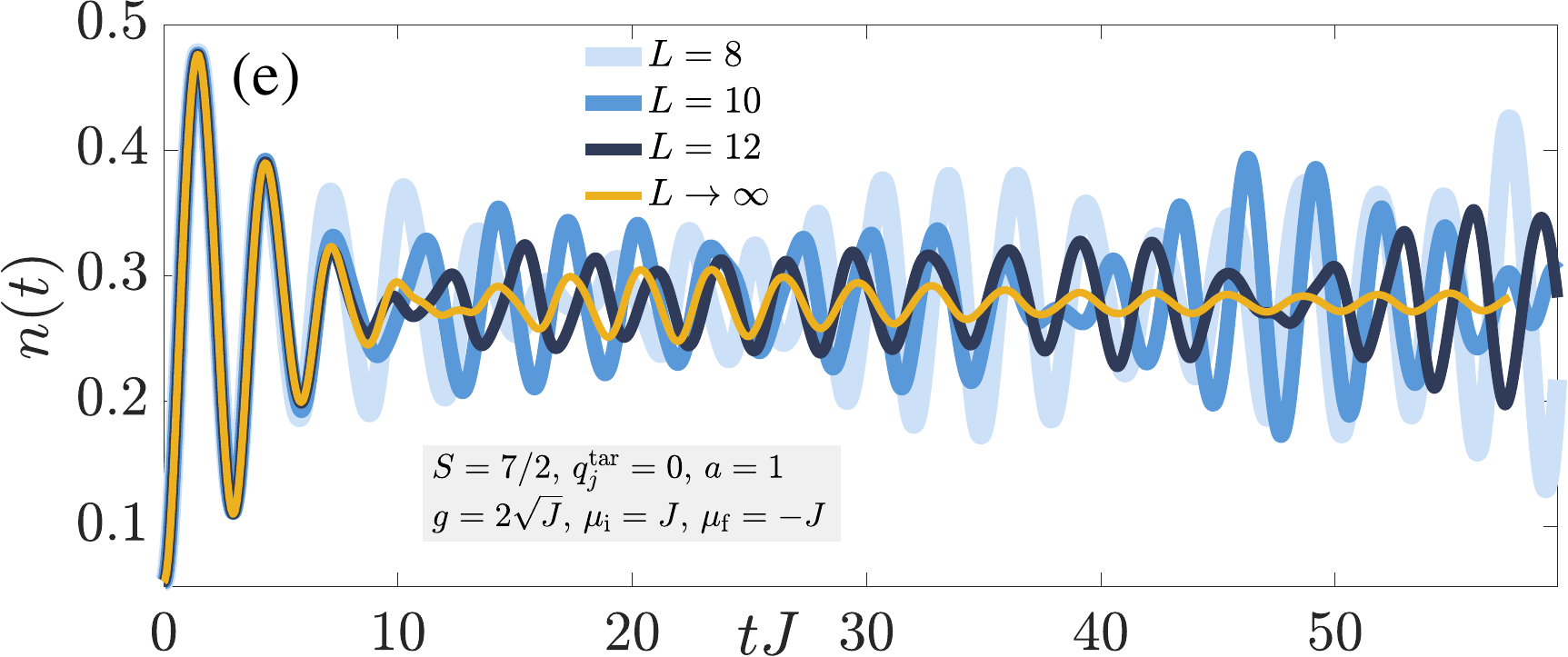}\quad\includegraphics[width=.48\textwidth]{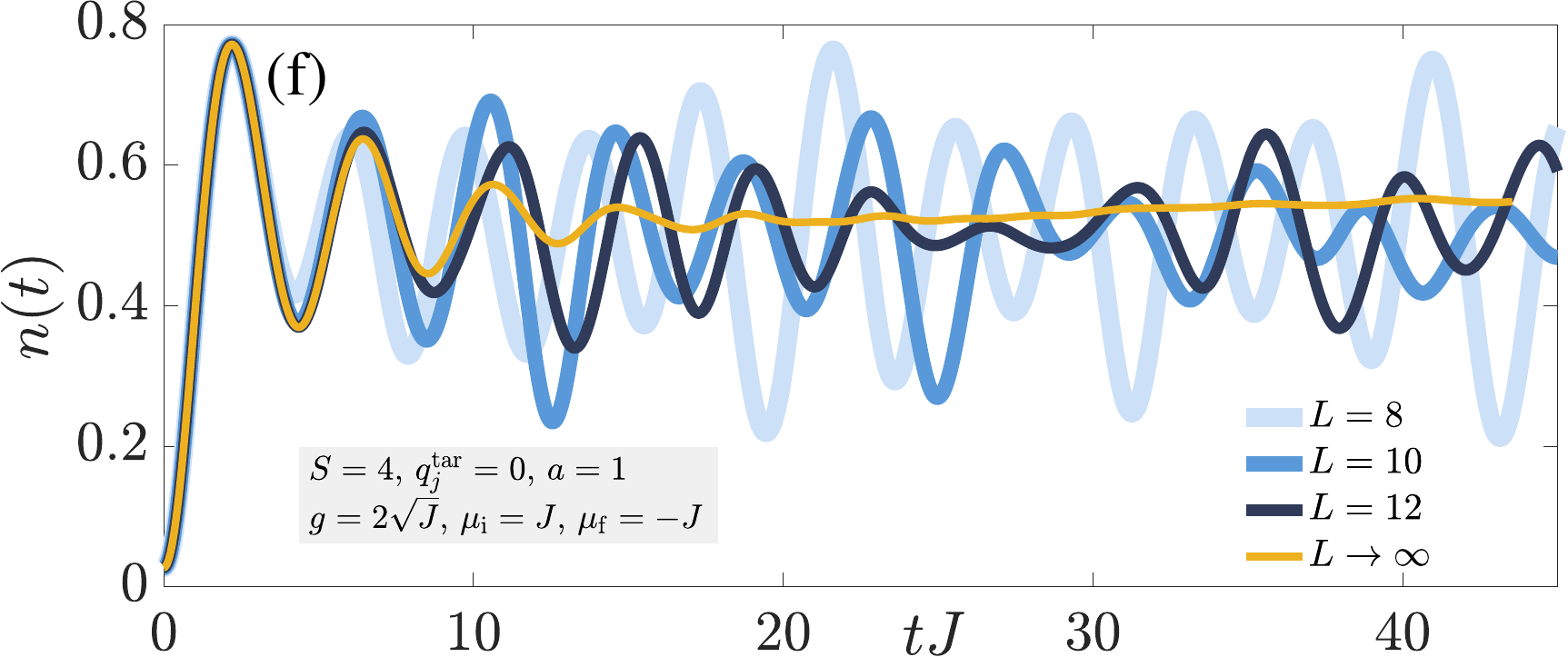}
	\caption{(Color online). Same as Fig.~\ref{fig:Sscaling_q0_g0.1_a1} but in the strong electric-field coupling regime with $g=2\sqrt{J}$. Compared to the weak-coupling regime, convergence to the Wilson--Kogut--Susskind limit is slightly slower for the case of half-integer $S$ (a,c) although still relatively quite good, and significantly faster for the case of integer $S$ (b,d). Indeed, in the latter case, we find that the return rate and condensate for $S=1$ is quantitatively almost identical to their counterparts for $S=4$ over most accessible evolution times. The return rate and chiral condensate for $S=7/2$ are vastly different from their counterparts for $S=4$, as indicated by the reference dotted black lines in (a-d), which represent the corresponding dynamics for $S=4$ in (a,c) and for $S=7/2$ in (b,d). The convergence to the thermodynamic limit is significantly slower than in the case of the weak-coupling regime, see Fig.~\ref{fig:Sscaling_q0_g0.1_a1} (e,f), for both half-integer and integer $S$.}
	\label{fig:Sscaling_q0_g2_a1} 
\end{figure*}

\section{Wilson--Kogut--Susskind limit $S\to\infty$}\label{sec:Kogut}
Let us first fix the lattice spacing to $a=1$, and study the effect of increasing the link spin length $S$ on the dynamics of the return rate~\eqref{eq:RR} and chiral condensate~\eqref{eq:n}. We will carry out this analysis in three different gauge superselection sectors corresponding to $q_j=0,1,2,\,\forall j$, where in each of these we will consider both the weak and strong electric-field coupling regimes. The reason we choose the first two target sectors is because they are the largest ones, while the third is chosen to further demonstrate the generality of our results.

\subsection{Target sector $q_j^\text{tar}=0$}\label{sec:q0}
We will now consider dynamics only within the superselection sector $q_j^\text{tar}=0,\,\forall j$, i.e., in the common eigenbasis $\{\ket{E_n}\}$ of $\hat{H}$ and $\hat{G}_j$ such that $\hat{H}\ket{E_n}=E_n\ket{E_n}$, we will consider only eigenstates satisfying $\hat{G}_j\ket{E_n}=0,\,\forall j$.

\subsubsection{Weak-coupling regime}\label{sec:q0_weak}

Let us first consider the weak-coupling regime and fix the electric-field coupling strength to $g=0.1\sqrt{J}$. Preparing our system in the ground states of Hamiltonian~\eqref{eq:H} at $\mu=J$ and suddenly quenching to $\mu=-J$, the ensuing quench dynamics of the return rate~\eqref{eq:RR} and chiral condensate~\eqref{eq:n} are shown in Fig.~\ref{fig:Sscaling_q0_g0.1_a1} for half-integer and integer link spin length $S$.

Focusing first on the quench dynamics of the return rate~\eqref{eq:RR}, we see in Fig.~\ref{fig:Sscaling_q0_g0.1_a1}(a,b) that for both half-integer $S$ and integer $S$, respectively, there is very fast convergence to the WKS limit, with the case of integer $S$ showing slightly faster convergence. Note that in both cases we have only used values of $S\leq4$, which comprise relatively small local Hilbert spaces for the gauge fields. Furthermore, there is excellent quantitative agreement between half-integer and integer $S$ near the converged limit, as indicated by the reference dotted black line for the return rate at $S=4$ in Fig.~\ref{fig:Sscaling_q0_g0.1_a1}(a) and that for the return rate at $S=7/2$ in Fig.~\ref{fig:Sscaling_q0_g0.1_a1}(b). The return rate is a global quantity, and its rapid convergence with $S$ to the WKS limit indicates that local observables should behave similarly or even better. Indeed, we find that this is also the case when we look at the quench dynamics of the chiral condensate~\eqref{eq:n} in Fig.~\ref{fig:Sscaling_q0_g0.1_a1}(c,d) in the case of half-integer and integer $S$, respectively. Once again, we find that the case of integer $S$ is slightly faster. As with the return rate, we find excellent quantitative agreement in the quench dynamics of the chiral condensate between half-integer and integer $S$ in the WKS limit. In Fig.~\ref{fig:Sscaling_q0_g0.1_a1}(c,d), we overlay the corresponding chiral-condensate quench dynamics for $S=4$ and $S=7/2$, respectively, where it shows great agreement with the WKS-limit converged results of half-integer and  integer $S$, respectively; see dotted black lines. This can be explained by noting that at small $g^2\ll1$, quantum fluctuations wash out the detailed ``low-energy'' structure of the link operators, and therefore the difference between the cases of half-integer and integer $S$ is no longer significant.

These results indicate that in the weak-coupling regime, we are able to achieve the WKS limit already at small values of $S$, and the behavior exhibits little dependence on whether $S$ is integer or half-integer. It is interesting now to investigate in ED how fast the quench dynamics converges to the thermodynamic limit in which iMPS directly works. In this vein, we focus on the quench dynamics of the chiral condensate for $S=7/2$ and $S=4$ in Fig.~\ref{fig:Sscaling_q0_g0.1_a1}(e,f), respectively, for various finite system sizes denoted by the number of matter sites $L$ (ED) and in the thermodynamic limit $L\to\infty$ (iMPS). For the case of half-integer $S$, we find that the convergence to the thermodynamic limit is considerably slower than that of integer $S$. Indeed, whereas for $S=7/2$ we do not see convergence to the thermodynamic limit already at early times for finite QLMs with $L=12$ matter sites, the chiral condensate in the case of $S=4$ shows very good convergence to the thermodynamic limit already for $L=12$ matter sites over most accessible evolution times. This is encouraging news for modern QSM implementations of LGTs restricted to a few building blocks \cite{Schweizer2019,Mil2020}, because our results indicate that within current accessible experimental lifetimes the WKS and thermodynamic limits can be achieved with a relatively small integer link spin length and a few matter sites.

\subsubsection{Strong-coupling regime}

We now repeat this quench in the strong electric-field coupling regime, setting $g=2\sqrt{J}$. The corresponding results for the return rate and chiral condensate for half-integer and integer $S$ are shown in Fig.~\ref{fig:Sscaling_q0_g2_a1}. Similarly to the case of the weak-coupling regime, we see very rapid convergence to the WKS limit in the quench dynamics of the return rate already at small values of half-integer and integer $S$, as shown in Fig.~\ref{fig:Sscaling_q0_g2_a1}(a,b), respectively. This convergence is particularly fast for the case of integer $S$, where already it occurs at $S=1$ throughout most of the accessible evolution times. A main difference from the weak-coupling regime is that now the dynamics for half-integer and integer $S$ agree neither quantitatively nor qualitatively. This can be seen by the dotted black lines in Fig.~\ref{fig:Sscaling_q0_g2_a1}(a,b) for the return rate at $S=4$ and $S=7/2$, respectively. 

\begin{figure*}[t!]
	\centering
	\includegraphics[width=.48\textwidth]{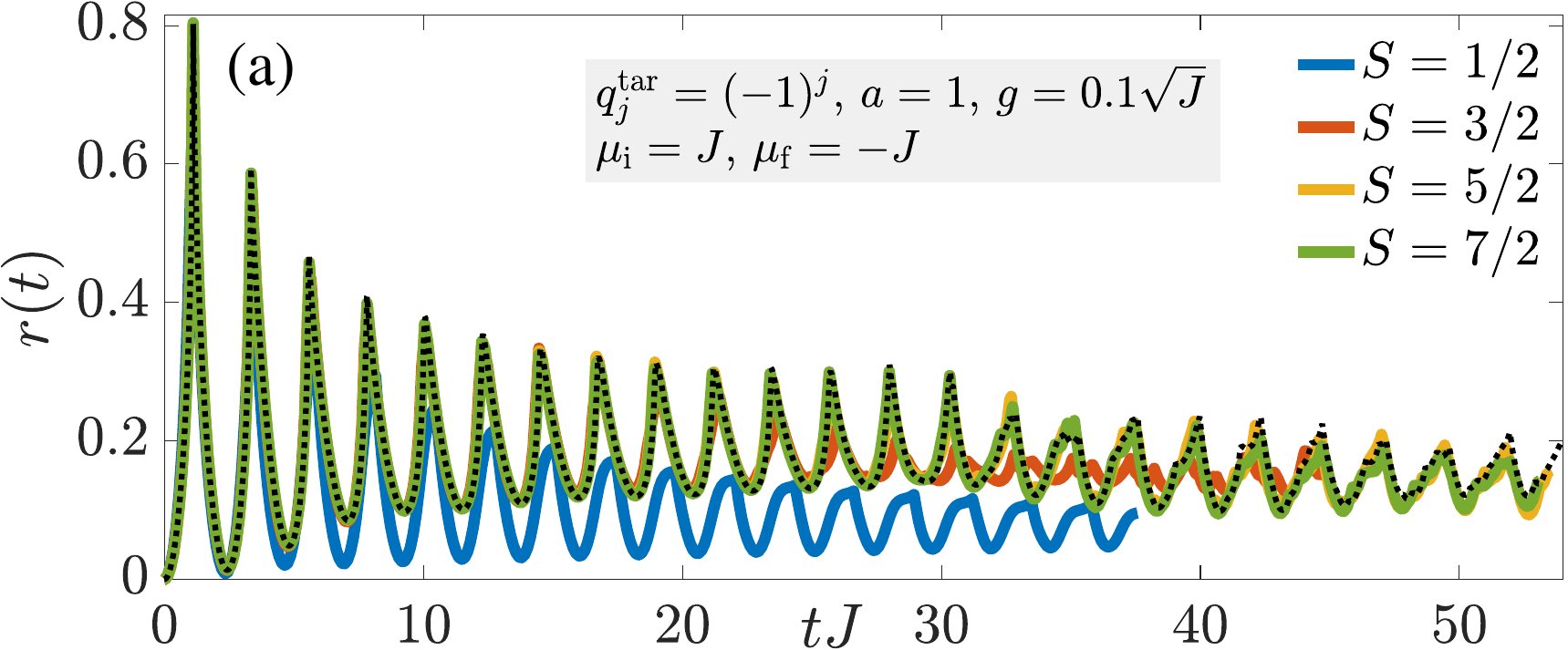}\quad\includegraphics[width=.48\textwidth]{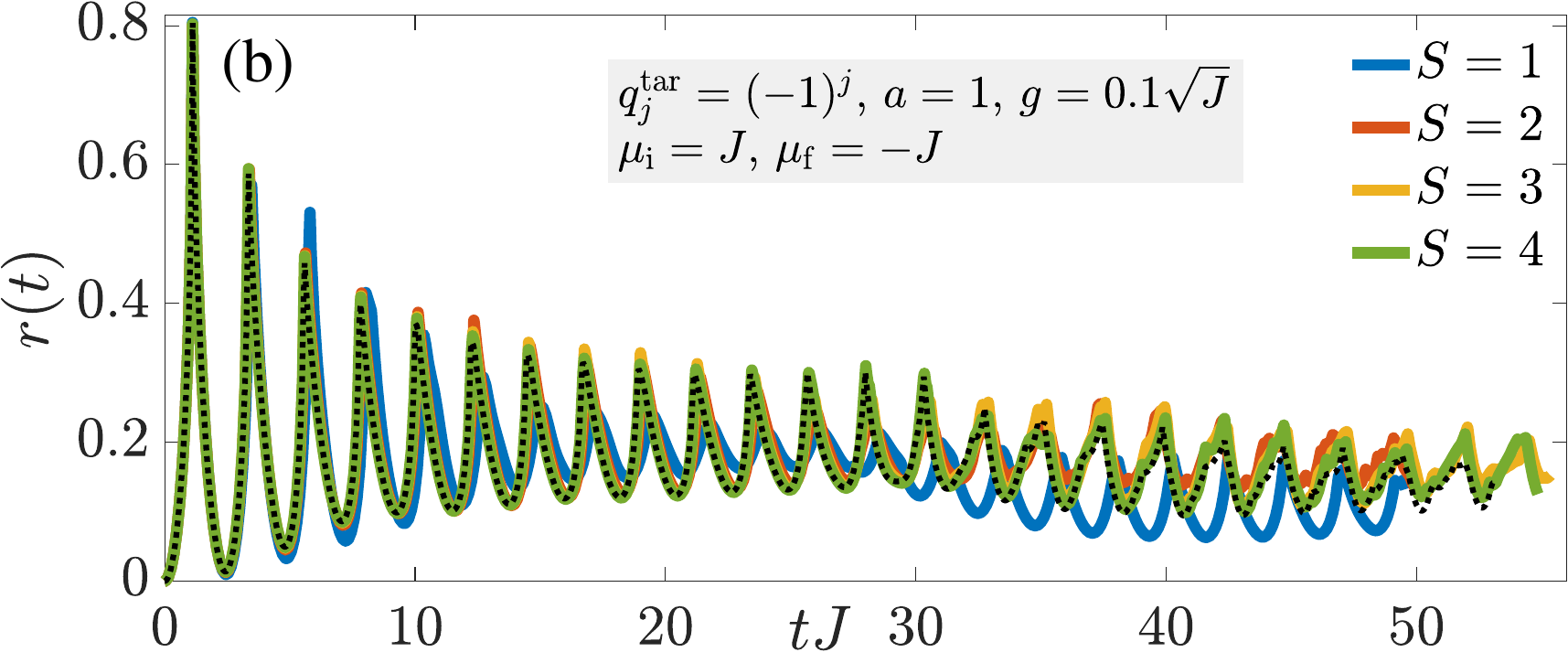}\\
	\vspace{1.1mm}
	\includegraphics[width=.48\textwidth]{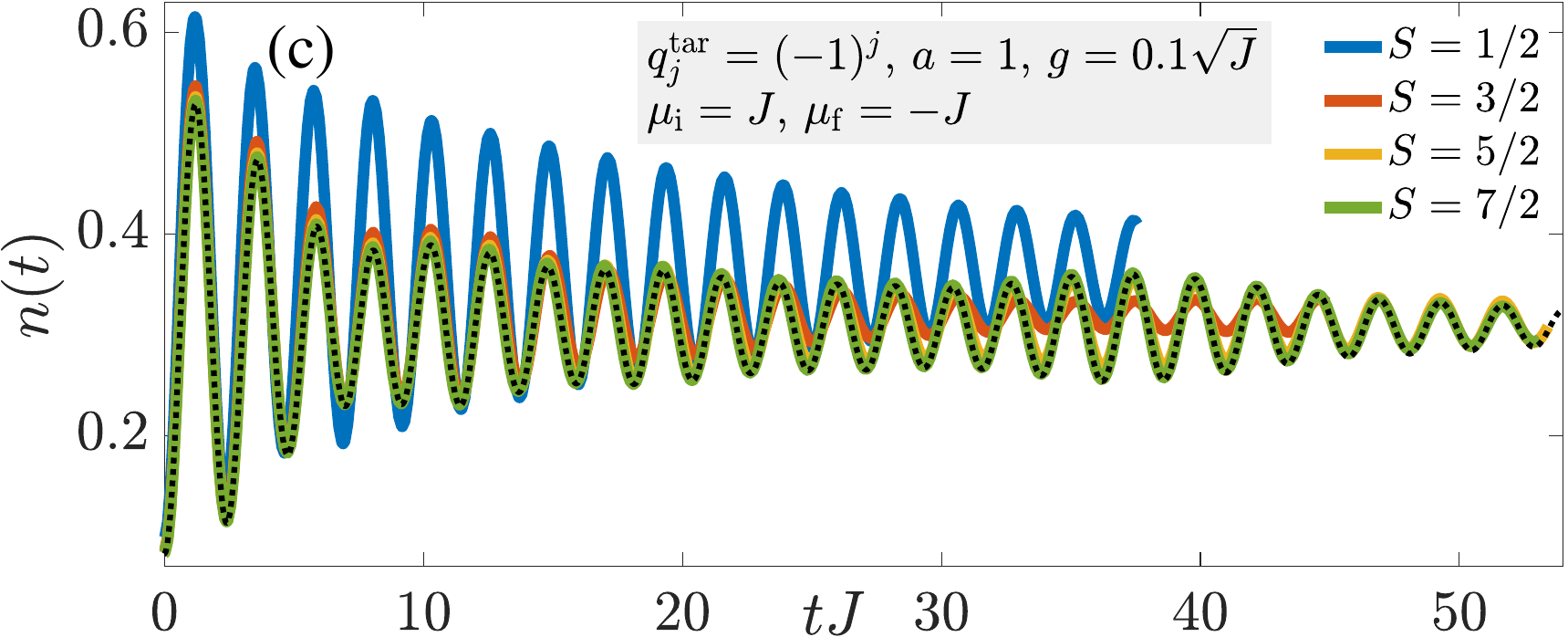}\quad\includegraphics[width=.48\textwidth]{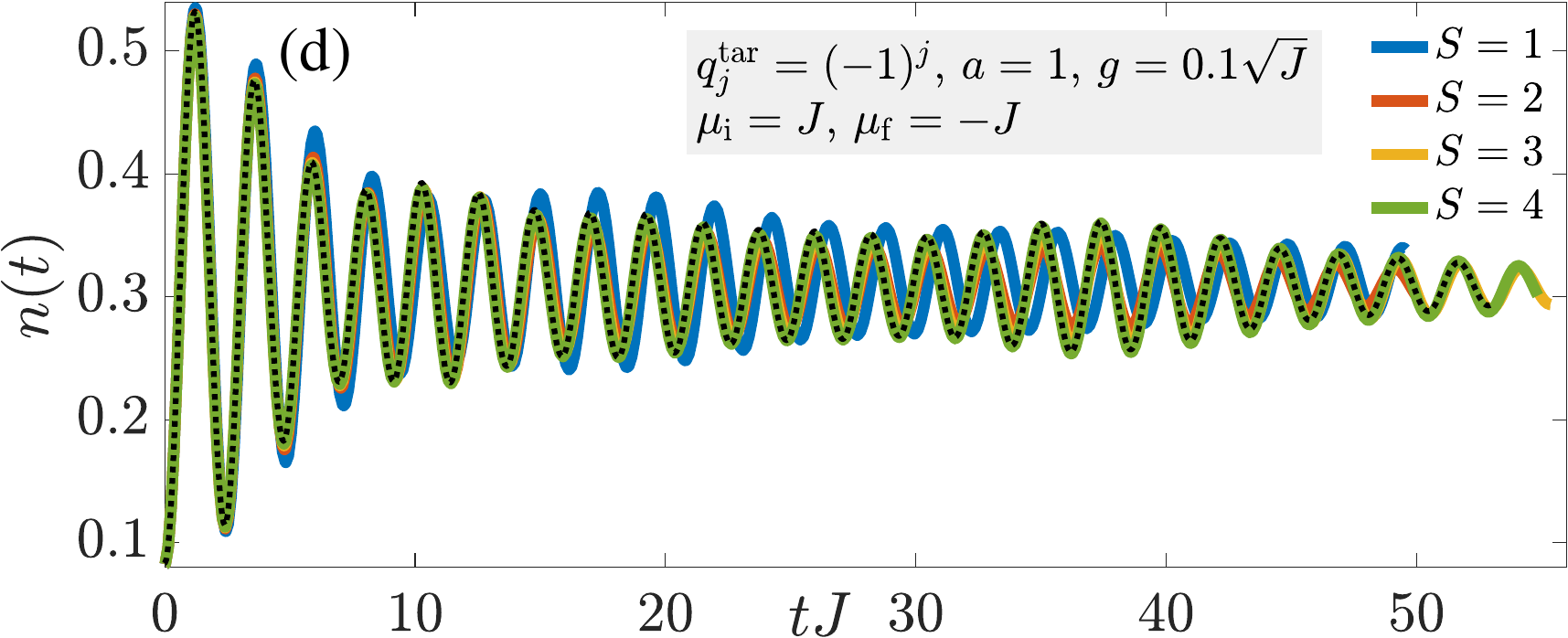}\\
	\vspace{1.1mm}
	\includegraphics[width=.48\textwidth]{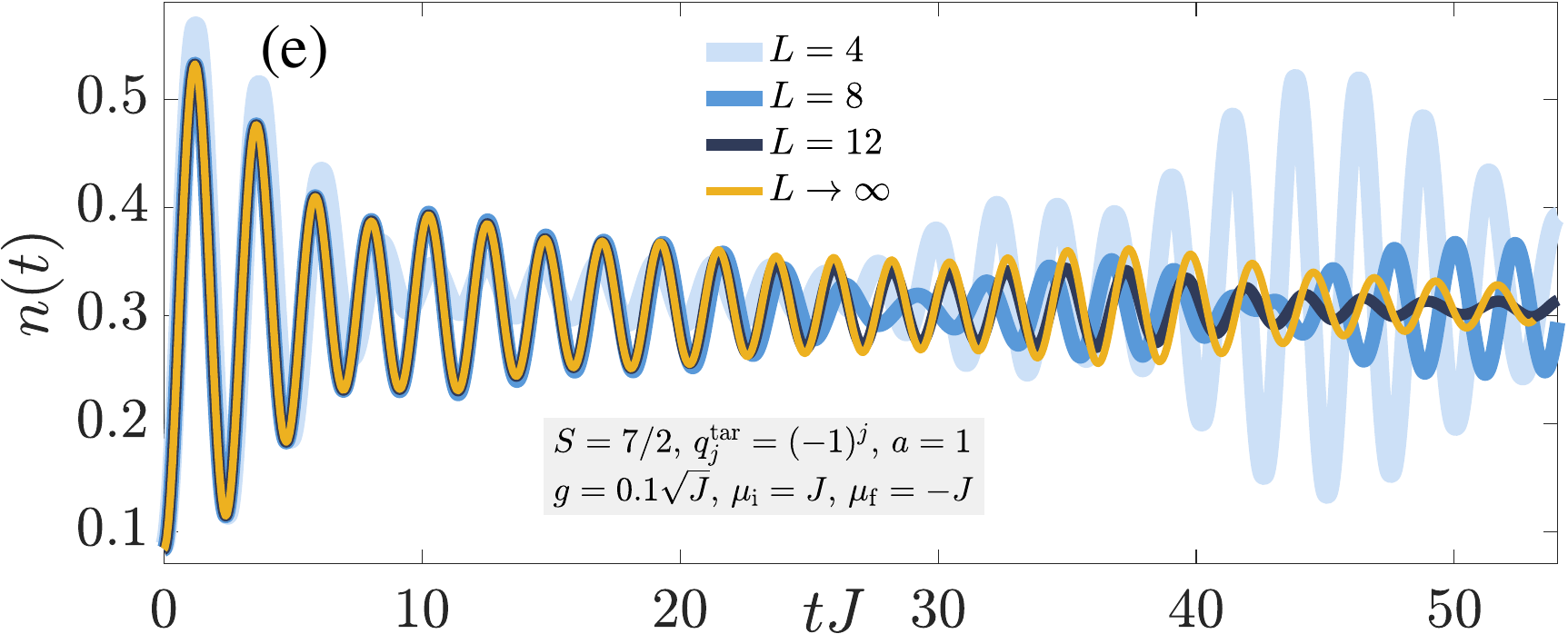}\quad\includegraphics[width=.48\textwidth]{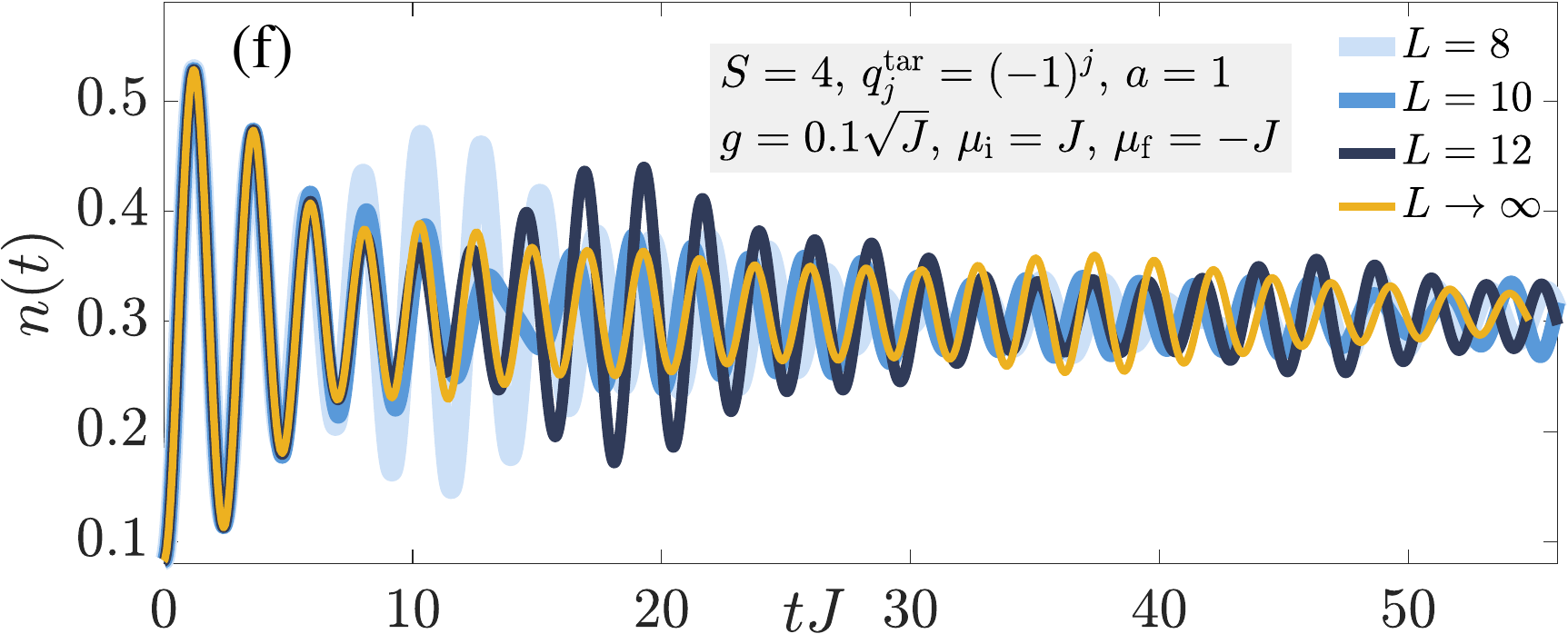}
	\caption{(Color online). Quench dynamics of the spin-$S$ $\mathrm{U}(1)$ quantum link model~\eqref{eq:H} starting in its ground state at $\mu=J$ and quenching the mass to a final value $\mu=-J$ in the weak-coupling regime at $g=0.1\sqrt{J}$ within the gauge superselection sector $q_j^\text{tar}=(-1)^j$. Similarly to the corresponding case of the superselection sector $q_j^\text{tar}=0$ shown in Fig.~\ref{fig:Sscaling_q0_g0.1_a1}, (a,b) the return rate and (c,d) the chiral condensate show very fast convergence to the Wilson--Kogut--Susskind limit for both (a,c) half-integer and (b,d) integer link spin length $S$, with the latter showing slightly faster convergence. Furthermore, the return rate and chiral condensate for $S=7/2$ and $S=4$ show excellent quantitative agreement over all accessible evolution times, as indicated by the reference dotted black lines in (a-d), which represent the corresponding dynamics for $S=4$ in (a,c) and for $S=7/2$ in (b,d). The main difference from the corresponding case of the superselection sector $q_j^\text{tar}=0$ shown in Fig.~\ref{fig:Sscaling_q0_g0.1_a1} is the convergence to the thermodynamic limit, where here the latter is approached much faster in the quench dynamics of the chiral condensate for (e) half-integer $S$ rather than (f) integer $S$. The finite-size results are obtained from exact diagonalization.}
	\label{fig:Sscaling_q1_g0.1_a1} 
\end{figure*}

The same picture repeats itself in the quench dynamics of the chiral condensate, shown in Fig.~\ref{fig:Sscaling_q0_g2_a1}(c,d) for the case of half-integer and integer $S$, respectively. Here, too, the WKS limit is achieved rapidly already at small half-integer and integer $S$, albeit in the latter case the convergence is much faster, with already $S=1$ achieving this limit for all accessible evolution times. This can be explained by noting that at large $g^2\gg1$, quantum fluctuations are suppressed, and the low-energy structure of the link operators is now revealed. In the case of integer $S$, the zero-eigenvalue state of the electric-field operator $\hat{s}^z_{j,j+1}$ dominates at any $S$, which leads to fast convergence to the WKS limit already at integer $S\gtrsim1$. Even though the chiral condensate behaves qualitatively the same for half-integer and integer spin, quantitatively they are much different, with the reference dotted black lines again highlighting this. This is again attributed to quantum fluctuations, which are not as suppressed in the case of half-integer $S$. 

As such, we see that in the target superselection sector $q_j^\text{tar}=0,\,\forall j$, the WKS limit is very rapidly achieved even in the strong-coupling regime. We turn now to the approach to the thermodynamic limit of the quench dynamics of the chiral condensate, shown in Fig.~\ref{fig:Sscaling_q0_g2_a1}(e,f) for the cases of $S=7/2$ and $S=4$, respectively. In both cases, we see a rather slow approach to the thermodynamic limit at intermediate to late evolution times, although at early times the largest system sizes that we employ in ED ($L=8,12$ matter sites) show rather good convergence to the thermodynamic limit. Therefore, we conclude that in this superselection sector the convergence to the thermodynamic limit of our considered quenched dynamics is slow in the strong-coupling regime compared to the case of weak coupling, where there at least for integer $S$ the convergence to the thermodynamic limit is very good.

\begin{figure*}[t!]
	\centering
	\includegraphics[width=.48\textwidth]{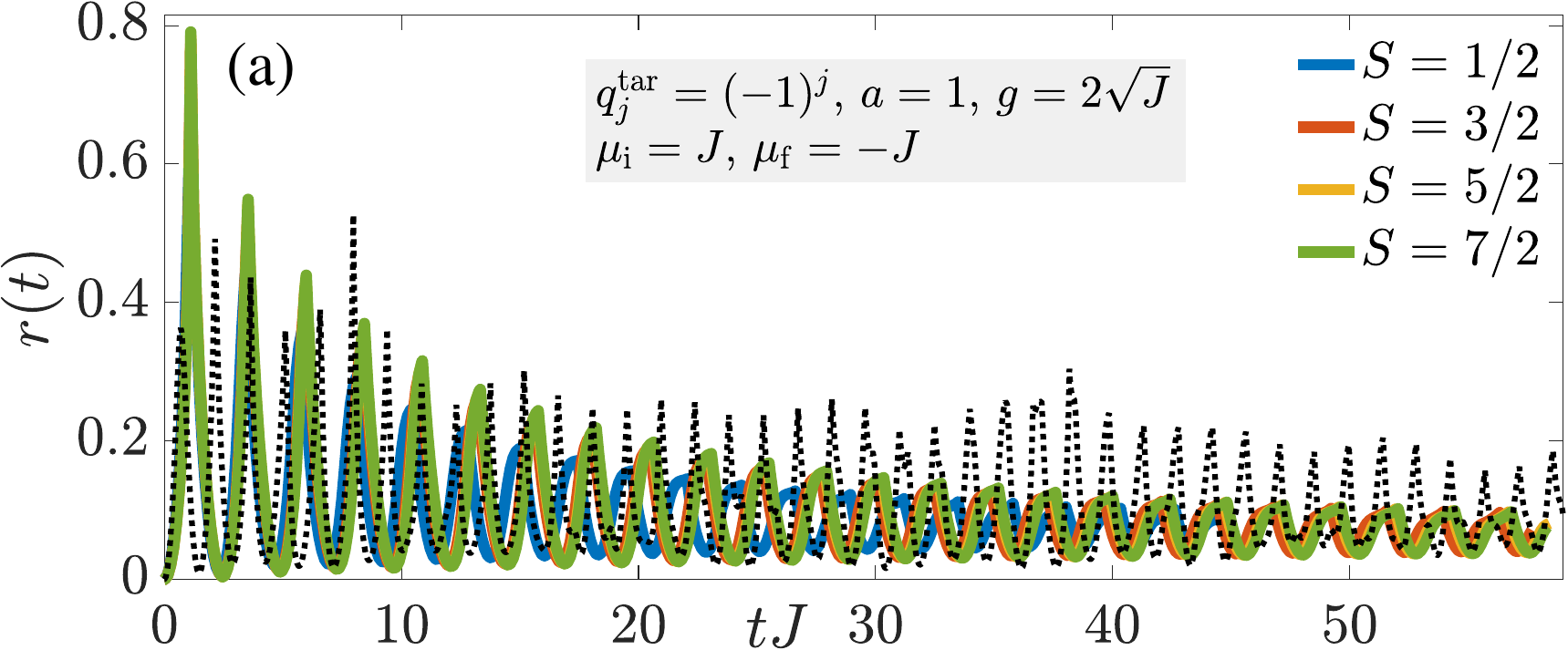}\quad\includegraphics[width=.48\textwidth]{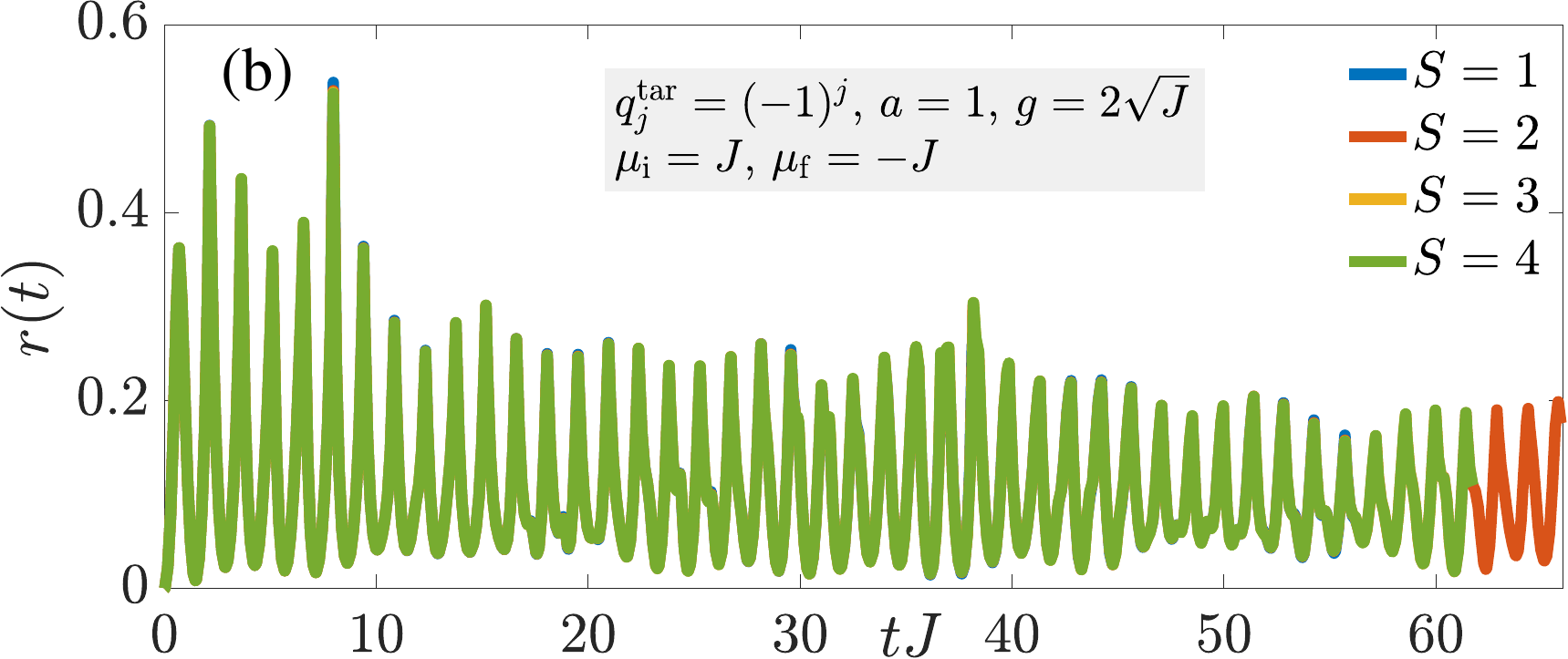}\\
	\vspace{1.1mm}
	\includegraphics[width=.48\textwidth]{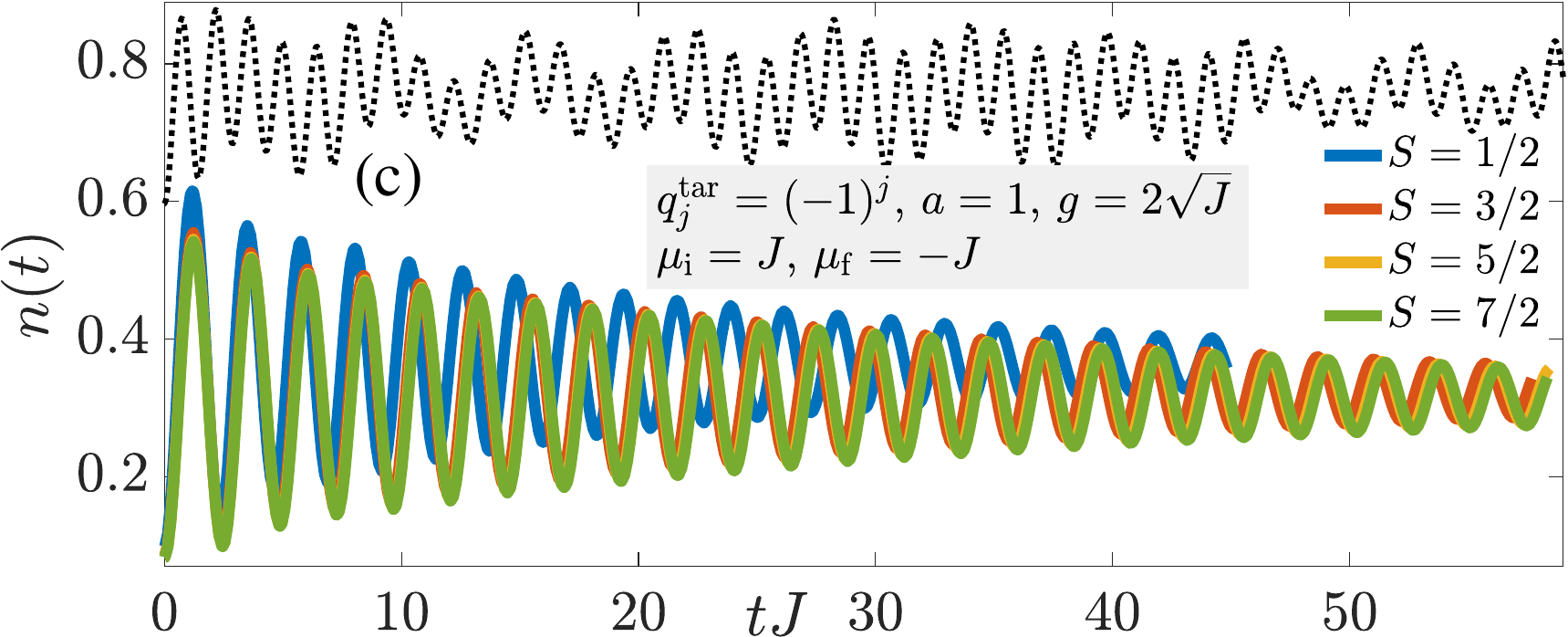}\quad\includegraphics[width=.48\textwidth]{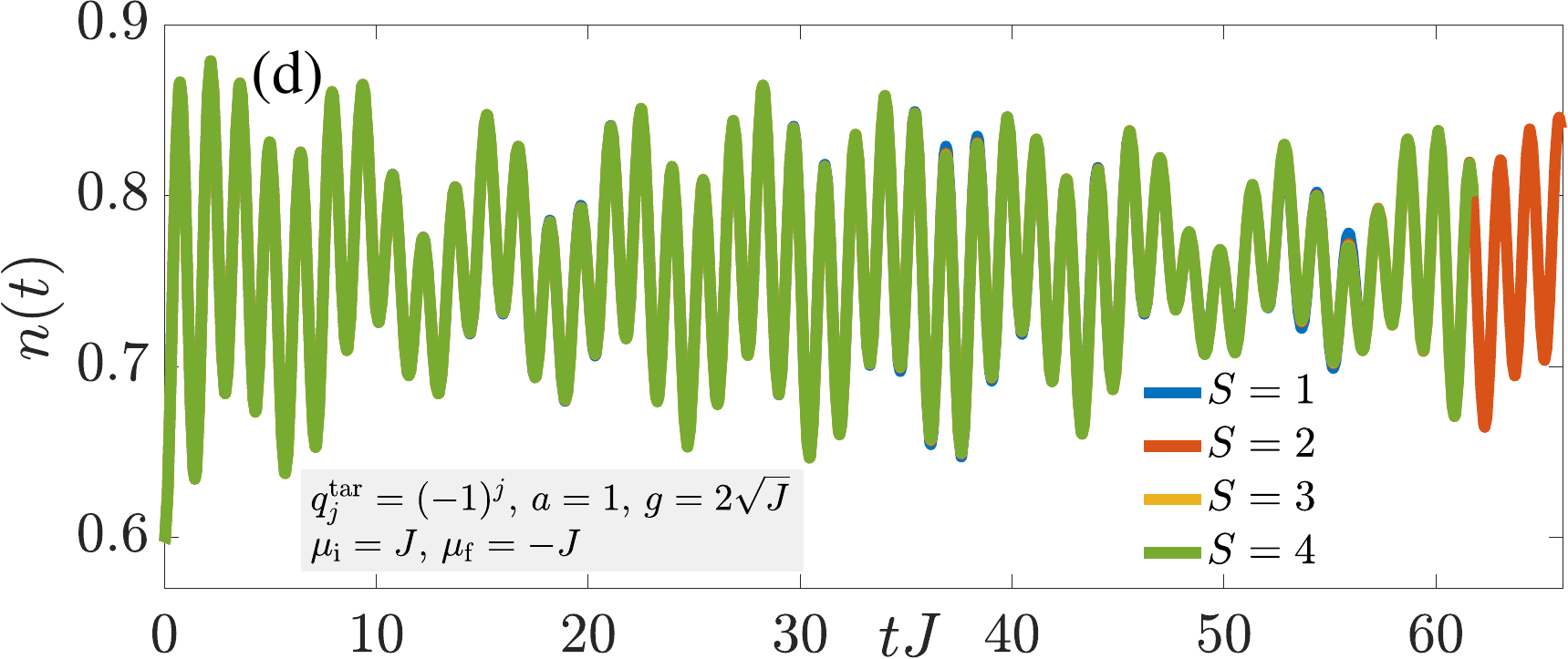}\\
	\vspace{1.1mm}
	\includegraphics[width=.48\textwidth]{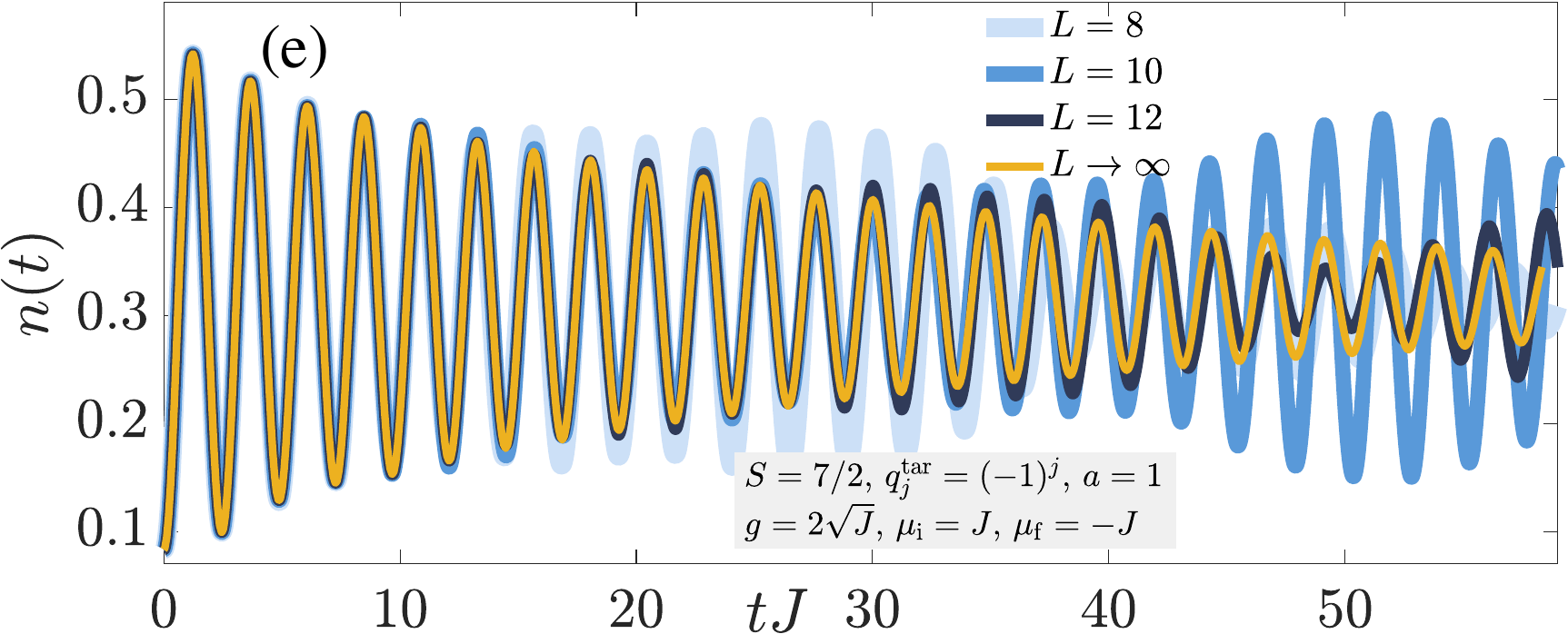}\quad\includegraphics[width=.48\textwidth]{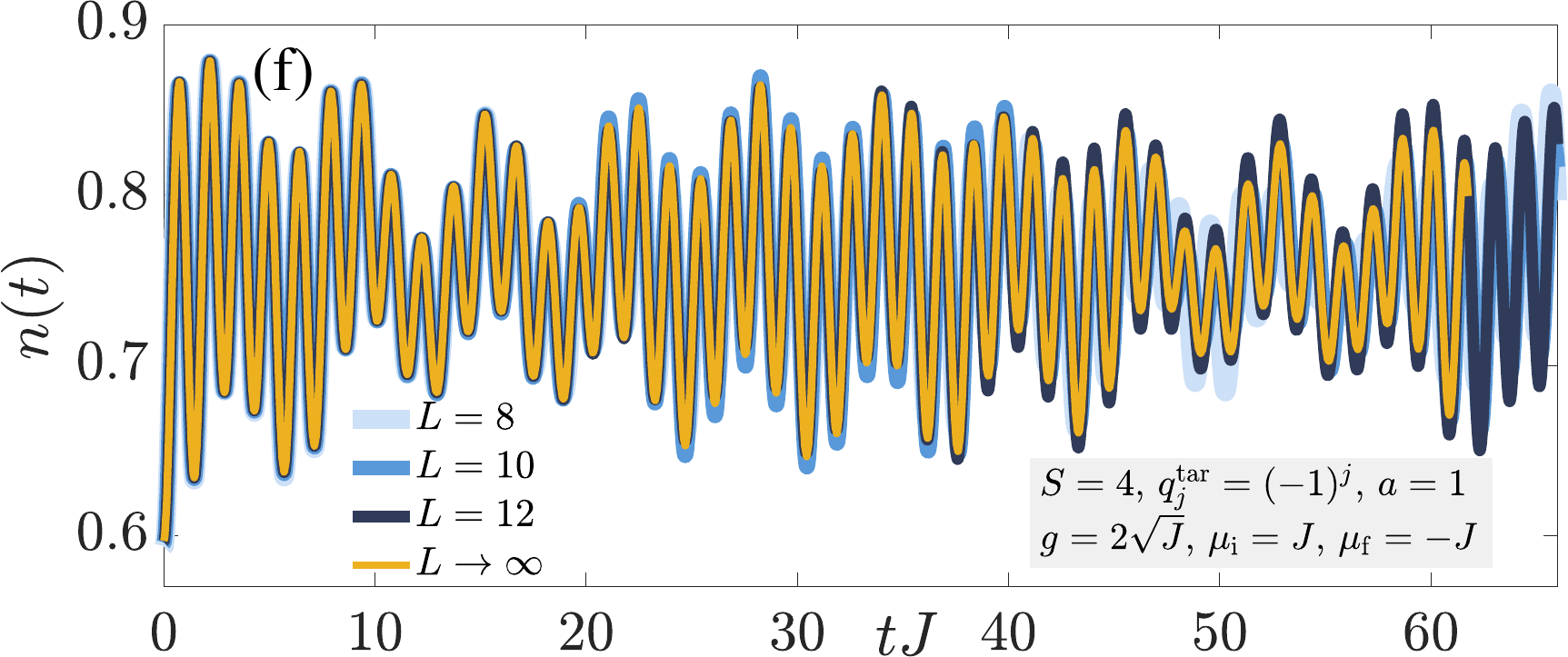}
	\caption{(Color online). Same as Fig.~\ref{fig:Sscaling_q1_g0.1_a1} but in the strong-coupling regime with $g=2\sqrt{J}$. As in all our results so far, we see very fast convergence to the Wilson--Kogut--Susskind limit for (a,b) the return rate and (c,d) the chiral condensate in the case of (a,c) half-integer and (b,d) integer link spin length $S$. Similarly to the corresponding case in the target sector $q_j^\text{tar}=0$ of Fig.~\ref{fig:Sscaling_q0_g2_a1}, also in this superselection sector $q_j^\text{tar}=(-1)^j$ the convergence to the Wilson--Kogut--Susskind limit is strikingly fast for integer $S$, where (b) the return rate and (d) the chiral condensate seem to achieve this limit already at $S=1$. Also here, we find that there is large quantitative disagreement between half-integer and integer $S$, as indicated by the reference dotted black lines, which represent the corresponding dynamics for $S=4$ in (a,c) and for $S=7/2$ in (b,d). However, differently from the corresponding case in the target sector $q_j^\text{tar}=0$, in this target sector $q_j^\text{tar}=(-1)^j$ the convergence to the thermodynamic limit is very fast for both half-integer and integer $S$, as shown in the quench dynamics of the chiral condensate in (e,f), respectively.}
	\label{fig:Sscaling_q1_g2_a1} 
\end{figure*}

\subsection{Target sector $q_j^\text{tar}=(-1)^j$}\label{sec:q1}
We now repeat the numerical calculations of Sec.~\ref{sec:q0} in a different gauge superselection sector. We now choose the target sector to be $q_j^\text{tar}=(-1)^j$, i.e., we consider the Hilbert subspace of eigenstates satisfying $\hat{G}_j\ket{E_n}=(-1)^j\ket{E_n}$.

\subsubsection{Weak-coupling regime}
We first consider the case of weak electric-field coupling and set $g=0.1\sqrt{J}$. The corresponding quench dynamics of the return rate and chiral condensate are presented in Fig.~\ref{fig:Sscaling_q1_g0.1_a1}. The time evolution of the return rate approaches the WKS limit very rapidly for the small values of $S\leq4$ we consider, as shown in Fig.~\ref{fig:Sscaling_q1_g0.1_a1}(a,b) for half-integer and integer $S$, respectively. We also find that in the WKS limit, the quench dynamics of the return rate are quantitatively identical whether $S$ is half-integer or integer. This can be seen by the dotted black lines in Fig.~\ref{fig:Sscaling_q1_g0.1_a1}(a,b) representing the return rate for $S=4$ and $S=7/2$, respectively.

Turning to the chiral-condensate quench dynamics in Fig.~\ref{fig:Sscaling_q1_g0.1_a1}(c,d) for half-integer and integer $S$, we deduce the same conclusions. The WKS limit is achieved over the small values of $S$ we consider, and there is great quantitative agreement in this limit between the case of half-integer and integer $S$, as indicated by the reference dotted black lines representing the chiral condensate for $S=4$ and $S=7/2$, respectively.

Therefore, we see that little has changed in terms of the approach to the WKS limit by changing the target sector from the superselection sector $q_j^\text{tar}=0$ to $q_j^\text{tar}=(-1)^j$. However, when it comes to the approach to the thermodynamic limit, it turns out that here the case of half-integer $S$ fares much better than its integer counterpart, as can be seen in Fig.~\ref{fig:Sscaling_q1_g0.1_a1}(e,f). The ED calculations of the quench dynamics of the chiral condensate in the case of half-integer $S$ shows excellent convergence to the thermodynamic limit up to very long evolution times already at $L=12$ matter sites in Fig.~\ref{fig:Sscaling_q1_g0.1_a1}(e). In the case of integer $S$, on the other hand, this convergence occurs only at early times after which deviations become significant, see Fig.~\ref{fig:Sscaling_q1_g0.1_a1}(f).

Interestingly, we find that the quench dynamics of the return rate and chiral condensate in Figs.~\ref{fig:Sscaling_q1_g0.1_a1} and~\ref{fig:Sscaling_q0_g0.1_a1} show great quantitative agreement in the WKS limit. This seems to indicate that at least in the weak-coupling regime, the quench dynamics in the WKS limit is independent of the choice of target superselection sector between $q_j=0$ and $q_j=(-1)^j$, and does not depend on whether $S$ is half-integer or integer.

\begin{figure*}[t!]
	\centering
	\includegraphics[width=.48\textwidth]{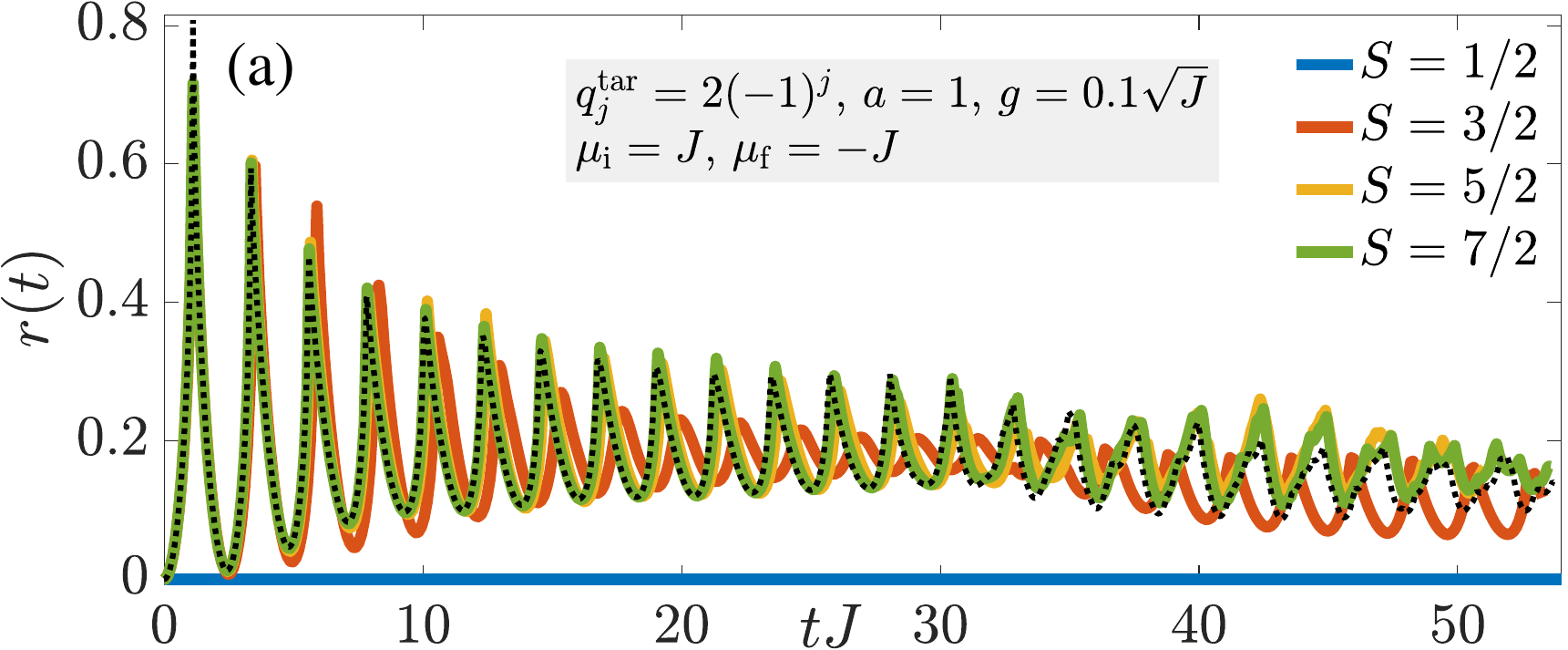}\quad\includegraphics[width=.48\textwidth]{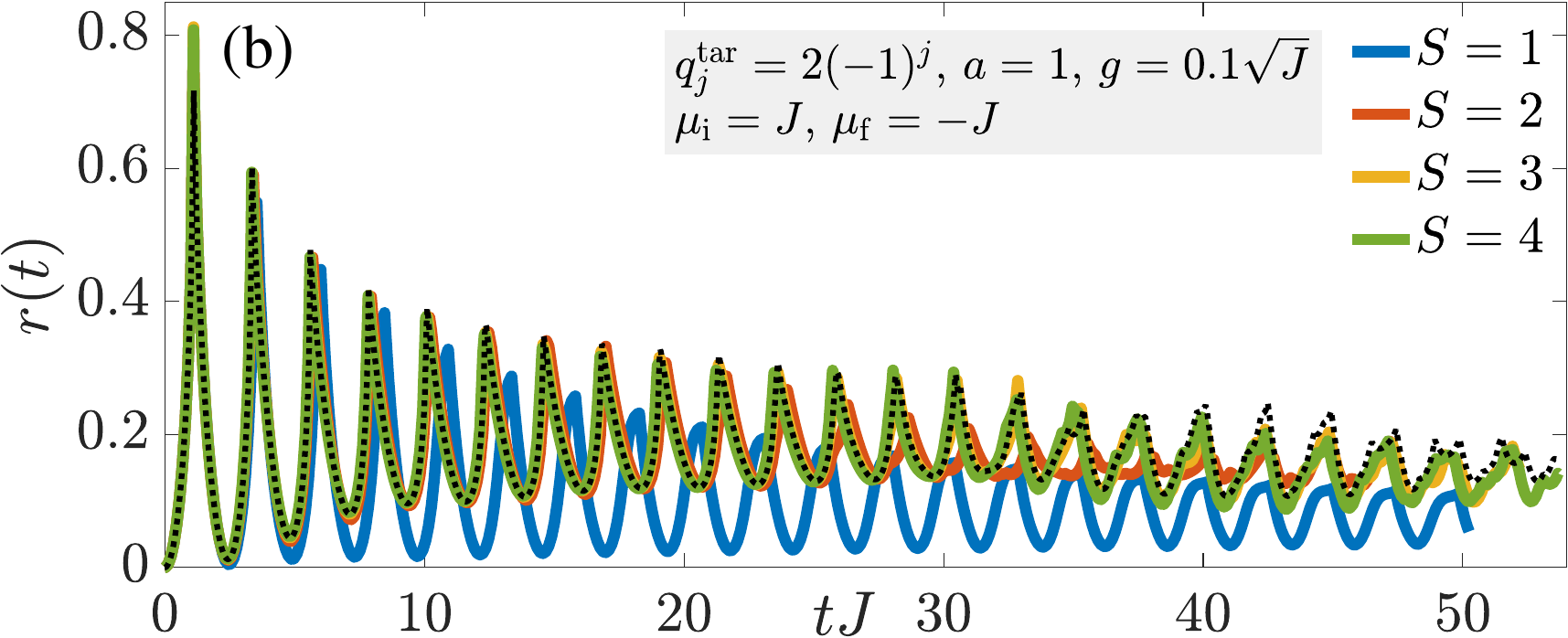}\\
	\vspace{1.1mm}
	\includegraphics[width=.48\textwidth]{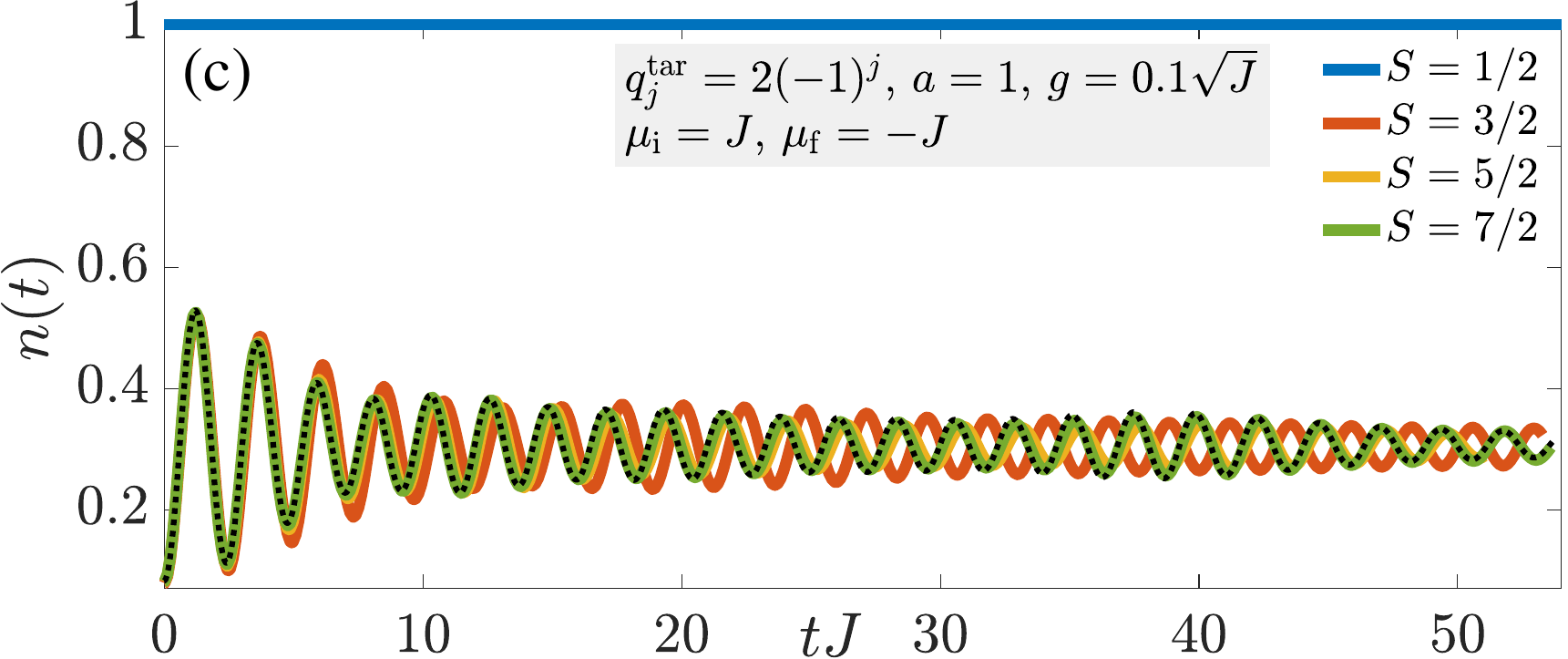}\quad\includegraphics[width=.48\textwidth]{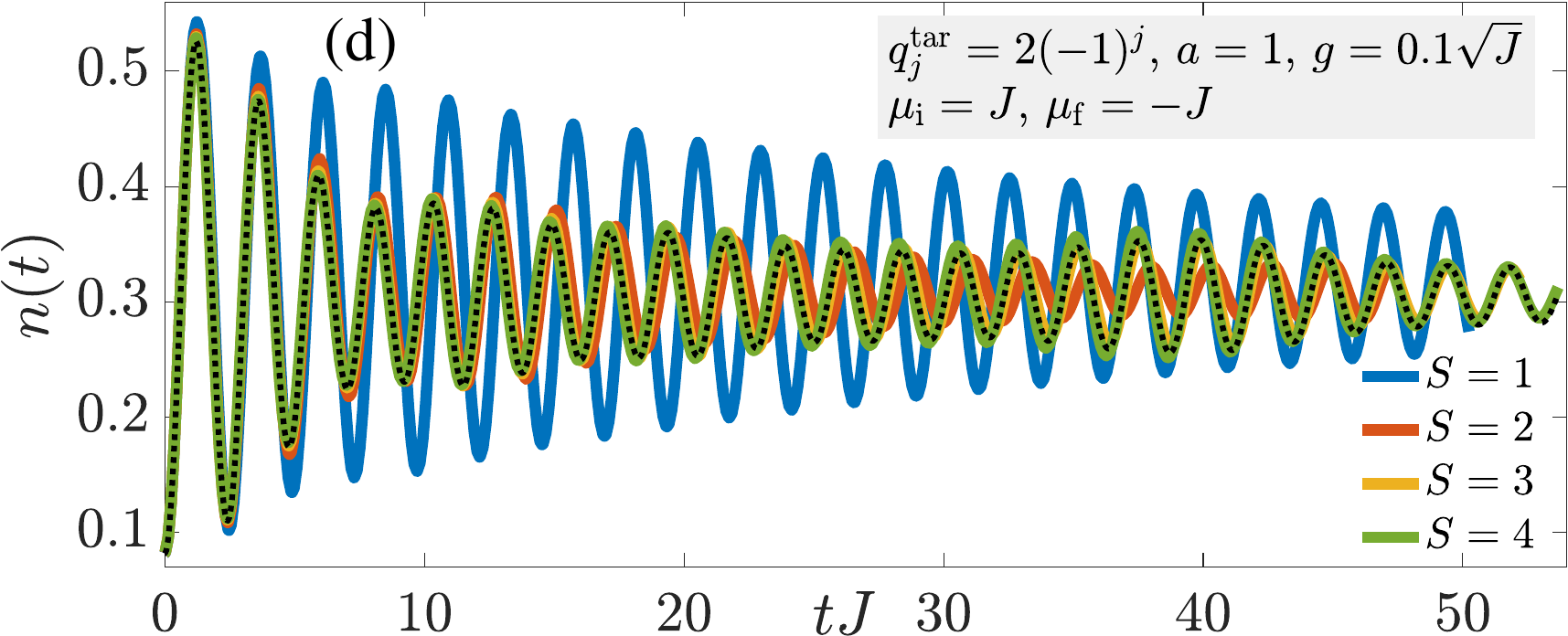}\\
	\vspace{1.1mm}
	\includegraphics[width=.48\textwidth]{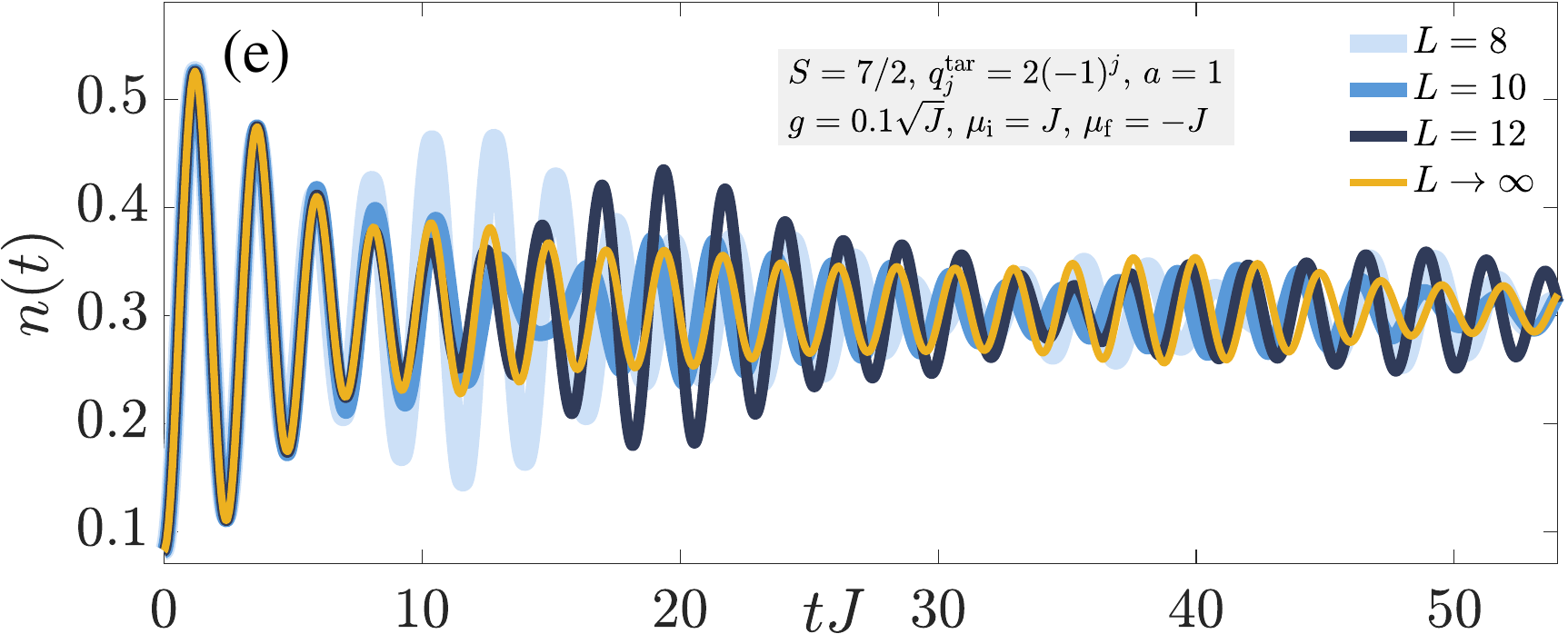}\quad\includegraphics[width=.48\textwidth]{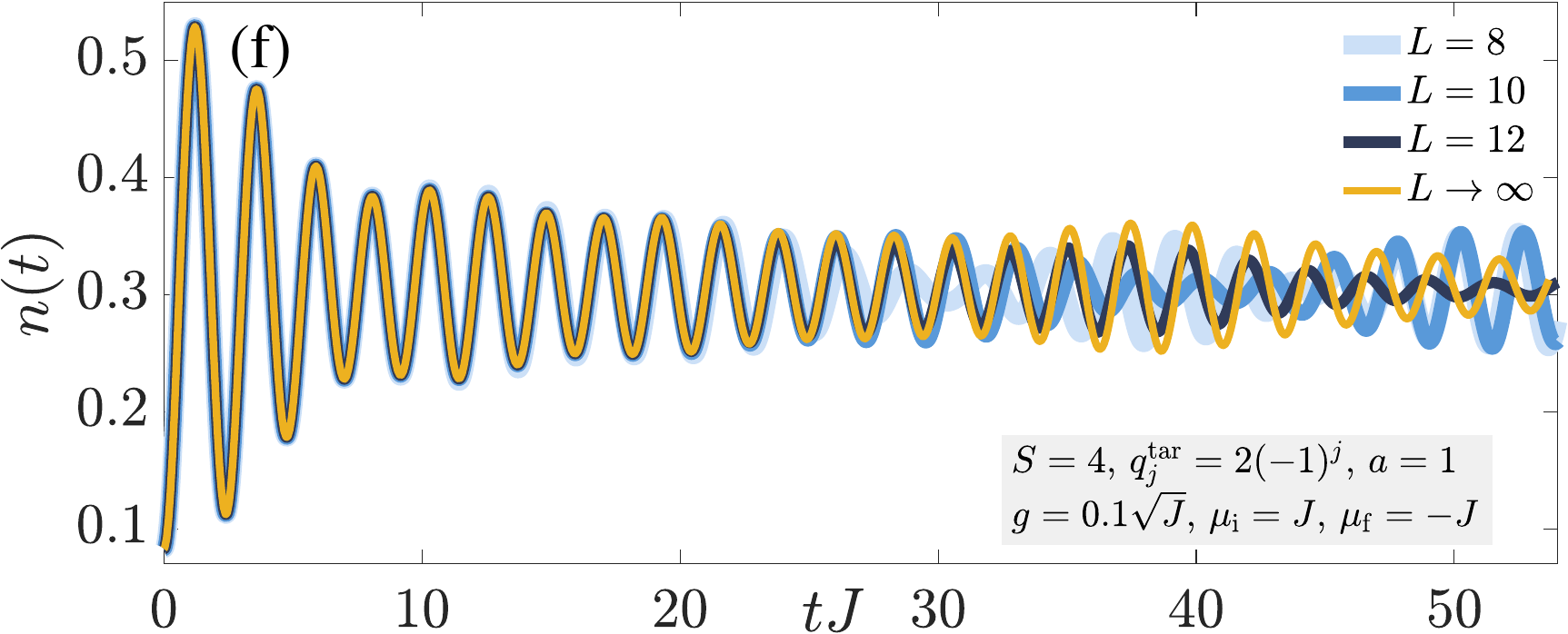}
	\caption{(Color online). Quench dynamics of the return rate~\eqref{eq:RR} and chiral condensate~\eqref{eq:n} in the spin-$S$ $\mathrm{U}(1)$ quantum link model starting in its ground state at $\mu=J$ in the weak-coupling regime with $g=0.1\sqrt{J}$ and quenching the mass to $\mu=-J$ within the gauge superselection sector $q_j^\text{tar}=2(-1)^j$. Similarly to our other conclusions thus far, we find that (a,b) the return rate and (c,d) the chiral condensate rapidly converge to the Wilson--Kogut--Susskind limit for both (a,c) half-integer and (b,d) integer link spin length $S$. In the case of $S=1/2$ we see no dynamics in either (a) the return rate or (c) the chiral condensate, and this is because in this superselection sector there is only one possible state in the case of $S=1/2$ (see text). As we have observed previously in the weak-coupling regime, also here there is very good quantitative agreement between the dynamics of half-integer and integer $S$, as indicated by the reference dotted black lines in (a-d), which represent the corresponding dynamics for $S=4$ in (a,c) and for $S=7/2$ in (b,d). The approach to the thermodynamic limit of the quench dynamics of the chiral condensate is shown in (e) for the case of half-integer $S$ and in (f) for the case of integer $S$, where the latter is much faster, similarly to the corresponding case of the superselection sector $q_j^\text{tar}=0$.}
	\label{fig:Sscaling_q2_g0.1_a1} 
\end{figure*}

\subsubsection{Strong-coupling regime}
We now repeat this quench in the strong-coupling regime at $g=2\sqrt{J}$, the results of which are shown in Fig.~\ref{fig:Sscaling_q1_g2_a1}. As in every case we have considered so far, we see very fast convergence to the WKS limit in the quench dynamics of the return rate~\eqref{eq:RR} for both half-integer and integer $S$, as shown in Fig.~\ref{fig:Sscaling_q1_g2_a1}(a,b), respectively. The convergence is again extremely rapid for integer $S$, where we see in Fig.~\ref{fig:Sscaling_q1_g2_a1}(b) that already at $S=1$ the WKS limit is achieved for all accessible evolution times. Similar to other cases in the strong-coupling regime, the quench dynamics of the return rate is very different depending on whether $S$ is half-integer or integer. This is made clear by overlaying the return rate for $S=4$ (dotted black line) in Fig.~\ref{fig:Sscaling_q1_g2_a1}(a), showing how it vastly differs from the WKS limit in the case of the half-integer $S$.

This picture is confirmed through the quench dynamics of the chiral condensate~\eqref{eq:n} in Fig.~\ref{fig:Sscaling_q1_g2_a1}(c,d) for both half-integer and integer $S$, respectively. In both cases, the WKS limit is quickly achieved within our small values of $S$, with the integer case already converging to that limit at $S=1$. As expected, the chiral condensate for $S=4$, depicted by a dotted black line in Fig.~\ref{fig:Sscaling_q1_g2_a1}(c), is drastically different from its counterpart for half-integer $S$.

We now investigate the approach to the thermodynamic limit in Fig.~\ref{fig:Sscaling_q1_g2_a1}(e,f) for half-integer and integer $S$, respectively. We find that in both cases, the convergence to this limit is excellent, with the finite-size result at $L=12$ matter sites, obtained from ED, faithfully reproducing the result in the thermodynamic limit, obtained from iMPS, over most of the accessible evolution times. This is sharp contrast to the strong-coupling regime for the superselection sector $q_j^\text{tar}=0$, see Fig.~\ref{fig:Sscaling_q0_g2_a1}(e,f), where the approach to the thermodynamic limit is quite slow regardless of whether $S$ is half-integer or integer.

\begin{figure*}[t!]
	\centering
	\includegraphics[width=.48\textwidth]{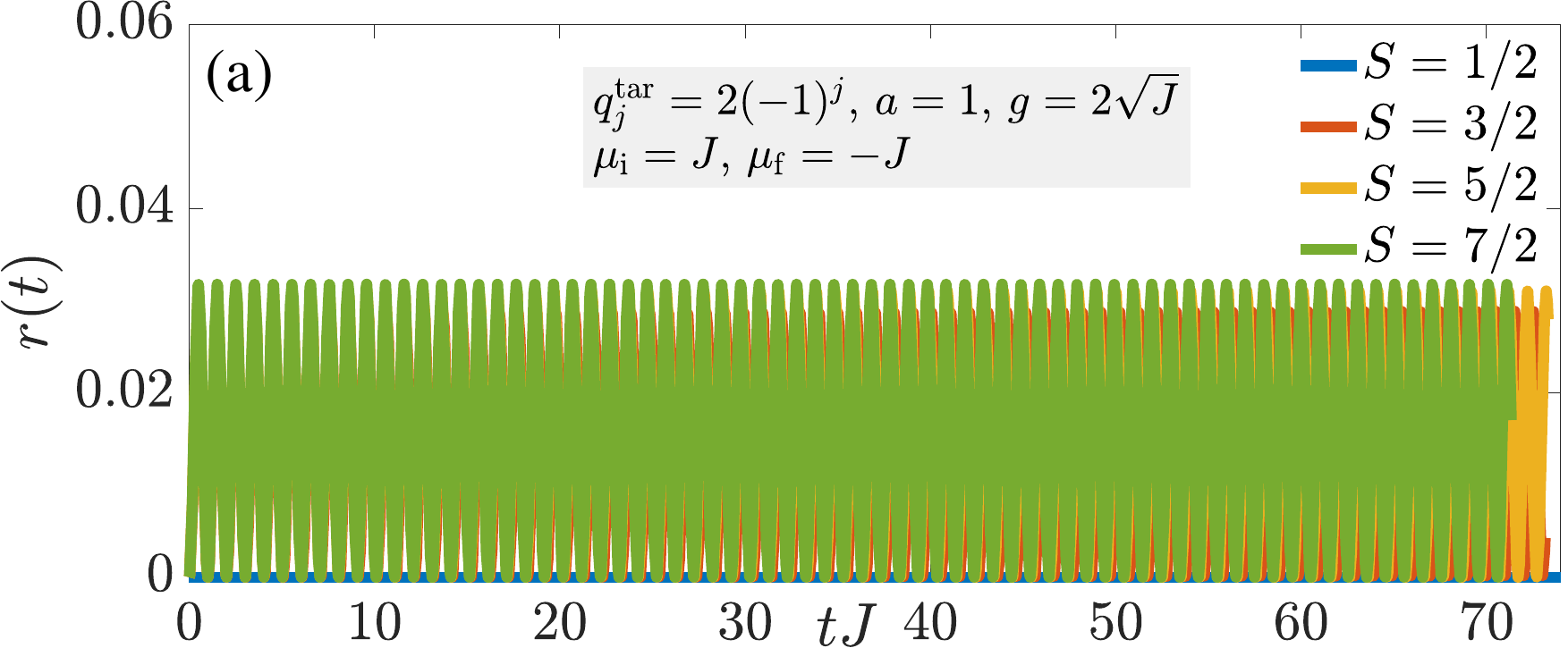}\quad\includegraphics[width=.48\textwidth]{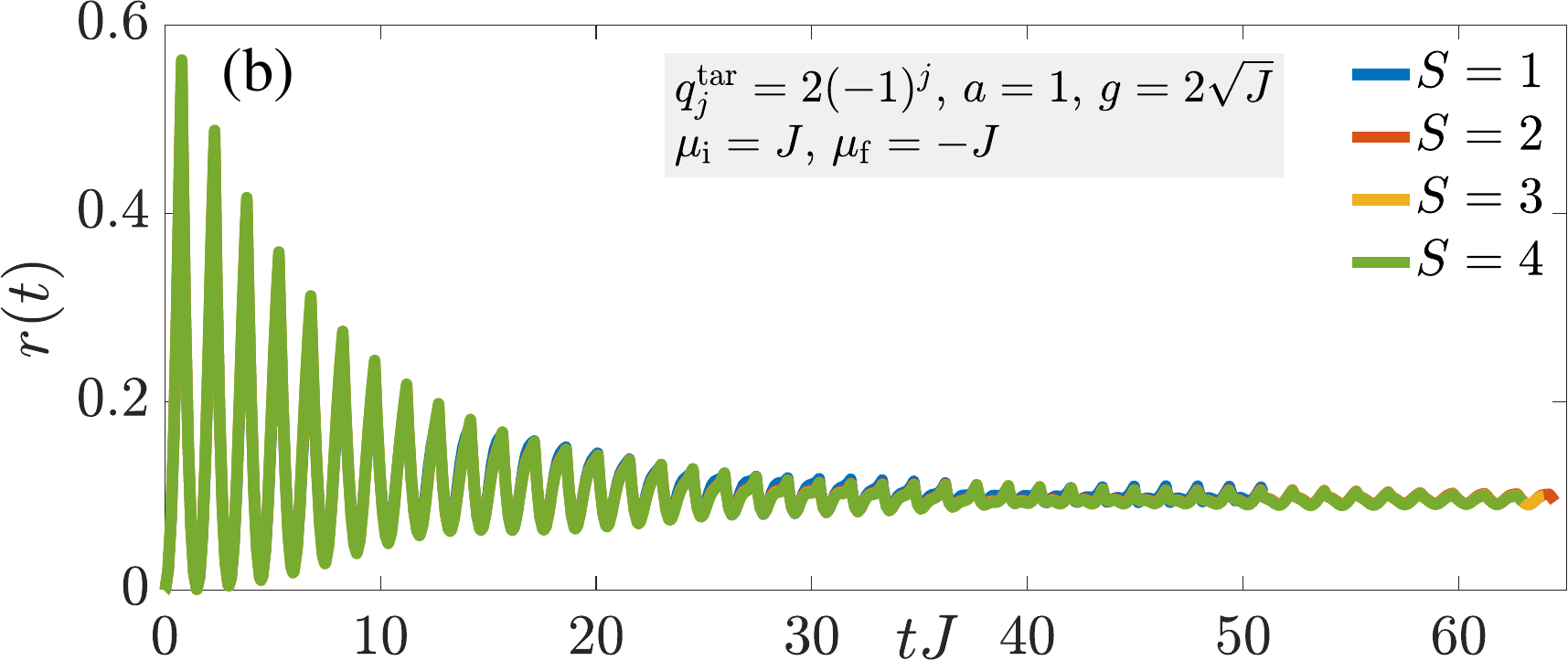}\\
	\vspace{1.1mm}
	\includegraphics[width=.48\textwidth]{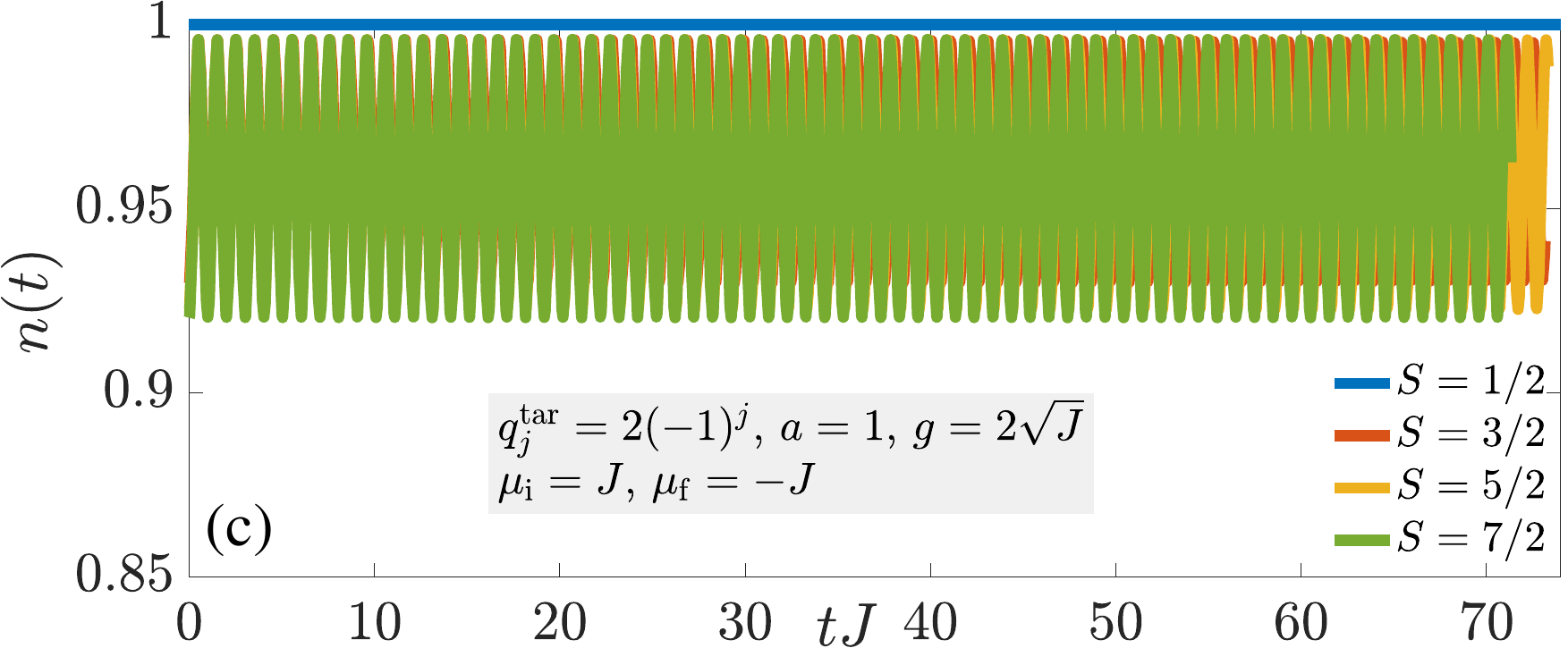}\quad\includegraphics[width=.48\textwidth]{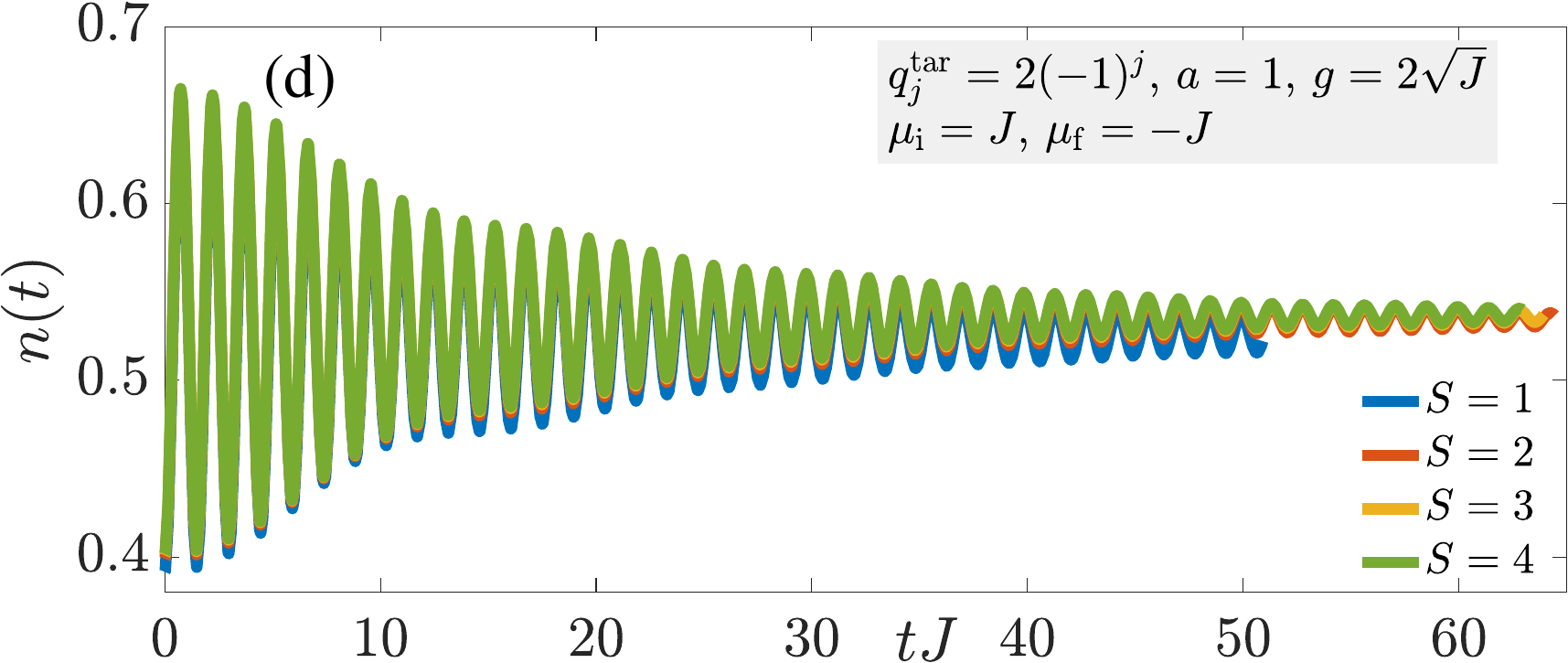}\\
	\vspace{1.1mm}
	\includegraphics[width=.48\textwidth]{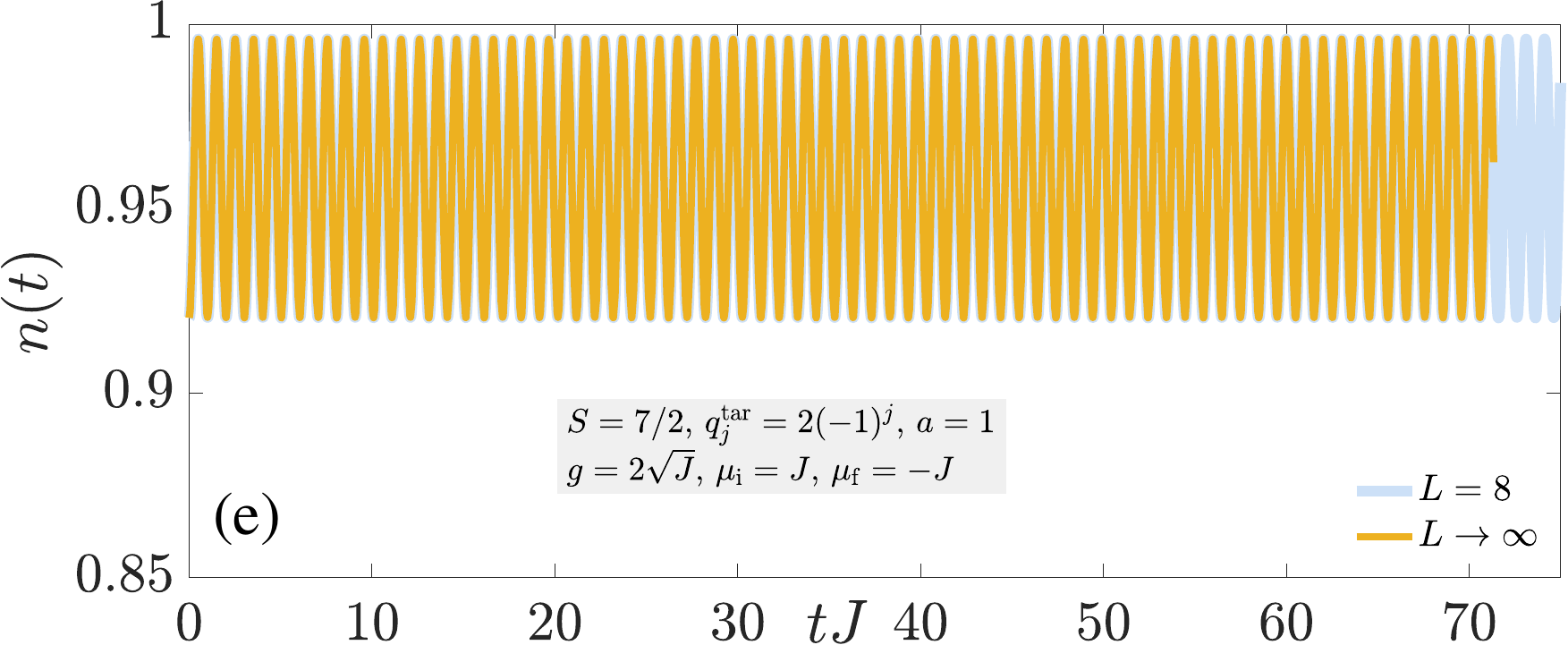}\quad\includegraphics[width=.48\textwidth]{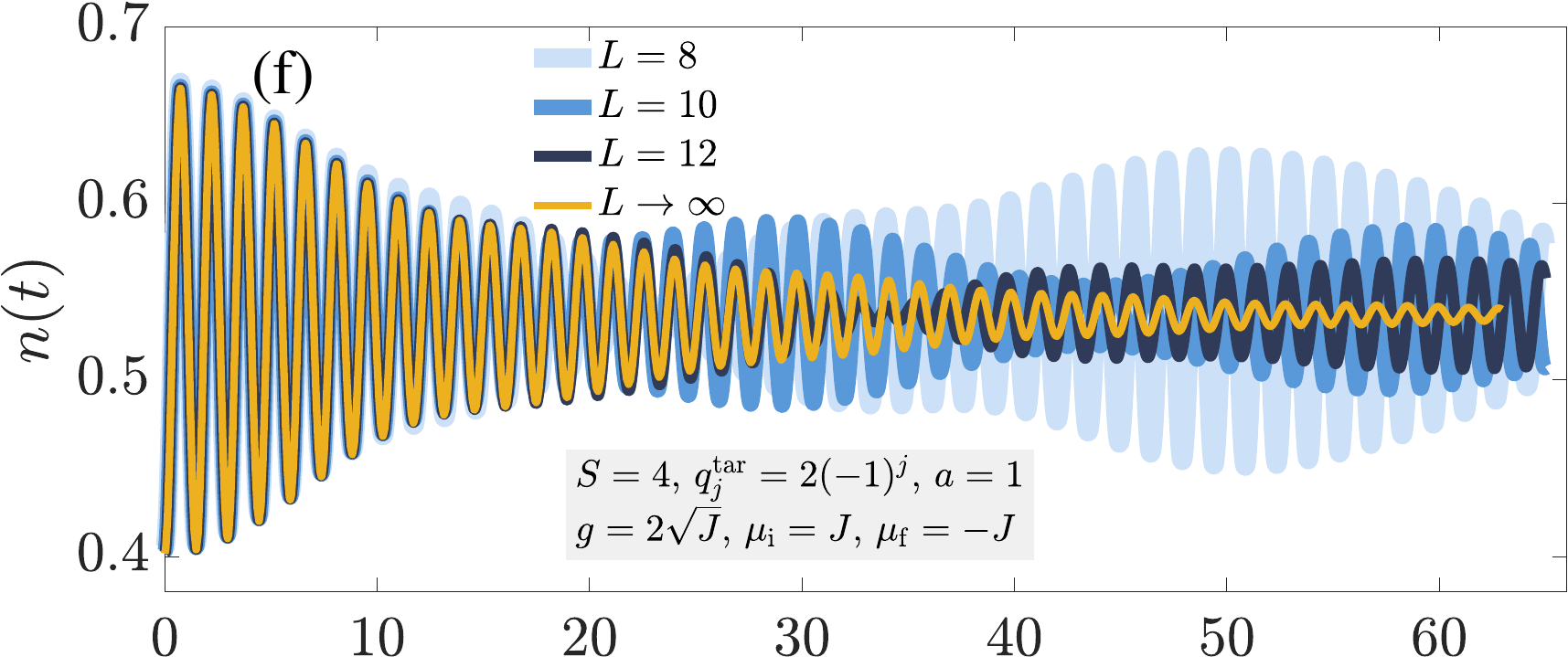}
	\caption{(Color online). Same as Fig.~\ref{fig:Sscaling_q2_g0.1_a1} but in the strong-coupling regime $g=2\sqrt{J}$. As has been the common conclusion of our results so far, the convergence to the Wilson--Kogut--Susskind limit is quite fast as seen in the quench dynamics of (a,b) the return rate and (c,d) the chiral condensate for both (a,c) half-integer and (b,d) integer $S$. However, as per our previous conclusions in this strong-coupling regime, the Wilson--Kogut--Susskind limit is converged to much faster for integer $S$. Also in agreement with previous observations in this regime, there is significant quantitative difference between the quench dynamics of (a,c) half-integer and (b,d) integer $S$. The approach to the thermodynamic limit looks excellent in the quench dynamics of the chiral condensate for (e) half-integer $S$ at all accessible evolution times, while for (f) integer $S$ the thermodynamic limit is achieved up to quite long evolution times already for a finite system size of $L=12$ matter sites. We note here, however, that the quench dynamics for the case of half-integer $S$ is somewhat trivial, since the initial state has large overlap with the ground state of the quench Hamiltonian.}
	\label{fig:Sscaling_q2_g2_a1} 
\end{figure*}

\subsection{Target sector $q_j^\text{tar}=2(-1)^j$}\label{sec:q2}
We now consider a third and final target superselection sector, which is $q_j^\text{tar}=2(-1)^j$. This means that we will consider only the Hilbert subspace of eigenstates satisfying $\hat{G}_j\ket{E_n}=2(-1)^j\ket{E_n}$.

\subsubsection{Weak-coupling regime}
The quench dynamics of the return rate and chiral condensate within this target superselection sector in the weak-coupling regime are shown in Fig.~\ref{fig:Sscaling_q2_g0.1_a1}. As in all cases considered so far, the convergence to the WKS limit is very fast over the small values of $S$ we consider in our numerical simulations, and this is evident in the quench dynamics of the return rate in Fig.~\ref{fig:Sscaling_q2_g0.1_a1}(a,b) and the chiral condensate in Fig.~\ref{fig:Sscaling_q2_g0.1_a1}(c,d) for both half-integer and integer $S$, respectively. Also in agreement with our earlier observations on the weak-coupling regime, in the WKS limit there is very good quantitative agreement in the quench dynamics between half-integer and integer $S$, as shown in the dotted black lines for $S=4$ in Fig.~\ref{fig:Sscaling_q2_g0.1_a1}(a,c) and for $S=7/2$ in Fig.~\ref{fig:Sscaling_q2_g0.1_a1}(b,d). It is worth noting here the lack of dynamics for the case of $S=1/2$ in Fig.~\ref{fig:Sscaling_q2_g0.1_a1}(a,c). This is because in this superselection sector the only allowed configuration at a local constraint defined by Eq.~\eqref{eq:Gj} is $\uparrow\bullet\uparrow$ in our particle-hole-transformed picture. In words, this configuration means that at every local constraint the matter site is occupied and the electric fields at the adjacent links are both pointing along the positive $z$-direction. Any other configuration will violate $\hat{G}_j\ket{E_n}=2(-1)^j\ket{E_n}$. As such, the system will always be in this initial state $\ket{\psi_0}=\ket{\ldots\uparrow\bullet\uparrow\bullet\uparrow\bullet\uparrow\ldots}$, leading to a zero return rate and a unity chiral condensate for all times.

We now study the approach to the thermodynamic limit through ED calculations in Fig.~\ref{fig:Sscaling_q2_g0.1_a1}(e,f), where we find that it depends on the nature of the link spin length $S$. When the latter is half-integer, we find that convergence to the thermodynamic limit is achieved by small system sizes of up to $L=12$ matter sites up to only relatively short evolution times, see Fig.~\ref{fig:Sscaling_q2_g0.1_a1}(e). In the case of integer $S$, on the other hand, we see that the same system size faithfully reproduces the thermodynamic limit up to considerably long evolution times, see Fig.~\ref{fig:Sscaling_q2_g0.1_a1}(f).

It is interesting to note that in the WKS and thermodynamic limits, the quench dynamics of the return rate and chiral condensate in the weak-coupling regime within this superselection sector are in excellent quantitative agreement with their counterparts in the weak-coupling regime within the superselection sectors $q_j^\text{tar}=0$ and $q_j^\text{tar}=(-1)^j$; see Figs.~\ref{fig:Sscaling_q0_g0.1_a1} and~\ref{fig:Sscaling_q1_g0.1_a1}. As such, this confirms our earlier conclusion that in the WKS and thermodynamic limits the quench dynamics in the weak-coupling regime are largely independent of the target gauge superselection sector.

\subsubsection{Strong-coupling regime}
We now repeat this quench in the strong-coupling regime with $g=2\sqrt{J}$, with the corresponding numerical results shown in Fig.~\ref{fig:Sscaling_q2_g2_a1}. As for all cases we have considered, we see very rapid convergence to the WKS limit for the small values of $S$ that we use in both the return rate, shown in Fig.~\ref{fig:Sscaling_q2_g2_a1}(a,b), and the chiral condensate, shown in Fig.~\ref{fig:Sscaling_q2_g2_a1}(c,d), for half-integer and integer $S$, respectively. The case of integer $S$, like previous results in the strong-coupling regime, shows excellent convergence to the WKS limit already at $S=1$. The case of half-integer $S$ also shows fast convergence, except that the dynamics of the return rate and chiral condensate both seem to be relatively trivial, and not just for $S=1/2$ where the corresponding Hilbert subspace is composed of a single state. The reason for this is that the initial state at which we start has a very large overlap with the ground state and a few low-lying eigenstates of the quench Hamiltonian. As for other cases we have considered within the strong-coupling regime, the quench dynamics of the return rate and the chiral condensate in the WKS limit are vastly different between half-integer and integer $S$.

Convergence to the thermodynamic limit is excellent for the case of half-integer $S$, as shown in Fig.~\ref{fig:Sscaling_q2_g2_a1}(e) for the chiral condensate, but we again note that this is in large part due to the trivial dynamics taking place in this case. For the case of integer $S$, which shows nontrivial dynamics, we find from ED that the thermodynamic limit is faithfully achieved up to very long evolution times already at $L=12$ matter sites.

\begin{figure}[t!]
	\centering
	\includegraphics[width=.48\textwidth]{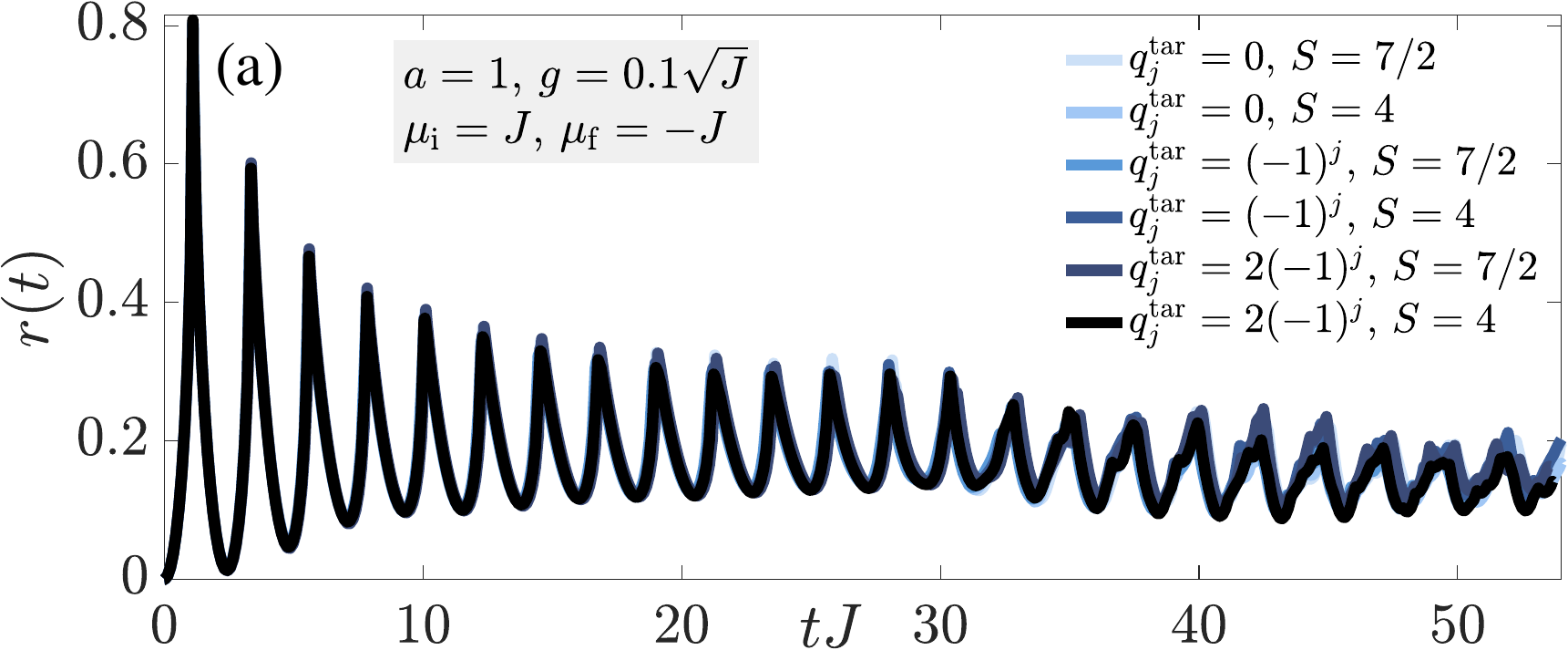}\\
	\vspace{1.1mm}
	\includegraphics[width=.48\textwidth]{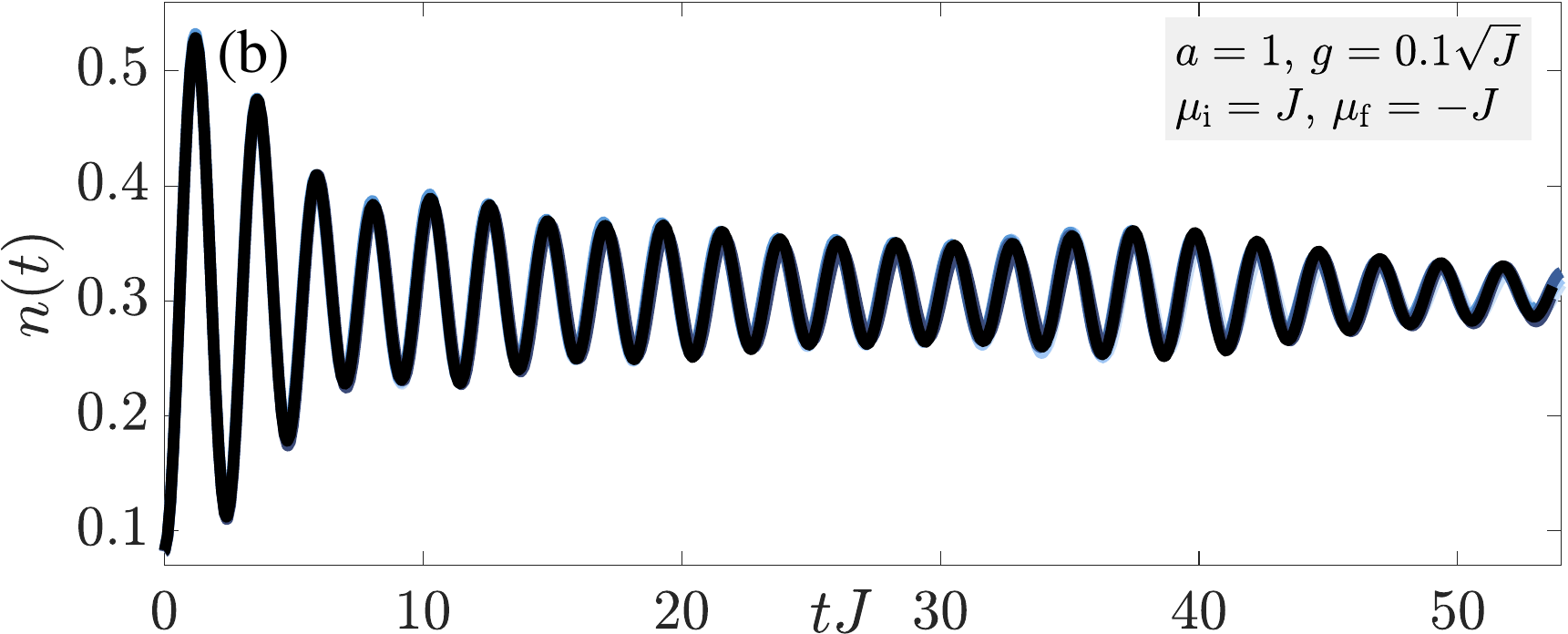}
	\caption{(Color online). Quench dynamics of (a) the return rate and (b) the chiral condensate in spin-$S$ $\mathrm{U}(1)$ quantum link model~\eqref{eq:H} starting in its ground state at $\mu=J$ and quenching the mass to $\mu=-J$ in the weak-coupling regime at $g=0.1\sqrt{J}$ within the target sector $q_j^\text{tar}$. Restricting to the Wilson--Kogut--Susskind limit ($S=7/2\,\text{or}\,4$), we see that the quench dynamics are identical across different choices of the target superselection sector for both the return rate and the chiral condensate, and are independent of whether $S$ is half-integer or integer.}
	\label{fig:WeakCoupling} 
\end{figure}

\subsection{Discussion and summary}
Let us now summarize the results presented in Secs.~\ref{sec:q0},~\ref{sec:q1}, and~\ref{sec:q2}, and put them into context. Four main conclusions can be drawn from them:
\begin{enumerate}
\item The WKS limit is achieved at relatively small values of half-integer and integer $S\lesssim4$ regardless of the target superselection sector and the electric-field coupling regime.
\item In the strong-coupling limit, the quench dynamics with an integer link spin length $S$ lead to faster convergence to the WKS limit than the case of half-integer $S$. We attribute this to a large $g$ suppressing quantum fluctuations in the case of integer $S$ more severely than in the half-integer case, because the former allows for a local electric-field eigenvalue of $0$.
\item This qualitative distinction also affects the quench dynamics in the strong-coupling limit, leading to the vastly different behavior observed for integer and half-integer spin $S$.
\item In the weak-coupling regime, quantum fluctuations dominate and the physics is no longer dominated by the (non)existence of a local zero electric field state.
For the quenches discussed in this work, the WKS limit is 
thus effectively
the same for half-integer and integer $S$, and it does not depend on the choice of target superselection sector. This conclusion is summarized in Fig.~\ref{fig:WeakCoupling}, where the quench dynamics of the return rate and the chiral condensate in the WKS limit ($S=7/2$ and $S=4$) show great quantitative agreement over all accessible evolution times in iMPS regardless of the chosen target superselection sector and independently of whether $S$ is half-integer or integer. Again, we attribute this to the fact that at small $g$ quantum fluctuations dominate, washing out the detailed ``low-energy'' structure of the link operators.
\end{enumerate}
These conclusions bode well for ongoing QSM experiments on LGTs. Indeed, smaller local Hilbert spaces of the gauge fields imply fewer degrees of freedom to be implemented experimentally. Furthermore, our work shows that it does not matter whether $S$ is half-integer or integer in the weak-coupling regime, and also that the choice of target superselection sector is not important. This further simplifies implementational requirements.

\section{Continuum limit $a\to0$}\label{sec:continuum}
In this section, we explore the possibility to reach the continuum limit $a\to 0$ for the quench dynamics discussed in Sec.~\ref{sec:Kogut} using the quantum link regularization with spin length $S$ sufficiently large to achieve the WKS limit. In equilibrium, it is well-known how the lattice Schwinger model approaches continuum quantum electrodynamics (QED)~\cite{banuls2013mass,banuls2016chiral}, which, due to its super-renormalizability, enables a relatively simple extrapolation of appropriately rescaled observables towards $a\to 0$. While the situation is less clear in nonequilibrium, our results presented below suggest that it is sufficient to perform a renormalization at the initial time $t=0$.

To be specific, we focus on the chiral condensate $\Sigma = \langle \bar{\psi} \psi \rangle$ in the following. As discussed, e.g., in~\cite{banuls2016chiral}, the vacuum expectation value of $\Sigma$ is logarithmically divergent in the thermodynamic limit. In equilibrium, this infrared divergence can be isolated analytically for any finite lattice spacing $a$, providing a regularized chiral condensate $\Sigma_\text{reg.}$, which in turn can be extrapolated to $a\rightarrow 0$. Since we start our quench dynamics with the ground state of \begin{align}\nonumber
	\frac{\hat{H}}{\mu}=&-\frac{1}{2a\mu\sqrt{S(S+1)}}\sum_{j=1}^{L-1}\big(\hat{\sigma}^-_j\hat{s}^+_{j,j+1}\hat{\sigma}^-_{j+1}+\text{H.c.}\big)\\\label{eq:H_lattice}
	&+\frac{1}{2}\sum_{j=1}^L\hat{\sigma}^z_j+\frac{(ga)^2}{2a\mu }\sum_{j=1}^{L-1}\big(\hat{s}^z_{j,j+1}\big)^2.
\end{align}
with $a\mu>0$, we apply the same procedure to the time-evolved chiral condensate extracted from our simulations. Note that we have set $J=1$.

\begin{figure}
	\centering{
	\includegraphics[scale=0.34]{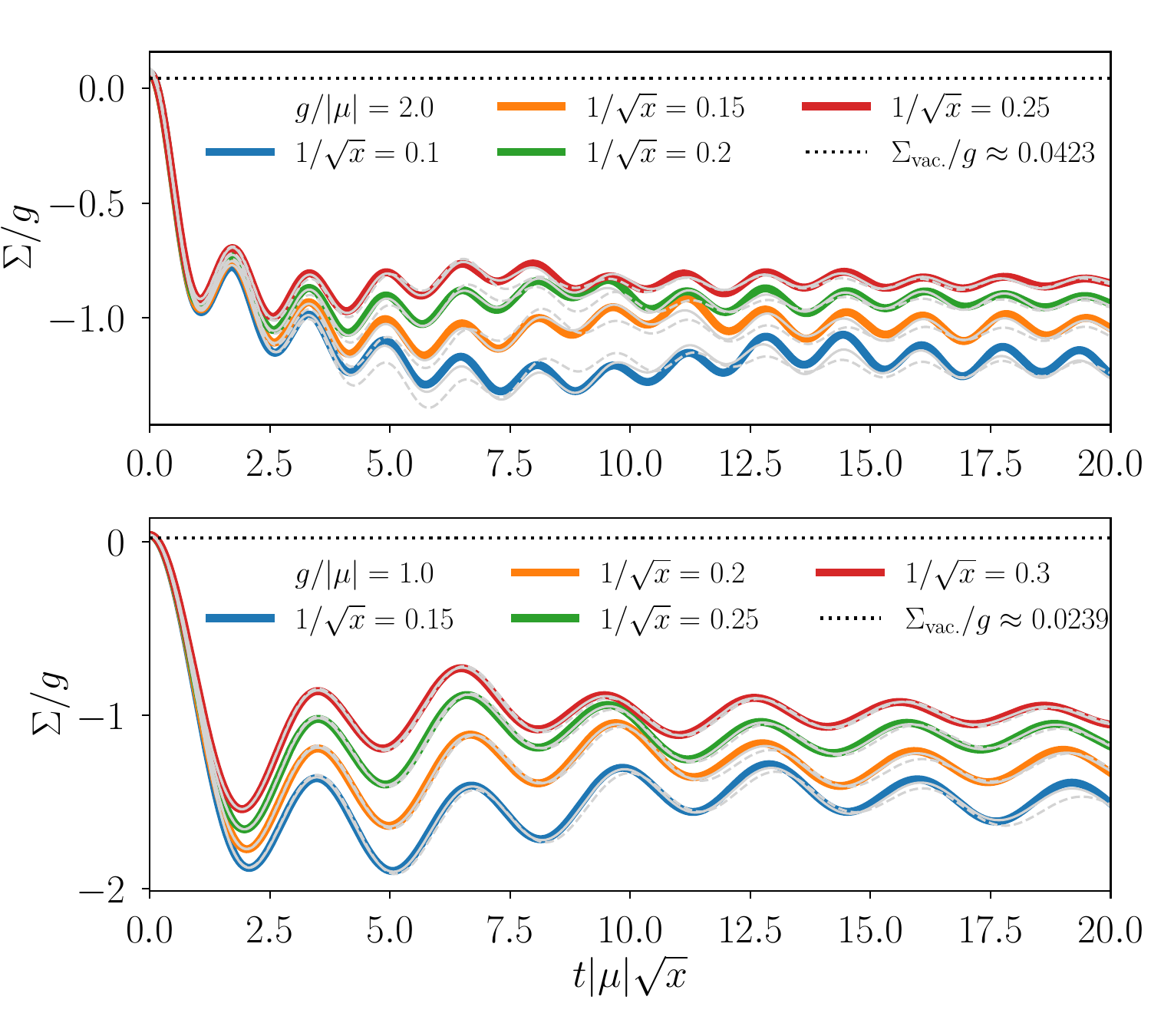}
\caption{\label{fig:CC_continuum} Quench dynamics of the regularized (see text) chiral condensate. The top(bottom) panel shows results for $g/|\mu| = 2(1)$ with thick colored lines (red to blue, top to bottom) approaching $\sqrt{x} \rightarrow \infty$ for $S=4$.  Light grey (solid resp. dashed) lines correspond to smaller spin lengths ($S=3$ resp. $2$). The dotted black lines indicate the expected initial vacuum expectation values of the chiral condensate~\cite{banuls2016chiral}.}
}
\end{figure}

Explicitly, we extract the regularized chiral condensate as
\begin{align}
	\Sigma (t)= g\sqrt{x} \left[\frac{1}{2} - n(t)\right]  - \Sigma_0,
\end{align}
where we defined the dimensionless parameter $x =1/(ag)^2$. The divergent contribution is given by~\cite{banuls2016chiral}
\begin{subequations}
\begin{align}
	\Sigma_0 &= \frac{\mu}{\pi\sqrt{\xi}} K \left(\frac{1}{\xi}\right),\\
	\xi &= 1 +  \frac{1}{x} \left(\frac{\mu}{g}\right)^2,
\end{align}
\end{subequations}
where $K$ denotes the complete elliptic integral of the first kind.
The resulting dynamics of the regularized chiral condensate is shown in Fig.~\ref{fig:CC_continuum} for $g/|\mu|=2$ and $1$. 

As anticipated, the initial condensate $\Sigma_\text{reg.} (0)$ approaches the known~\cite{banuls2016chiral} vacuum expectation values as $\sqrt{x} \rightarrow \infty$. Remarkably, we find fast convergence of the real-time dynamics for relatively small spins ($S \le 4$). In contrast to the equilibrium case, this convergence appears to be faster for smaller values of $g/|\mu|$, at least for the two values accessible with our numerics. At early times, our simulations further exhibit a smooth approach to a continuum limit with increasing $\sqrt{x}$. 
These findings suggest that the nonequilibrium dynamics of QED can indeed be probed by careful extrapolations of the quantum link model, requiring only a renormalization of the chiral condensate at initial time.

For the quench considered here, the chiral condensate exhibits a very strong response with a much larger magnitude and opposite sign than its initial vacuum expectation value. This behaviour indicates a strong breaking of chiral symmetry, which stands in contrast to the smooth chiral symmetry restoration in QED at (in)finite temperature~\cite{banuls2016chiral}, implying a small value of the chiral condensate at late times if the system thermalizes to some finite temperature. As a possible explanation, we speculate that the large response of the chiral condensate at early times is a transient, truly nonequilibrium phenomenon that disappears at later times, something which could be tested in a quantum simulation of this model.

It is worth making a comment on numerical resources here. The main limitation for convergence to the continuum comes from the sequence of extrapolations, $S \rightarrow \infty$ and $x \rightarrow \infty$. For fixed $x$, we first have to extrapolate to large spin lengths S, which leads to a larger local Hilbert space, and hence to numerically more costly MPS calculations. Given finite computational resources, we therefore had to limit ourselves to relatively small spin lengths $S$, which in turn limited the size of accessible $x$. We emphasize that due the order of extrapolations, it is not useful to obtain more data at larger $x$ without increasing $S$ for obtaining the continuum limit.

\section{Conclusion}\label{sec:conc}
In this work, we have extended to the far-from-equilibrium regime our equilibrium study \cite{zache2021achieving} of the approach of quantum link model regularizations of lattice gauge theories to their quantum field theory limit. We have found that in the thermodynamic limit, the Wilson--Kogut--Susskind limit is achieved over all considered parameter regimes and superselection sectors at relatively small values of the link spin length $S\lesssim4$, regardless of whether $S$ is half-integer or integer.

In the weak electric-field coupling regime, the quench dynamics in the Wilson--Kogut--Susskind limit are quantitatively almost identical regardless of whether $S$ is half-integer or integer, and are independent of the target superselection sector. We attribute this to strong quantum fluctuations in the weak-coupling regime ($g^2\ll1$), which wash out the detailed low-energy structure of the link operators. In the strong electric-field coupling regime, quantum fluctuations are suppressed, and even though the Wilson--Kogut--Susskind limit is still very rapidly achieved, and extremely so for integer $S$, there is a qualitative difference in the dynamics depending on whether $S$ is half-integer or integer. Indeed, in the case of integer $S$, the zero-eigenvalue state of the link operator is dominant at large $g^2\gg1$, leading to extremely fast convergence to the Wilson--Kogut--Susskind limit already at $S\gtrsim1$. In the case of half-integer $S$, on the other hand, quantum fluctuations are not as suppressed, and this leads to qualitative and quantitative differences in the dynamics with respect to the case of integer $S$. In other words, the detailed low-energy structure of the link operators is revealed in the strong-coupling regime.

Concerning the approach to the continuum limit ($a\rightarrow 0$), where we have focused on the dynamics of the chiral condensate in the $q_j=0$ sector, our findings demonstrate fast convergence at early times. In general, however, the situation dynamically depends on the coupling $g/|\mu|$---smaller couplings and later times require finer lattices to reach the continuum limit, although relatively small spin lengths appear to be sufficient for the quenches considered in this work. 

Our results are relevant for ongoing quantum synthetic matter experiments on lattice gauge theories seeking to achieve the quantum field theory limit. They show that the resources required to converge to this limit will not scale uncontrollably, and that already small finite values of the link spin length $S\lesssim4$ are sufficient for achieving this limit up to all experimentally relevant lifetimes in the quench dynamics. These findings may further guide the design of future quantum simulation experiments that target the quantum field theory limit of the present model. An interesting avenue for such future studies is, for example, the fate of the chiral condensate in the long-time limit, which is beyond the reach of our classical simulations.

\begin{acknowledgments}
This work is part of and supported by Provincia Autonoma di Trento, the ERC Starting Grant StrEnQTh (project ID 804305), the Google Research Scholar Award ProGauge, and Q@TN — Quantum Science and Technology in Trento, Research Foundation Flanders (G0E1520N, G0E1820N), and ERC grants QUTE (647905) and ERQUAF (715861). This work was supported by the Simons Collaboration on UltraQuantum Matter, which is a grant from the Simons Foundation (651440, P.Z.).
\end{acknowledgments}

\appendix

\begin{figure*}[t!]
	\centering
	\includegraphics[width=.48\textwidth]{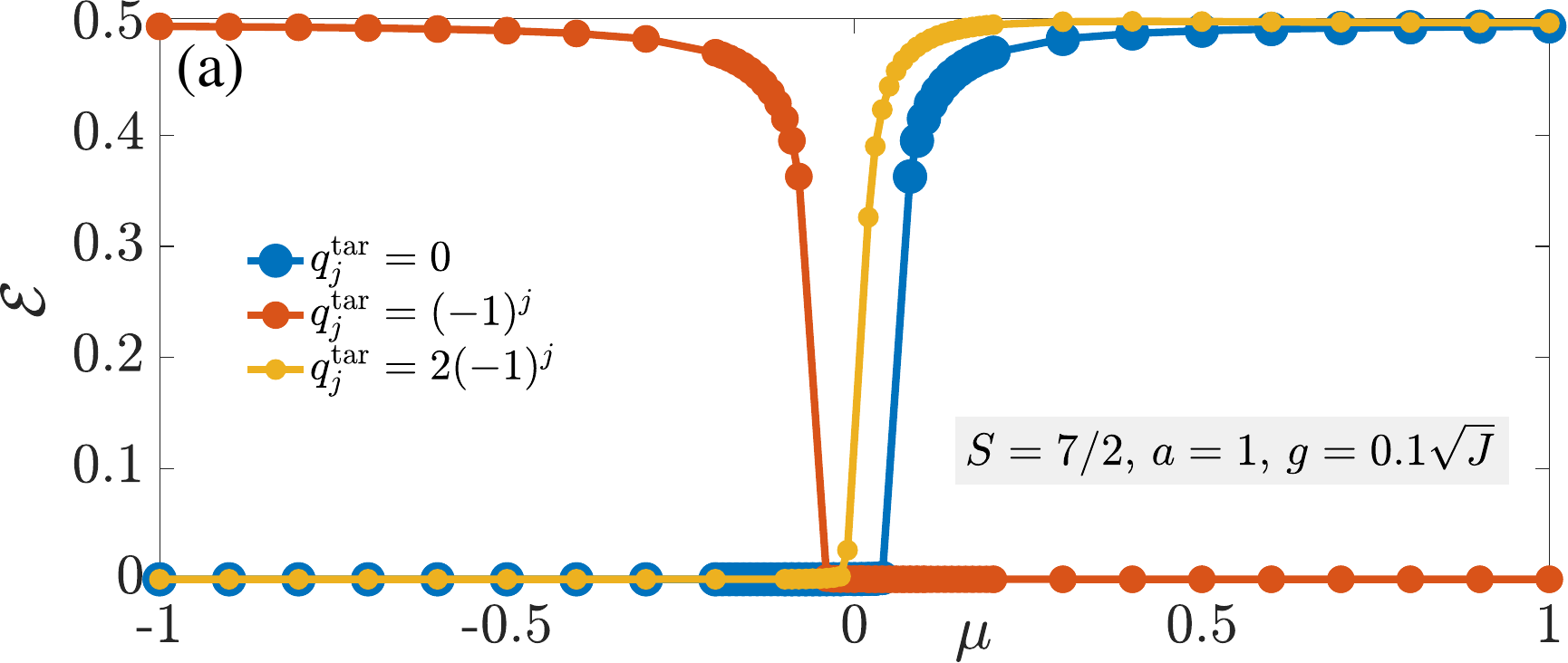}\quad\includegraphics[width=.48\textwidth]{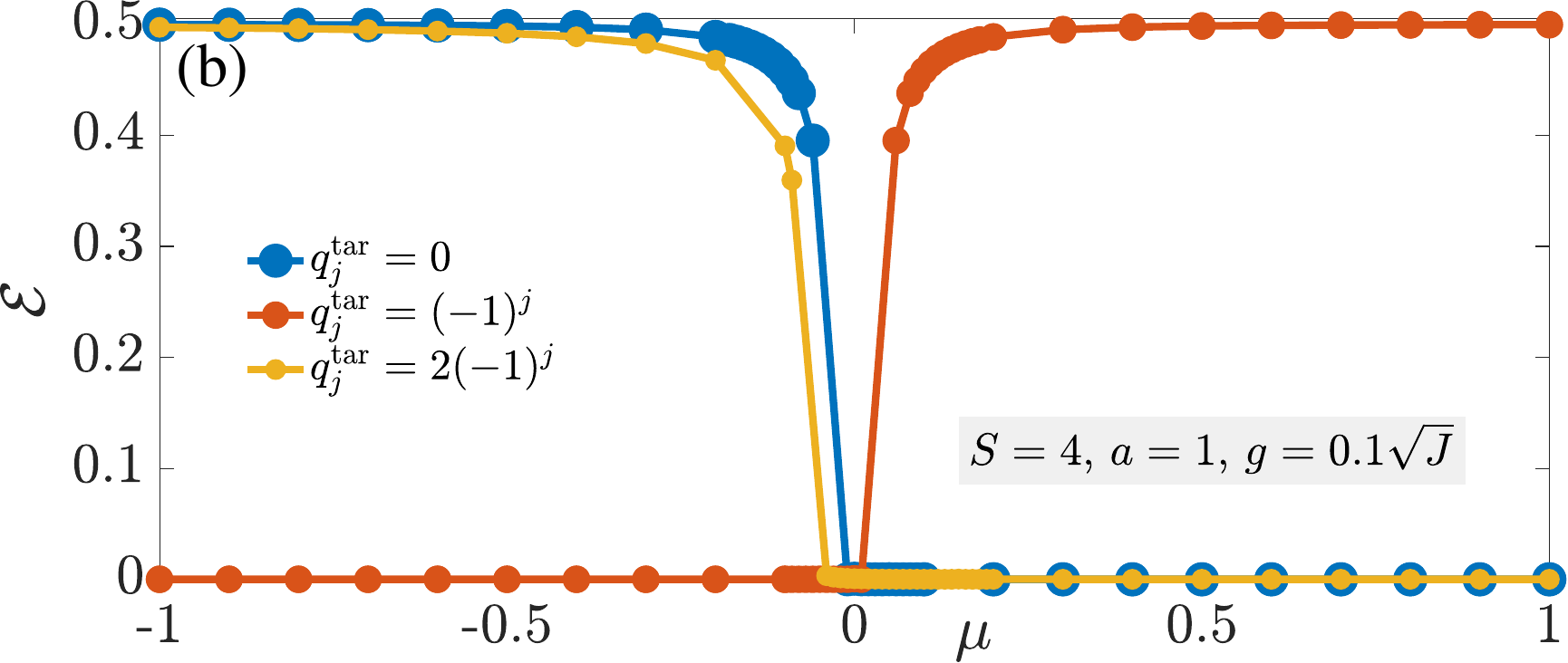}\\
	\vspace{1.1mm}
	\includegraphics[width=.48\textwidth]{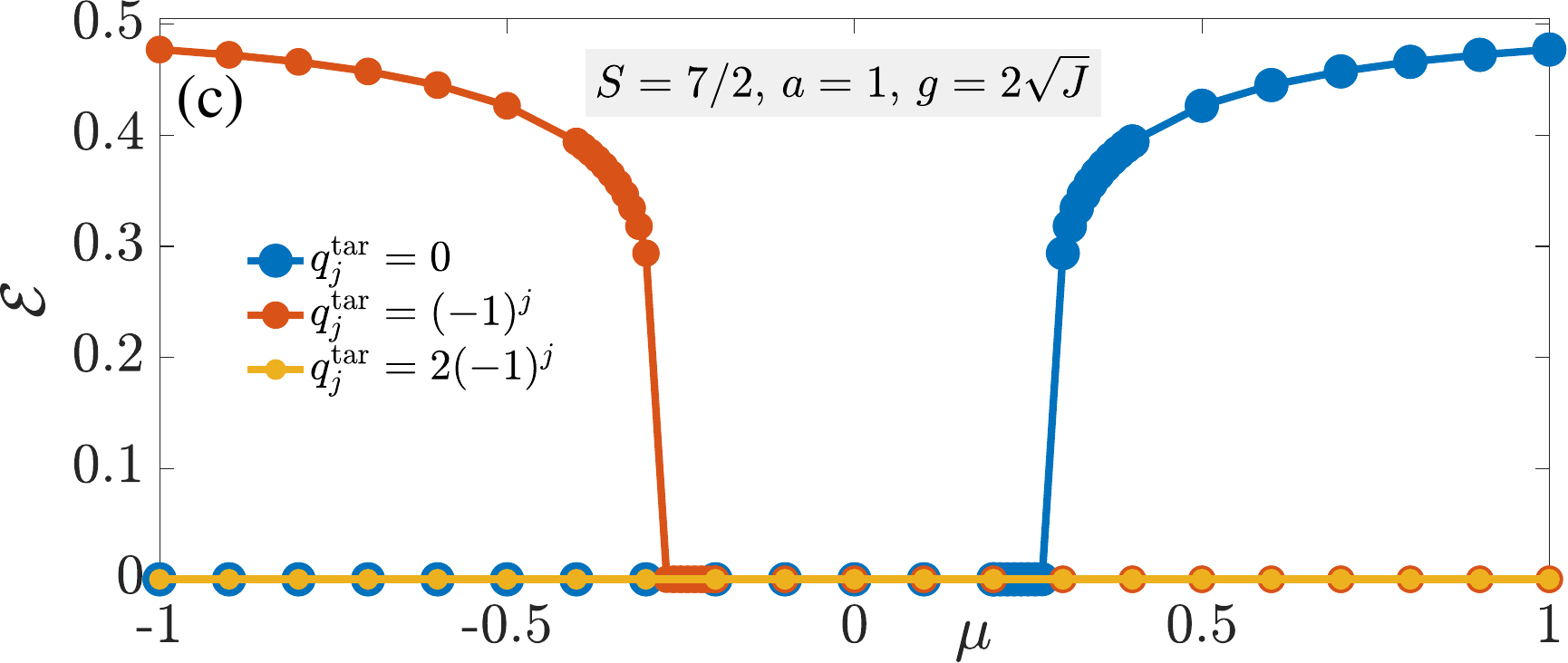}\quad\includegraphics[width=.48\textwidth]{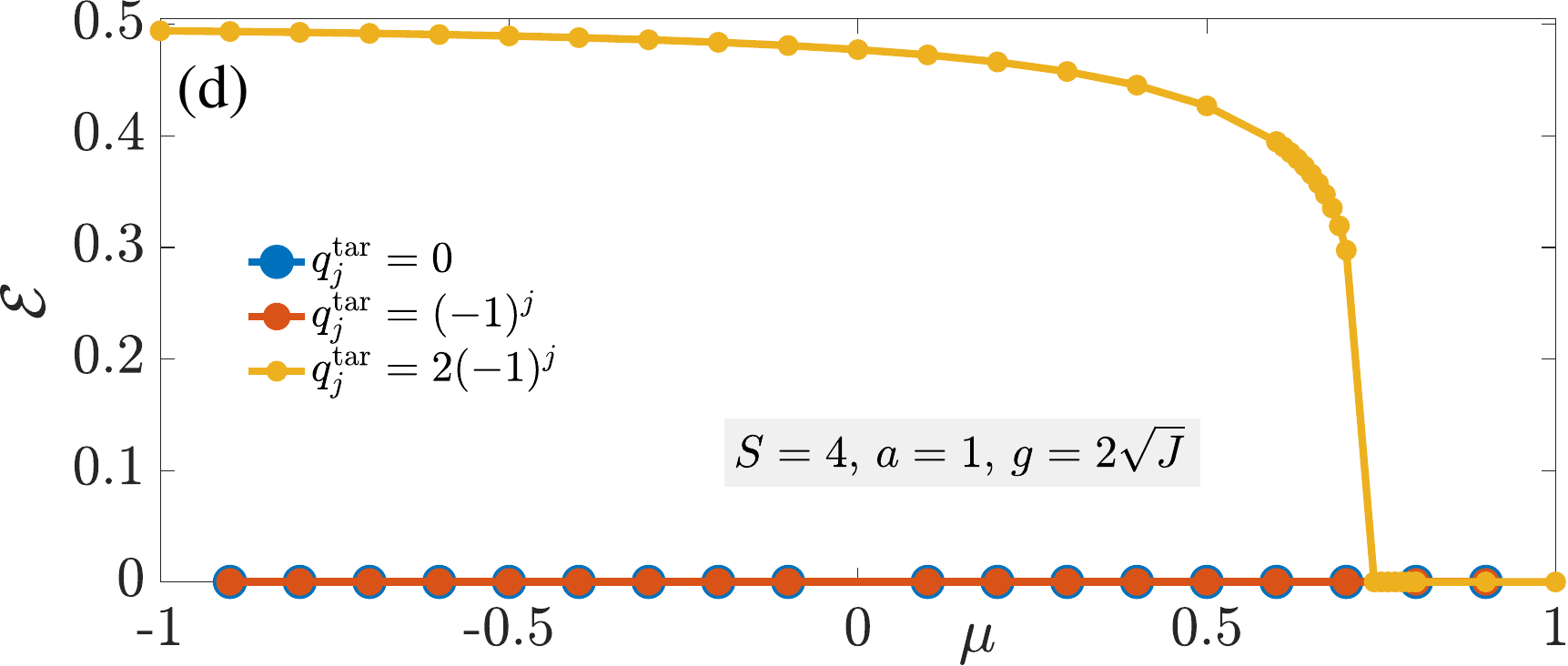}
	\caption{(Color online). Ground-state phase diagram of the spin-$S$ $\mathrm{U}(1)$ QLM for in (a,b) the weak-coupling and (c,d) strong-coupling regimes at link spin lengths (a,c) $S=7/2$ and (b,d) $S=4$. These numerical results are obtained from iMPS.}
	\label{fig:GS} 
\end{figure*}

\section{Particle-hole transformation}\label{app:PH_trafo}
In this Appendix, we write down the particle-hole transformation that leads to Eqs.~\eqref{eq:H} and~\eqref{eq:Gj} in the main text. We consider the spin-$S$ $\mathrm{U}(1)$ QLM described by the Hamiltonian
\begin{align}\nonumber
	\hat{H}=&-\frac{J}{2a\sqrt{S(S+1)}}\sum_{j=1}^{L-1}\big(\hat{\sigma}^+_j\hat{s}^+_{j,j+1}\hat{\sigma}^-_{j+1}+\text{H.c.}\big)\\\label{eq:H_noPH}
	&+\frac{\mu}{2}\sum_{j=1}^L(-1)^j\hat{\sigma}^z_j+\frac{g^2a}{2}\sum_{j=1}^{L-1}\big(\hat{s}^z_{j,j+1}\big)^2,
\end{align}
and whose gauge-symmetry generator is
\begin{align}
\hat{G}_j=&\frac{\hat{\sigma}^z_j+(-1)^j}{2}+\hat{s}^z_{j-1,j}-\hat{s}^z_{j,j+1}.
\end{align}
Here, the on-site chiral condensate is 
\begin{align}
\hat{n}_j=\frac{(-1)^j\hat{\sigma}^z_j+\mathds{1}}{2}.
\end{align}
Employing the particle-hole transformation
\begin{subequations}
\begin{align}
&\hat{\sigma}^{z(y)}_j\to(-1)^j\hat{\sigma}^{z(y)}_j,\\
&\hat{s}^{z(y)}_{j,j+1}\to(-1)^{j+1}\hat{s}^{z(y)}_{j,j+1},
\end{align}
\end{subequations}
we get
\begin{subequations}
\begin{align}\nonumber
	\hat{H}=&-\frac{J}{2a\sqrt{S(S+1)}}\sum_{j=1}^{L-1}\big(\hat{\sigma}^-_j\hat{s}^+_{j,j+1}\hat{\sigma}^-_{j+1}+\text{H.c.}\big)\\
	&+\frac{\mu}{2}\sum_{j=1}^L\hat{\sigma}^z_j+\frac{g^2a}{2}\sum_{j=1}^{L-1}\big(\hat{s}^z_{j,j+1}\big)^2,\\
	\hat{G}_j=&(-1)^j\big(\hat{n}_j+\hat{s}^z_{j-1,j}+\hat{s}^z_{j,j+1}\big),\\
	\hat{n}_j=&\frac{\hat{\sigma}^z_j+\mathds{1}}{2}.
\end{align}
\end{subequations}

\begin{figure}[t!]
	\centering
	\includegraphics[width=.33\textwidth]{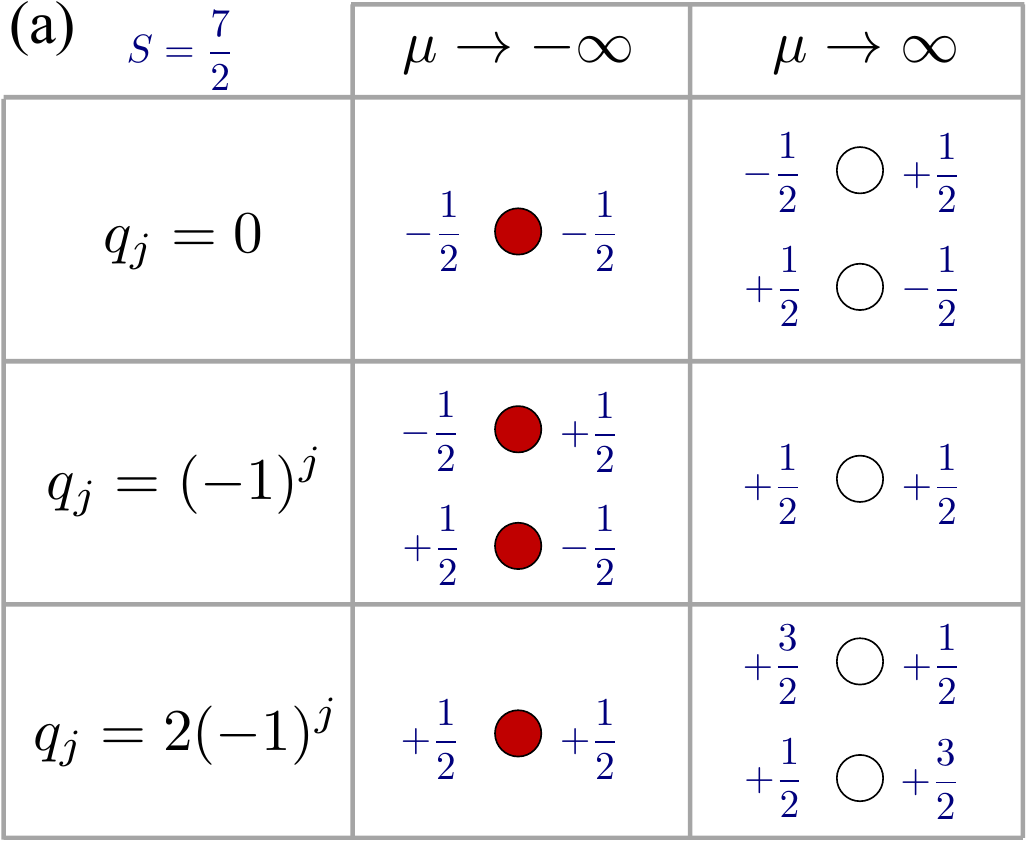}\\
	\vspace{3.1mm}
	\includegraphics[width=.33\textwidth]{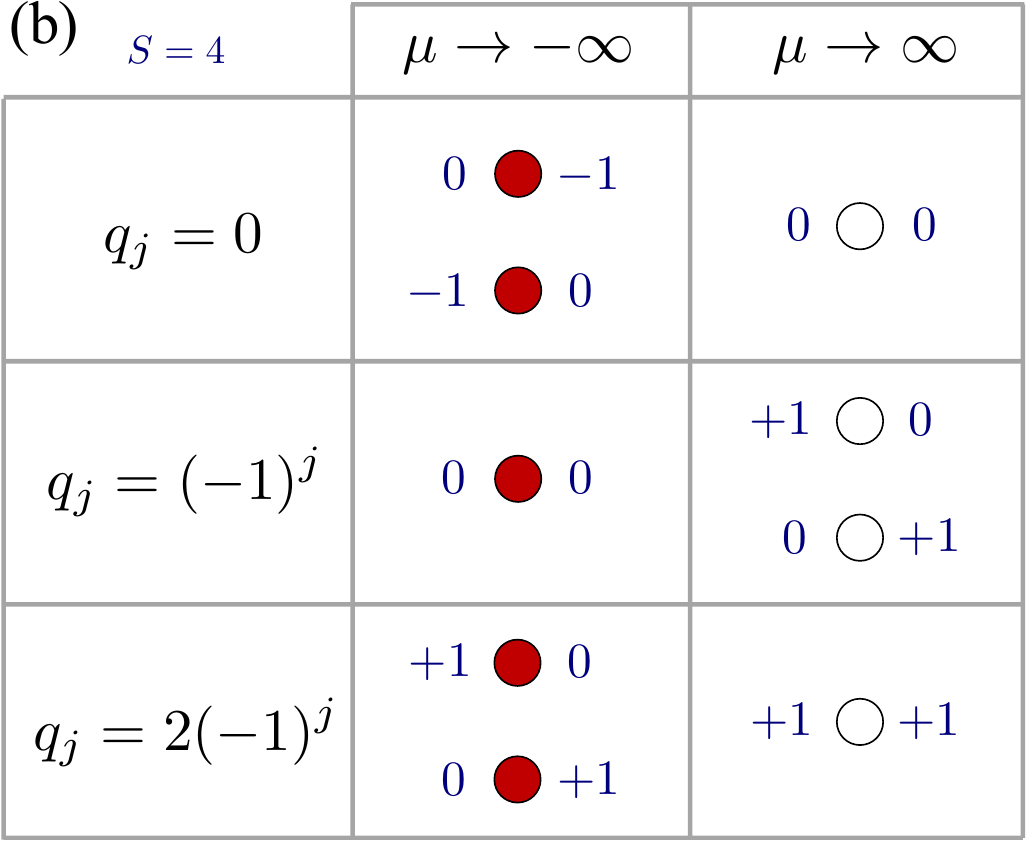}
	\caption{(Color online). Lowest-energy configurations of the local constraint at $\mu\to\pm\infty$ and a finite electric-field coupling strength $g$ for (a) $S=7/2$ and (b) $S=4$. Recalling that a local constraint consists of a link-site-link configuration on the lattice, we will indicate the matter occupation on the site with a circle, where a white circle indicates an empty site, and a red circle indicates an occupied one. The numbers on either side of the circle indicate the eigenvalues $m_z\in\{-S,\ldots,S\}$ of the electric-field operators on the adjacent links. These configurations represent the matter and electric-field configurations allowed by Gauss's law in the computational basis.}
	\label{fig:configs} 
\end{figure}

\section{Ground-state phase diagram}\label{app:GS}
We now consider the ground-state phase diagrams of the spin-$7/2$ and spin-$4$ $\mathrm{U}(1)$ QLMs, shown in Fig.~\ref{fig:GS}, in the various target superselection sectors and electric-field coupling regimes we have considered. The ground-state simulations are obtained using a mixture of the variational uniform matrix product state (VUMPS) algorithm \cite{vumps} and the infinite-size density matrix renormalization group (iDMRG) method \cite{McCulloch2008}, which was used to dynamically grow the bond dimension up to some chosen singular value cutoff. The gauge symmetry was imposed by ``approximating it'' as a global symmetry that works alternatingly on the even and odd bonds. One can then choose the symmetry sectors in the virtual space of the underlying matrix product state in such a way that this approximation becomes exact, in the sense that the state truly satisfies a local gauge symmetry.

The underlying phase transition is connected to the spontaneous breaking of a global $\mathbb{Z}_2$ symmetry, characterized by the order parameter represented by the electric flux
\begin{align}
    \mathcal{E}=\frac{1}{L}\Big\lvert\sum_{j=1}^L(-1)^j\bra{\psi_0}\hat{s}^z_{j,j+1}\ket{\psi_0}\Big\rvert,
\end{align}
where $\ket{\psi_0}$ denotes the ground state of the $\mathrm{U}(1)$ QLM Hamiltonian~\eqref{eq:H} at a given value of the mass $\mu$. This is plotted in Fig.~\ref{fig:GS}.

Focusing on the case of $S=7/2$, we see that at large $\mu$ there has to be $\mathbb{Z}_2$ symmetry-breaking in the superselection sector $q_j=0$, since in the limit $\mu\to\infty$ the lowest-energy configuration at every local constraint is either $(+1/2)\circ(-1/2)$ or $(-1/2)\circ(+1/2)$ at finite $g^2>0$, see Fig.~\ref{fig:configs}(a). In other words, at a local constraint the matter site will have to be empty and its neighboring electric fields will be eigenstates of $\hat{s}^z_{j,j+1}$ with opposite eigenvalues $\pm1/2$. On the other hand, when $\mu\to-\infty$, the matter site will be occupied, and therefore the electric fields must be identical (eigenvalues $-1/2$) on the adjacent links, leading to a phase that does not break charge conjugation (or global $\mathbb{Z}_2$) symmetry. The picture is reversed for $q_j=(-1)^j$, since at $\mu\to\infty$ the electric-field eigenvalues must sum to unity, rendering the $\mathbb{Z}_2$-symmetric configuration $(+1/2)\circ(+1/2)$ energetically most favorable at finite $g^2>0$. For $\mu\to-\infty$, the electric fields on adjacent links of the occupied matter site must align in the opposite direction, leading to a $\mathbb{Z}_2$ symmetry-broken phase. For $q_j=2(-1)^j$ at finite $g^2>0$, the limit $\mu\to\infty$ will result in $(+1/2)\circ(+3/2)$ or $(+3/2)\circ(+1/2)$ as the lowest-energy configurations, both of which correspond to a $\mathbb{Z}_2$ symmetry-broken phase. However, when $\mu\to-\infty$, then the lowest-energy configuration is $(+1/2)\bullet(+1/2)$, which corresponds to the $\mathbb{Z}_2$-symmetric phase. This is exactly what we see in Fig.~\ref{fig:GS}(a) for the weak-coupling regime, and partially in Fig.~\ref{fig:GS}(c) for the strong-coupling regime. In contrast to the weak-coupling regime, when $g$ is large the $\mathbb{Z}_2$ symmetry-broken configuration in the superselection sector $q_j=2(-1)^j$ incurs a larger energy penalty than the $\mathbb{Z}_2$-symmetric configurations due to the term $\propto g^2\sum_j\big(\hat{s}^z_{j,j+1}\big)^2$ in Eq.~\eqref{eq:H}. This results in shifting the critical point to the right and extending the $\mathbb{Z}_2$-symmetric phase, such that in the considered mass regime only the $\mathbb{Z}_2$-symmetric phase is seen in Fig.~\ref{fig:GS}(c).

For the case of $S=4$ at finite $g^2>0$, we find that in the superselection sector $q_j=0$ at $\mu\to-\infty$ the lowest-energy configurations are $(0)\bullet(-1)$ and $(-1)\bullet(0)$, both of which are $\mathbb{Z}_2$ symmetry-broken, while at $\mu\to\infty$ the lowest-energy configuration is $(0)\circ(0)$, which is $\mathbb{Z}_2$-symmetric, see Fig.~\ref{fig:configs}(b). At large $g$, the $\mathbb{Z}_2$ symmetry-broken state will be penalized while the $\mathbb{Z}_2$-symmetric state will not, which explains why going from the weak-coupling to the strong-coupling regime in Fig.~\ref{fig:GS}(b,d), we see how only the $\mathbb{Z}_2$-symmetric phase remains in the mass regime we consider. In the superselection sector $q_j=(-1)^j$ at finite $g^2>0$, the lowest-energy configuration at $\mu\to-\infty$ is $(0)\bullet(0)$, which is $\mathbb{Z}_2$-symmetric, while at $\mu\to\infty$ the lowest-energy configurations are $(1)\circ(0)$ and $(0)\circ(1)$, both of which are $\mathbb{Z}_2$ symmetry-broken. Once again, the $\mathbb{Z}_2$ symmetry-broken states will be penalized at large $g$ but not the $\mathbb{Z}_2$-symmetric state, leading to only the $\mathbb{Z}_2$-symmetric phase remaining in the mass regime we consider in Fig.~\ref{fig:GS}(d). In the superselection sector $q_j=2(-1)^j$ at finite $g^2>0$, the lowest-energy configurations at $\mu\to-\infty$ are $(1)\bullet(0)$ and $(0)\bullet(1)$, both of which are $\mathbb{Z}_2$ symmetry-broken, while the lowest-energy configuration at $\mu\to\infty$ is $(1)\circ(1)$, which is $\mathbb{Z}_2$-symmetric. In this case, a large value of $g$ will penalize the $\mathbb{Z}_2$-symmetric state more than its $\mathbb{Z}_2$ symmetry-broken counterpart, which is why the $\mathbb{Z}_2$ symmetry-broken phase is extended going from the weak-coupling regime in Fig.~\ref{fig:GS}(b) to the strong-coupling regime in Fig.~\ref{fig:GS}(d).

It is interesting to note here that the superselection sectors $q_j=0$ and $q_j=(-1)^j$ are related by the particle-hole transformation
\begin{subequations}
\begin{align}
    &\hat{n}_j\to\mathds{1}-\hat{n}_j,\\
    &\hat{s}^z_{j,j+1}\to-\hat{s}^z_{j,j+1},
\end{align}
\end{subequations}
and this is why the phase diagrams for these two sectors are mirror-symmetric to each other with respect to $\mu=0$ in Fig.~\ref{fig:GS}. In terms of quench dynamics, this means that quenching the mass from $\mu_\text{i}$ to $\mu_\text{f}$ in the superselection sector $q_j=0$ is equivalent to quenching the mass from $-\mu_\text{i}$ to $-\mu_\text{f}$ in the superselection sector $q_j=(-1)^j$, and vice versa.

Relating Fig.~\ref{fig:GS} to our results on quench dynamics in Sec.~\ref{sec:Kogut}, we can make several observations. We have remarked that at strong electric-field coupling, the convergence to the WKS limit is extremely fast for integer $S$, where $S=1$ is already sufficient to achieve this limit, see Figs.~\ref{fig:Sscaling_q0_g2_a1}(b,d) and~\ref{fig:Sscaling_q1_g2_a1}(b,d) for the target superselection sectors $q_j=0$ and $q_j=(-1)^j$, respectively. Looking at Fig.~\ref{fig:GS}(d), we see that in these sectors the considered quench takes place only within the $\mathbb{Z}_2$-symmetric phase. This further supports our explanation that quantum fluctuations are strongly suppressed in this case. Note how the same quench for half-integer $S$ crosses the corresponding phase transition point.

We have also remarked that the quench dynamics in the WKS and thermodynamic limits in the superselection sector $q_j=2(-1)^j$ is trivial in the case of half-integer $S$. Looking at Fig.~\ref{fig:GS}(c), we see that this quench occurs within the $\mathbb{Z}_2$-symmetric phase of the corresponding phase diagram. This is not surprising, since we have checked in ED that the initial state we start in at $\mu=J$ has large overlap with the ground state of Eq.~\eqref{eq:H} at $\mu=-J$. This leads to very slow and trivial dynamics during the evolution times we can access in iMPS.

\bibliography{DQPT_LGT_biblio}

\begin{thebibliography}{10}

\bibitem{Bloch2008}
Immanuel Bloch, Jean Dalibard, and Wilhelm Zwerger.
\newblock ``Many-body physics with ultracold gases''.
\newblock \href{https://dx.doi.org/10.1103/RevModPhys.80.885}{Rev. Mod. Phys.
  {\bf 80}, 885--964}~(2008).

\bibitem{Lewenstein_book}
M.~Lewenstein, A.~Sanpera, and V.~Ahufinger.
\newblock ``Ultracold atoms in optical lattices: Simulating quantum many-body
  systems''.
\newblock OUP Oxford. ~(2012).
\newblock  url:~\url{https://books.google.de/books?id=Wpl91RDxV5IC}.

\bibitem{Blatt_review}
R.~Blatt and C.~F. Roos.
\newblock ``Quantum simulations with trapped ions''.
\newblock \href{https://dx.doi.org/10.1038/nphys2252}{Nature Physics {\bf 8},
  277--284}~(2012).

\bibitem{Hauke2012}
Philipp Hauke, Fernando~M Cucchietti, Luca Tagliacozzo, Ivan Deutsch, and
  Maciej Lewenstein.
\newblock ``Can one trust quantum simulators?''.
\newblock \href{https://dx.doi.org/10.1088/0034-4885/75/8/082401}{Reports on
  Progress in Physics {\bf 75}, 082401}~(2012).

\bibitem{Jurcevic2017}
P.~Jurcevic, H.~Shen, P.~Hauke, C.~Maier, T.~Brydges, C.~Hempel, B.~P. Lanyon,
  M.~Heyl, R.~Blatt, and C.~F. Roos.
\newblock ``Direct observation of dynamical quantum phase transitions in an
  interacting many-body system''.
\newblock \href{https://dx.doi.org/10.1103/PhysRevLett.119.080501}{Phys. Rev.
  Lett. {\bf 119}, 080501}~(2017).

\bibitem{Zhang2017dpt}
J.~Zhang, G.~Pagano, P.~W. Hess, A.~Kyprianidis, P.~Becker, H.~Kaplan, A.~V.
  Gorshkov, Z.-X. Gong, and C.~Monroe.
\newblock ``Observation of a many-body dynamical phase transition with a
  53-qubit quantum simulator''.
\newblock \href{https://dx.doi.org/10.1038/nature24654}{Nature {\bf 551},
  601--604}~(2017).

\bibitem{Flaeschner2018}
N.~Fl{\"a}schner, D.~Vogel, M.~Tarnowski, B.~S. Rem, D.-S. L{\"u}hmann,
  M.~Heyl, J.~C. Budich, L.~Mathey, K.~Sengstock, and C.~Weitenberg.
\newblock ``Observation of dynamical vortices after quenches in a system with
  topology''.
\newblock Nature Physics {\bf 14}, 265--268~(2018).
\newblock  url:~\url{https://doi.org/10.1038/s41567-017-0013-8}.

\bibitem{Gring2012}
M.~Gring, M.~Kuhnert, T.~Langen, T.~Kitagawa, B.~Rauer, M.~Schreitl, I.~Mazets,
  D.~Adu Smith, E.~Demler, and J.~Schmiedmayer.
\newblock ``Relaxation and prethermalization in an isolated quantum system''.
\newblock \href{https://dx.doi.org/10.1126/science.1224953}{Science {\bf 337},
  1318--1322}~(2012).

\bibitem{Langen2015}
Tim Langen, Sebastian Erne, Remi Geiger, Bernhard Rauer, Thomas Schweigler,
  Maximilian Kuhnert, Wolfgang Rohringer, Igor~E. Mazets, Thomas Gasenzer, and
  J{\"o}rg Schmiedmayer.
\newblock ``Experimental observation of a generalized gibbs ensemble''.
\newblock \href{https://dx.doi.org/10.1126/science.1257026}{Science {\bf 348},
  207--211}~(2015).

\bibitem{Neyenhuis2017}
Brian Neyenhuis, Jiehang Zhang, Paul~W. Hess, Jacob Smith, Aaron~C. Lee, Phil
  Richerme, Zhe-Xuan Gong, Alexey~V. Gorshkov, and Christopher Monroe.
\newblock ``Observation of prethermalization in long-range interacting spin
  chains''.
\newblock \href{https://dx.doi.org/10.1126/sciadv.1700672}{Science Advances{\bf
  3}}~(2017).

\bibitem{Schreiber2015}
Michael Schreiber, Sean~S. Hodgman, Pranjal Bordia, Henrik~P. L{\"u}schen,
  Mark~H. Fischer, Ronen Vosk, Ehud Altman, Ulrich Schneider, and Immanuel
  Bloch.
\newblock ``Observation of many-body localization of interacting fermions in a
  quasirandom optical lattice''.
\newblock \href{https://dx.doi.org/10.1126/science.aaa7432}{Science {\bf 349},
  842--845}~(2015).

\bibitem{Choi2016}
Jae-yoon Choi, Sebastian Hild, Johannes Zeiher, Peter Schau{\ss}, Antonio
  Rubio-Abadal, Tarik Yefsah, Vedika Khemani, David~A. Huse, Immanuel Bloch,
  and Christian Gross.
\newblock ``Exploring the many-body localization transition in two
  dimensions''.
\newblock \href{https://dx.doi.org/10.1126/science.aaf8834}{Science {\bf 352},
  1547--1552}~(2016).

\bibitem{Smith2016}
J.~Smith, A.~Lee, P.~Richerme, B.~Neyenhuis, P.~W. Hess, P.~Hauke, M.~Heyl,
  D.~A. Huse, and C.~Monroe.
\newblock ``Many-body localization in a quantum simulator with programmable
  random disorder''.
\newblock \href{https://dx.doi.org/10.1038/nphys3783}{Nature Physics {\bf 12},
  907--911}~(2016).

\bibitem{Kaplan2020}
Harvey~B. Kaplan, Lingzhen Guo, Wen~Lin Tan, Arinjoy De, Florian Marquardt,
  Guido Pagano, and Christopher Monroe.
\newblock ``Many-body dephasing in a trapped-ion quantum simulator''.
\newblock \href{https://dx.doi.org/10.1103/PhysRevLett.125.120605}{Phys. Rev.
  Lett. {\bf 125}, 120605}~(2020).

\bibitem{semeghini2021probing}
G.~Semeghini, H.~Levine, A.~Keesling, S.~Ebadi, T.~T. Wang, D.~Bluvstein,
  R.~Verresen, H.~Pichler, M.~Kalinowski, R.~Samajdar, A.~Omran, S.~Sachdev,
  A.~Vishwanath, M.~Greiner, V.~Vuletić, and M.~D. Lukin.
\newblock ``Probing topological spin liquids on a programmable quantum
  simulator''.
\newblock \href{https://dx.doi.org/10.1126/science.abi8794}{Science {\bf 374},
  1242--1247}~(2021).

\bibitem{Satzinger2021}
K.~J. Satzinger, Y.-J Liu, A.~Smith, C.~Knapp, M.~Newman, C.~Jones, Z.~Chen,
  C.~Quintana, X.~Mi, A.~Dunsworth, C.~Gidney, I.~Aleiner, F.~Arute, K.~Arya,
  J.~Atalaya, R.~Babbush, J.~C. Bardin, R.~Barends, J.~Basso, A.~Bengtsson,
  A.~Bilmes, M.~Broughton, B.~B. Buckley, D.~A. Buell, B.~Burkett, N.~Bushnell,
  B.~Chiaro, R.~Collins, W.~Courtney, S.~Demura, A.~R. Derk, D.~Eppens,
  C.~Erickson, L.~Faoro, E.~Farhi, A.~G. Fowler, B.~Foxen, M.~Giustina,
  A.~Greene, J.~A. Gross, M.~P. Harrigan, S.~D. Harrington, J.~Hilton, S.~Hong,
  T.~Huang, W.~J. Huggins, L.~B. Ioffe, S.~V. Isakov, E.~Jeffrey, Z.~Jiang,
  D.~Kafri, K.~Kechedzhi, T.~Khattar, S.~Kim, P.~V. Klimov, A.~N. Korotkov,
  F.~Kostritsa, D.~Landhuis, P.~Laptev, A.~Locharla, E.~Lucero, O.~Martin,
  J.~R. McClean, M.~McEwen, K.~C. Miao, M.~Mohseni, S.~Montazeri,
  W.~Mruczkiewicz, J.~Mutus, O.~Naaman, M.~Neeley, C.~Neill, M.~Y. Niu, T.~E.
  O’Brien, A.~Opremcak, B.~Pató, A.~Petukhov, N.~C. Rubin, D.~Sank,
  V.~Shvarts, D.~Strain, M.~Szalay, B.~Villalonga, T.~C. White, Z.~Yao, P.~Yeh,
  J.~Yoo, A.~Zalcman, H.~Neven, S.~Boixo, A.~Megrant, Y.~Chen, J.~Kelly,
  V.~Smelyanskiy, A.~Kitaev, M.~Knap, F.~Pollmann, and P.~Roushan.
\newblock ``Realizing topologically ordered states on a quantum processor''.
\newblock \href{https://dx.doi.org/10.1126/science.abi8378}{Science {\bf 374},
  1237--1241}~(2021).

\bibitem{Mi2021}
Xiao Mi, Matteo Ippoliti, Chris Quintana, Ami Greene, Zijun Chen, Jonathan
  Gross, Frank Arute, Kunal Arya, Juan Atalaya, Ryan Babbush, Joseph~C. Bardin,
  Joao Basso, Andreas Bengtsson, Alexander Bilmes, Alexandre Bourassa, Leon
  Brill, Michael Broughton, Bob~B. Buckley, David~A. Buell, Brian Burkett,
  Nicholas Bushnell, Benjamin Chiaro, Roberto Collins, William Courtney, Dripto
  Debroy, Sean Demura, Alan~R. Derk, Andrew Dunsworth, Daniel Eppens, Catherine
  Erickson, Edward Farhi, Austin~G. Fowler, Brooks Foxen, Craig Gidney, Marissa
  Giustina, Matthew~P. Harrigan, Sean~D. Harrington, Jeremy Hilton, Alan Ho,
  Sabrina Hong, Trent Huang, Ashley Huff, William~J. Huggins, L.~B. Ioffe,
  Sergei~V. Isakov, Justin Iveland, Evan Jeffrey, Zhang Jiang, Cody Jones, Dvir
  Kafri, Tanuj Khattar, Seon Kim, Alexei Kitaev, Paul~V. Klimov, Alexander~N.
  Korotkov, Fedor Kostritsa, David Landhuis, Pavel Laptev, Joonho Lee, Kenny
  Lee, Aditya Locharla, Erik Lucero, Orion Martin, Jarrod~R. McClean, Trevor
  McCourt, Matt McEwen, Kevin~C. Miao, Masoud Mohseni, Shirin Montazeri,
  Wojciech Mruczkiewicz, Ofer Naaman, Matthew Neeley, Charles Neill, Michael
  Newman, Murphy~Yuezhen Niu, Thomas~E. O'Brien, Alex Opremcak, Eric Ostby,
  Balint Pato, Andre Petukhov, Nicholas~C. Rubin, Daniel Sank, Kevin~J.
  Satzinger, Vladimir Shvarts, Yuan Su, Doug Strain, Marco Szalay, Matthew~D.
  Trevithick, Benjamin Villalonga, Theodore White, Z.~Jamie Yao, Ping Yeh,
  Juhwan Yoo, Adam Zalcman, Hartmut Neven, Sergio Boixo, Vadim Smelyanskiy,
  Anthony Megrant, Julian Kelly, Yu~Chen, S.~L. Sondhi, Roderich Moessner,
  Kostyantyn Kechedzhi, Vedika Khemani, and Pedram Roushan.
\newblock ``Time-crystalline eigenstate order on a quantum processor''.
\newblock \href{https://dx.doi.org/10.1038/s41586-021-04257-w}{Nature {\bf
  601}, 531--536}~(2022).

\bibitem{Martinez2016}
Esteban~A. Martinez, Christine~A. Muschik, Philipp Schindler, Daniel Nigg,
  Alexander Erhard, Markus Heyl, Philipp Hauke, Marcello Dalmonte, Thomas Monz,
  Peter Zoller, and Rainer Blatt.
\newblock ``Real-time dynamics of lattice gauge theories with a few-qubit
  quantum computer''.
\newblock \href{https://dx.doi.org/10.1038/nature18318}{Nature {\bf 534},
  516--519}~(2016).

\bibitem{Klco2018}
N.~Klco, E.~F. Dumitrescu, A.~J. McCaskey, T.~D. Morris, R.~C. Pooser, M.~Sanz,
  E.~Solano, P.~Lougovski, and M.~J. Savage.
\newblock ``Quantum-classical computation of schwinger model dynamics using
  quantum computers''.
\newblock \href{https://dx.doi.org/10.1103/PhysRevA.98.032331}{Phys. Rev. A
  {\bf 98}, 032331}~(2018).

\bibitem{Kokail2019}
C.~Kokail, C.~Maier, R.~van Bijnen, T.~Brydges, M.~K. Joshi, P.~Jurcevic, C.~A.
  Muschik, P.~Silvi, R.~Blatt, C.~F. Roos, and P.~Zoller.
\newblock ``Self-verifying variational quantum simulation of lattice models''.
\newblock \href{https://dx.doi.org/10.1038/s41586-019-1177-4}{Nature {\bf 569},
  355--360}~(2019).

\bibitem{Klco2020}
Natalie Klco, Martin~J. Savage, and Jesse~R. Stryker.
\newblock ``Su(2) non-abelian gauge field theory in one dimension on digital
  quantum computers''.
\newblock \href{https://dx.doi.org/10.1103/PhysRevD.101.074512}{Phys. Rev. D
  {\bf 101}, 074512}~(2020).

\bibitem{Lu2019}
Hsuan-Hao Lu, Natalie Klco, Joseph~M. Lukens, Titus~D. Morris, Aaina Bansal,
  Andreas Ekstr\"om, Gaute Hagen, Thomas Papenbrock, Andrew~M. Weiner,
  Martin~J. Savage, and Pavel Lougovski.
\newblock ``Simulations of subatomic many-body physics on a quantum frequency
  processor''.
\newblock \href{https://dx.doi.org/10.1103/PhysRevA.100.012320}{Phys. Rev. A
  {\bf 100}, 012320}~(2019).

\bibitem{Goerg2019}
Frederik G{\"o}rg, Kilian Sandholzer, Joaqu{\'\i}n Minguzzi, R{\'e}mi
  Desbuquois, Michael Messer, and Tilman Esslinger.
\newblock ``Realization of density-dependent peierls phases to engineer
  quantized gauge fields coupled to ultracold matter''.
\newblock \href{https://dx.doi.org/10.1038/s41567-019-0615-4}{Nature Physics
  {\bf 15}, 1161--1167}~(2019).

\bibitem{Schweizer2019}
Christian Schweizer, Fabian Grusdt, Moritz Berngruber, Luca Barbiero, Eugene
  Demler, Nathan Goldman, Immanuel Bloch, and Monika Aidelsburger.
\newblock ``Floquet approach to $\mathbb{Z}2$ lattice gauge theories with
  ultracold atoms in optical lattices''.
\newblock \href{https://dx.doi.org/10.1038/s41567-019-0649-7}{Nature Physics
  {\bf 15}, 1168--1173}~(2019).

\bibitem{Mil2020}
Alexander Mil, Torsten~V. Zache, Apoorva Hegde, Andy Xia, Rohit~P. Bhatt,
  Markus~K. Oberthaler, Philipp Hauke, J{\"u}rgen Berges, and Fred
  Jendrzejewski.
\newblock ``A scalable realization of local u(1) gauge invariance in cold
  atomic mixtures''.
\newblock \href{https://dx.doi.org/10.1126/science.aaz5312}{Science {\bf 367},
  1128--1130}~(2020).

\bibitem{Yang2020}
Bing Yang, Hui Sun, Robert Ott, Han-Yi Wang, Torsten~V. Zache, Jad~C. Halimeh,
  Zhen-Sheng Yuan, Philipp Hauke, and Jian-Wei Pan.
\newblock ``Observation of gauge invariance in a 71-site bose--hubbard quantum
  simulator''.
\newblock \href{https://dx.doi.org/10.1038/s41586-020-2910-8}{Nature {\bf 587},
  392--396}~(2020).

\bibitem{zhou2021thermalization}
Zhao-Yu Zhou, Guo-Xian Su, Jad~C. Halimeh, Robert Ott, Hui Sun, Philipp Hauke,
  Bing Yang, Zhen-Sheng Yuan, Jürgen Berges, and Jian-Wei Pan.
\newblock ``Thermalization dynamics of a gauge theory on a quantum simulator''.
\newblock \href{https://dx.doi.org/10.1126/science.abl6277}{Science {\bf 377},
  311--314}~(2022).

\bibitem{Wiese_review}
U.-J. Wiese.
\newblock ``Ultracold quantum gases and lattice systems: quantum simulation of
  lattice gauge theories''.
\newblock \href{https://dx.doi.org/10.1002/andp.201300104}{Annalen der Physik
  {\bf 525}, 777--796}~(2013).

\bibitem{Zohar2015}
Erez Zohar, J~Ignacio Cirac, and Benni Reznik.
\newblock ``Quantum simulations of lattice gauge theories using ultracold atoms
  in optical lattices''.
\newblock \href{https://dx.doi.org/10.1088/0034-4885/79/1/014401}{Reports on
  Progress in Physics {\bf 79}, 014401}~(2015).

\bibitem{Dalmonte2016}
M.~Dalmonte and S.~Montangero.
\newblock ``Lattice gauge theory simulations in the quantum information era''.
\newblock \href{https://dx.doi.org/10.1080/00107514.2016.1151199}{Contemporary
  Physics {\bf 57}, 388--412}~(2016).

\bibitem{MariCarmen2019}
Mari~Carmen Ba{\~n}uls, Rainer Blatt, Jacopo Catani, Alessio Celi, Juan~Ignacio
  Cirac, Marcello Dalmonte, Leonardo Fallani, Karl Jansen, Maciej Lewenstein,
  Simone Montangero, Christine~A. Muschik, Benni Reznik, Enrique Rico, Luca
  Tagliacozzo, Karel Van~Acoleyen, Frank Verstraete, Uwe-Jens Wiese, Matthew
  Wingate, Jakub Zakrzewski, and Peter Zoller.
\newblock ``Simulating lattice gauge theories within quantum technologies''.
\newblock \href{https://dx.doi.org/10.1140/epjd/e2020-100571-8}{The European
  Physical Journal D {\bf 74}, 165}~(2020).

\bibitem{Alexeev_review}
Yuri Alexeev, Dave Bacon, Kenneth~R. Brown, Robert Calderbank, Lincoln~D. Carr,
  Frederic~T. Chong, Brian DeMarco, Dirk Englund, Edward Farhi, Bill Fefferman,
  Alexey~V. Gorshkov, Andrew Houck, Jungsang Kim, Shelby Kimmel, Michael Lange,
  Seth Lloyd, Mikhail~D. Lukin, Dmitri Maslov, Peter Maunz, Christopher Monroe,
  John Preskill, Martin Roetteler, Martin~J. Savage, and Jeff Thompson.
\newblock ``Quantum computer systems for scientific discovery''.
\newblock \href{https://dx.doi.org/10.1103/PRXQuantum.2.017001}{PRX Quantum
  {\bf 2}, 017001}~(2021).

\bibitem{aidelsburger2021cold}
Monika Aidelsburger, Luca Barbiero, Alejandro Bermudez, Titas Chanda, Alexandre
  Dauphin, Daniel González-Cuadra, Przemysław~R. Grzybowski, Simon Hands,
  Fred Jendrzejewski, Johannes Jünemann, Gediminas Juzeliūnas, Valentin
  Kasper, Angelo Piga, Shi-Ju Ran, Matteo Rizzi, Germán Sierra, Luca
  Tagliacozzo, Emanuele Tirrito, Torsten~V. Zache, Jakub Zakrzewski, Erez
  Zohar, and Maciej Lewenstein.
\newblock ``Cold atoms meet lattice gauge theory''.
\newblock \href{https://dx.doi.org/10.1098/rsta.2021.0064}{Philosophical
  Transactions of the Royal Society A: Mathematical, Physical and Engineering
  Sciences {\bf 380}, 20210064}~(2022).

\bibitem{zohar2021quantum}
Erez Zohar.
\newblock ``Quantum simulation of lattice gauge theories in more than one space
  dimension—requirements, challenges and methods''.
\newblock \href{https://dx.doi.org/10.1098/rsta.2021.0069}{Philosophical
  Transactions of the Royal Society A: Mathematical, Physical and Engineering
  Sciences {\bf 380}, 20210069}~(2022).

\bibitem{klco2021standard}
Natalie Klco, Alessandro Roggero, and Martin~J Savage.
\newblock ``Standard model physics and the digital quantum revolution: thoughts
  about the interface''.
\newblock \href{https://dx.doi.org/10.1088/1361-6633/ac58a4}{Reports on
  Progress in Physics {\bf 85}, 064301}~(2022).

\bibitem{Weinberg_book}
S.~Weinberg.
\newblock ``The quantum theory of fields''.
\newblock Vol. 2: Modern Applications. Cambridge University Press. ~(1995).
\newblock  url:~\url{https://books.google.de/books?id=doeDB3\_WLvwC}.

\bibitem{Gattringer_book}
C.~Gattringer and C.~Lang.
\newblock ``Quantum chromodynamics on the lattice: An introductory
  presentation''.
\newblock Lecture Notes in Physics. Springer Berlin Heidelberg. ~(2009).
\newblock  url:~\url{https://books.google.de/books?id=l2hZKnlYDxoC}.

\bibitem{Zee_book}
A.~Zee.
\newblock ``Quantum field theory in a nutshell''.
\newblock Princeton University Press. ~(2003).
\newblock  url:~\url{https://books.google.de/books?id=85G9QgAACAAJ}.

\bibitem{Bernien2017}
Hannes Bernien, Sylvain Schwartz, Alexander Keesling, Harry Levine, Ahmed
  Omran, Hannes Pichler, Soonwon Choi, Alexander~S. Zibrov, Manuel Endres,
  Markus Greiner, Vladan Vuleti{\'c}, and Mikhail~D. Lukin.
\newblock ``Probing many-body dynamics on a 51-atom quantum simulator''.
\newblock \href{https://dx.doi.org/10.1038/nature24622}{Nature {\bf 551},
  579--584}~(2017).

\bibitem{Surace2020}
Federica~M. Surace, Paolo~P. Mazza, Giuliano Giudici, Alessio Lerose, Andrea
  Gambassi, and Marcello Dalmonte.
\newblock ``Lattice gauge theories and string dynamics in rydberg atom quantum
  simulators''.
\newblock \href{https://dx.doi.org/10.1103/PhysRevX.10.021041}{Phys. Rev. X
  {\bf 10}, 021041}~(2020).

\bibitem{Banerjee2021}
Debasish Banerjee and Arnab Sen.
\newblock ``Quantum scars from zero modes in an abelian lattice gauge theory on
  ladders''.
\newblock \href{https://dx.doi.org/10.1103/PhysRevLett.126.220601}{Phys. Rev.
  Lett. {\bf 126}, 220601}~(2021).

\bibitem{Aramthottil2022}
Adith~Sai Aramthottil, Utso Bhattacharya, Daniel Gonz\'alez-Cuadra, Maciej
  Lewenstein, Luca Barbiero, and Jakub Zakrzewski.
\newblock ``Scar states in deconfined $\mathbb{Z}_2$ lattice gauge theories''.
\newblock \href{https://dx.doi.org/10.1103/PhysRevB.106.L041101}{Phys. Rev. B
  {\bf 106}, L041101}~(2022).

\bibitem{Desaules2022a}
Jean-Yves {Desaules}, Debasish {Banerjee}, Ana {Hudomal}, Zlatko {Papi{\'c}},
  Arnab {Sen}, and Jad~C. {Halimeh}.
\newblock ``{Weak Ergodicity Breaking in the Schwinger Model}''~(2022).
\newblock  \href{http://arxiv.org/abs/2203.08830}{arXiv:2203.08830}.

\bibitem{Desaules2022b}
Jean-Yves {Desaules}, Ana {Hudomal}, Debasish {Banerjee}, Arnab {Sen}, Zlatko
  {Papi{\'c}}, and Jad~C. {Halimeh}.
\newblock ``{Prominent quantum many-body scars in a truncated Schwinger
  model}''~(2022).
\newblock  \href{http://arxiv.org/abs/2204.01745}{arXiv:2204.01745}.

\bibitem{Smith2017}
A.~Smith, J.~Knolle, D.~L. Kovrizhin, and R.~Moessner.
\newblock ``Disorder-free localization''.
\newblock \href{https://dx.doi.org/10.1103/PhysRevLett.118.266601}{Phys. Rev.
  Lett. {\bf 118}, 266601}~(2017).

\bibitem{Brenes2018}
Marlon Brenes, Marcello Dalmonte, Markus Heyl, and Antonello Scardicchio.
\newblock ``Many-body localization dynamics from gauge invariance''.
\newblock \href{https://dx.doi.org/10.1103/PhysRevLett.120.030601}{Phys. Rev.
  Lett. {\bf 120}, 030601}~(2018).

\bibitem{smith2017absence}
A.~Smith, J.~Knolle, R.~Moessner, and D.~L. Kovrizhin.
\newblock ``Absence of ergodicity without quenched disorder: From quantum
  disentangled liquids to many-body localization''.
\newblock \href{https://dx.doi.org/10.1103/PhysRevLett.119.176601}{Phys. Rev.
  Lett. {\bf 119}, 176601}~(2017).

\bibitem{Metavitsiadis2017}
Alexandros Metavitsiadis, Angelo Pidatella, and Wolfram Brenig.
\newblock ``Thermal transport in a two-dimensional $\mathbb{Z}_2$ spin
  liquid''.
\newblock \href{https://dx.doi.org/10.1103/PhysRevB.96.205121}{Phys. Rev. B
  {\bf 96}, 205121}~(2017).

\bibitem{Smith2018}
Adam Smith, Johannes Knolle, Roderich Moessner, and Dmitry~L. Kovrizhin.
\newblock ``Dynamical localization in $\mathbb{Z}_2$ lattice gauge theories''.
\newblock \href{https://dx.doi.org/10.1103/PhysRevB.97.245137}{Phys. Rev. B
  {\bf 97}, 245137}~(2018).

\bibitem{Russomanno2020}
Angelo Russomanno, Simone Notarnicola, Federica~Maria Surace, Rosario Fazio,
  Marcello Dalmonte, and Markus Heyl.
\newblock ``Homogeneous floquet time crystal protected by gauge invariance''.
\newblock \href{https://dx.doi.org/10.1103/PhysRevResearch.2.012003}{Phys. Rev.
  Research {\bf 2}, 012003}~(2020).

\bibitem{Papaefstathiou2020}
Irene Papaefstathiou, Adam Smith, and Johannes Knolle.
\newblock ``Disorder-free localization in a simple $u(1)$ lattice gauge
  theory''.
\newblock \href{https://dx.doi.org/10.1103/PhysRevB.102.165132}{Phys. Rev. B
  {\bf 102}, 165132}~(2020).

\bibitem{karpov2021disorder}
P.~Karpov, R.~Verdel, Y.-P. Huang, M.~Schmitt, and M.~Heyl.
\newblock ``Disorder-free localization in an interacting 2d lattice gauge
  theory''.
\newblock \href{https://dx.doi.org/10.1103/PhysRevLett.126.130401}{Phys. Rev.
  Lett. {\bf 126}, 130401}~(2021).

\bibitem{hart2021logarithmic}
Oliver Hart, Sarang Gopalakrishnan, and Claudio Castelnovo.
\newblock ``Logarithmic entanglement growth from disorder-free localization in
  the two-leg compass ladder''.
\newblock \href{https://dx.doi.org/10.1103/PhysRevLett.126.227202}{Phys. Rev.
  Lett. {\bf 126}, 227202}~(2021).

\bibitem{Zhu2021}
Guo-Yi Zhu and Markus Heyl.
\newblock ``Subdiffusive dynamics and critical quantum correlations in a
  disorder-free localized kitaev honeycomb model out of equilibrium''.
\newblock \href{https://dx.doi.org/10.1103/PhysRevResearch.3.L032069}{Phys.
  Rev. Research {\bf 3}, L032069}~(2021).

\bibitem{Zohar2011}
Erez Zohar and Benni Reznik.
\newblock ``Confinement and lattice quantum-electrodynamic electric flux tubes
  simulated with ultracold atoms''.
\newblock \href{https://dx.doi.org/10.1103/PhysRevLett.107.275301}{Phys. Rev.
  Lett. {\bf 107}, 275301}~(2011).

\bibitem{Zohar2012}
Erez Zohar, J.~Ignacio Cirac, and Benni Reznik.
\newblock ``Simulating compact quantum electrodynamics with ultracold atoms:
  Probing confinement and nonperturbative effects''.
\newblock \href{https://dx.doi.org/10.1103/PhysRevLett.109.125302}{Phys. Rev.
  Lett. {\bf 109}, 125302}~(2012).

\bibitem{Banerjee2012}
D.~Banerjee, M.~Dalmonte, M.~M\"uller, E.~Rico, P.~Stebler, U.-J. Wiese, and
  P.~Zoller.
\newblock ``Atomic quantum simulation of dynamical gauge fields coupled to
  fermionic matter: From string breaking to evolution after a quench''.
\newblock \href{https://dx.doi.org/10.1103/PhysRevLett.109.175302}{Phys. Rev.
  Lett. {\bf 109}, 175302}~(2012).

\bibitem{Zohar2013}
Erez Zohar, J.~Ignacio Cirac, and Benni Reznik.
\newblock ``Simulating ($2+1$)-dimensional lattice qed with dynamical matter
  using ultracold atoms''.
\newblock \href{https://dx.doi.org/10.1103/PhysRevLett.110.055302}{Phys. Rev.
  Lett. {\bf 110}, 055302}~(2013).

\bibitem{Hauke2013}
P.~Hauke, D.~Marcos, M.~Dalmonte, and P.~Zoller.
\newblock ``Quantum simulation of a lattice schwinger model in a chain of
  trapped ions''.
\newblock \href{https://dx.doi.org/10.1103/PhysRevX.3.041018}{Phys. Rev. X {\bf
  3}, 041018}~(2013).

\bibitem{Stannigel2014}
K~Stannigel, Philipp Hauke, David Marcos, Mohammad Hafezi, S~Diehl, M~Dalmonte,
  and P~Zoller.
\newblock ``Constrained dynamics via the zeno effect in quantum simulation:
  Implementing non-abelian lattice gauge theories with cold atoms''.
\newblock \href{https://dx.doi.org/10.1103/PhysRevLett.112.120406}{Physical
  review letters {\bf 112}, 120406}~(2014).

\bibitem{Kuehn2014}
Stefan K\"uhn, J.~Ignacio Cirac, and Mari-Carmen Ba\~nuls.
\newblock ``Quantum simulation of the schwinger model: A study of
  feasibility''.
\newblock \href{https://dx.doi.org/10.1103/PhysRevA.90.042305}{Phys. Rev. A
  {\bf 90}, 042305}~(2014).

\bibitem{Kuno2017}
Yoshihito Kuno, Shinya Sakane, Kenichi Kasamatsu, Ikuo Ichinose, and Tetsuo
  Matsui.
\newblock ``Quantum simulation of ($1+1$)-dimensional u(1) gauge-higgs model on
  a lattice by cold bose gases''.
\newblock \href{https://dx.doi.org/10.1103/PhysRevD.95.094507}{Phys. Rev. D
  {\bf 95}, 094507}~(2017).

\bibitem{Yang2016}
Dayou Yang, Gouri~Shankar Giri, Michael Johanning, Christof Wunderlich, Peter
  Zoller, and Philipp Hauke.
\newblock ``Analog quantum simulation of $(1+1)$-dimensional lattice qed with
  trapped ions''.
\newblock \href{https://dx.doi.org/10.1103/PhysRevA.94.052321}{Phys. Rev. A
  {\bf 94}, 052321}~(2016).

\bibitem{Negretti2017}
A.~S. Dehkharghani, E.~Rico, N.~T. Zinner, and A.~Negretti.
\newblock ``Quantum simulation of abelian lattice gauge theories via
  state-dependent hopping''.
\newblock \href{https://dx.doi.org/10.1103/PhysRevA.96.043611}{Phys. Rev. A
  {\bf 96}, 043611}~(2017).

\bibitem{Dutta2017}
Omjyoti Dutta, Luca Tagliacozzo, Maciej Lewenstein, and Jakub Zakrzewski.
\newblock ``Toolbox for abelian lattice gauge theories with synthetic matter''.
\newblock \href{https://dx.doi.org/10.1103/PhysRevA.95.053608}{Phys. Rev. A
  {\bf 95}, 053608}~(2017).

\bibitem{Barros2019}
Jo{\~a}o~C. {Pinto Barros}, Michele {Burrello}, and Andrea {Trombettoni}.
\newblock ``Gauge theories with ultracold atoms''~(2019).
\newblock  \href{http://arxiv.org/abs/1911.06022}{arXiv:1911.06022}.

\bibitem{Halimeh2020a}
Jad~C. Halimeh and Philipp Hauke.
\newblock ``Reliability of lattice gauge theories''.
\newblock \href{https://dx.doi.org/10.1103/PhysRevLett.125.030503}{Phys. Rev.
  Lett. {\bf 125}, 030503}~(2020).

\bibitem{Lamm2020}
Henry Lamm, Scott Lawrence, and Yukari Yamauchi.
\newblock ``Suppressing coherent gauge drift in quantum simulations''~(2020).
\newblock  \href{http://arxiv.org/abs/2005.12688}{arXiv:2005.12688}.

\bibitem{Halimeh2020e}
Jad~C. Halimeh, Haifeng Lang, Julius Mildenberger, Zhang Jiang, and Philipp
  Hauke.
\newblock ``Gauge-symmetry protection using single-body terms''.
\newblock \href{https://dx.doi.org/10.1103/PRXQuantum.2.040311}{PRX Quantum
  {\bf 2}, 040311}~(2021).

\bibitem{kasper2021nonabelian}
Valentin Kasper, Torsten~V. Zache, Fred Jendrzejewski, Maciej Lewenstein, and
  Erez Zohar.
\newblock ``Non-abelian gauge invariance from dynamical decoupling''~(2021).
\newblock  \href{http://arxiv.org/abs/2012.08620}{arXiv:2012.08620}.

\bibitem{vandamme2021reliability}
Maarten~Van Damme, Haifeng Lang, Philipp Hauke, and Jad~C. Halimeh.
\newblock ``Reliability of lattice gauge theories in the thermodynamic
  limit''~(2021).
\newblock  \href{http://arxiv.org/abs/2104.07040}{arXiv:2104.07040}.

\bibitem{halimeh2021gauge}
Jad~C Halimeh, Haifeng Lang, and Philipp Hauke.
\newblock ``Gauge protection in non-abelian lattice gauge theories''.
\newblock \href{https://dx.doi.org/10.1088/1367-2630/ac5564}{New Journal of
  Physics {\bf 24}, 033015}~(2022).

\bibitem{halimeh2021stabilizing}
Jad~C. Halimeh, Lukas Homeier, Christian Schweizer, Monika Aidelsburger,
  Philipp Hauke, and Fabian Grusdt.
\newblock ``Stabilizing lattice gauge theories through simplified local
  pseudogenerators''.
\newblock \href{https://dx.doi.org/10.1103/PhysRevResearch.4.033120}{Phys. Rev.
  Research {\bf 4}, 033120}~(2022).

\bibitem{vandamme2021suppressing}
Maarten~Van Damme, Julius Mildenberger, Fabian Grusdt, Philipp Hauke, and
  Jad~C. Halimeh.
\newblock ``Suppressing nonperturbative gauge errors in the thermodynamic limit
  using local pseudogenerators''~(2021).
\newblock  \href{http://arxiv.org/abs/2110.08041}{arXiv:2110.08041}.

\bibitem{halimeh2021stabilizingDFL}
Jad~C. Halimeh, Hongzheng Zhao, Philipp Hauke, and Johannes Knolle.
\newblock ``Stabilizing disorder-free localization''~(2021).
\newblock  \href{http://arxiv.org/abs/2111.02427}{arXiv:2111.02427}.

\bibitem{halimeh2021enhancing}
Jad~C. Halimeh, Lukas Homeier, Hongzheng Zhao, Annabelle Bohrdt, Fabian Grusdt,
  Philipp Hauke, and Johannes Knolle.
\newblock ``Enhancing disorder-free localization through dynamically emergent
  local symmetries''.
\newblock \href{https://dx.doi.org/10.1103/PRXQuantum.3.020345}{PRX Quantum
  {\bf 3}, 020345}~(2022).

\bibitem{Chandrasekharan1997}
S~Chandrasekharan and U.-J Wiese.
\newblock ``Quantum link models: A discrete approach to gauge theories''.
\newblock
  \href{https://dx.doi.org/https://doi.org/10.1016/S0550-3213(97)80041-7}{Nuclear
  Physics B {\bf 492}, 455 -- 471}~(1997).

\bibitem{Buyens2017}
Boye Buyens, Simone Montangero, Jutho Haegeman, Frank Verstraete, and Karel
  Van~Acoleyen.
\newblock ``Finite-representation approximation of lattice gauge theories at
  the continuum limit with tensor networks''.
\newblock \href{https://dx.doi.org/10.1103/PhysRevD.95.094509}{Phys. Rev. D
  {\bf 95}, 094509}~(2017).

\bibitem{zache2021achieving}
Torsten~V. Zache, Maarten Van~Damme, Jad~C. Halimeh, Philipp Hauke, and
  Debasish Banerjee.
\newblock ``Toward the continuum limit of a $(1+1)\mathrm{D}$ quantum link
  schwinger model''.
\newblock \href{https://dx.doi.org/10.1103/PhysRevD.106.L091502}{Phys. Rev. D
  {\bf 106}, L091502}~(2022).

\bibitem{Kasper2017}
V~Kasper, F~Hebenstreit, F~Jendrzejewski, M~K Oberthaler, and J~Berges.
\newblock ``Implementing quantum electrodynamics with ultracold atomic
  systems''.
\newblock \href{https://dx.doi.org/10.1088/1367-2630/aa54e0}{New Journal of
  Physics {\bf 19}, 023030}~(2017).

\bibitem{Zache2019}
T.~V. Zache, N.~Mueller, J.~T. Schneider, F.~Jendrzejewski, J.~Berges, and
  P.~Hauke.
\newblock ``Dynamical topological transitions in the massive schwinger model
  with a $\ensuremath{\theta}$ term''.
\newblock \href{https://dx.doi.org/10.1103/PhysRevLett.122.050403}{Phys. Rev.
  Lett. {\bf 122}, 050403}~(2019).

\bibitem{PhysRevLett.38.1440}
R.~D. Peccei and Helen~R. Quinn.
\newblock ``$\mathrm{CP}$ conservation in the presence of pseudoparticles''.
\newblock \href{https://dx.doi.org/10.1103/PhysRevLett.38.1440}{Phys. Rev.
  Lett. {\bf 38}, 1440--1443}~(1977).

\bibitem{Heyl2013}
M.~Heyl, A.~Polkovnikov, and S.~Kehrein.
\newblock ``Dynamical quantum phase transitions in the transverse-field ising
  model''.
\newblock \href{https://dx.doi.org/10.1103/PhysRevLett.110.135704}{Phys. Rev.
  Lett. {\bf 110}, 135704}~(2013).

\bibitem{Heyl_review}
Markus Heyl.
\newblock ``Dynamical quantum phase transitions: a review''.
\newblock \href{https://dx.doi.org/10.1088/1361-6633/aaaf9a}{Reports on
  Progress in Physics {\bf 81}, 054001}~(2018).

\bibitem{Huang2019}
Yi-Ping Huang, Debasish Banerjee, and Markus Heyl.
\newblock ``Dynamical quantum phase transitions in u(1) quantum link models''.
\newblock \href{https://dx.doi.org/10.1103/PhysRevLett.122.250401}{Phys. Rev.
  Lett. {\bf 122}, 250401}~(2019).

\bibitem{Haegeman2011}
Jutho Haegeman, J.~Ignacio Cirac, Tobias~J. Osborne, Iztok
  Pi\ifmmode~\check{z}\else \v{z}\fi{}orn, Henri Verschelde, and Frank
  Verstraete.
\newblock ``Time-dependent variational principle for quantum lattices''.
\newblock \href{https://dx.doi.org/10.1103/PhysRevLett.107.070601}{Phys. Rev.
  Lett. {\bf 107}, 070601}~(2011).

\bibitem{Haegeman2016}
Jutho Haegeman, Christian Lubich, Ivan Oseledets, Bart Vandereycken, and Frank
  Verstraete.
\newblock ``Unifying time evolution and optimization with matrix product
  states''.
\newblock \href{https://dx.doi.org/10.1103/PhysRevB.94.165116}{Phys. Rev. B
  {\bf 94}, 165116}~(2016).

\bibitem{Vanderstraeten2019}
Laurens Vanderstraeten, Jutho Haegeman, and Frank Verstraete.
\newblock ``{Tangent-space methods for uniform matrix product states}''.
\newblock \href{https://dx.doi.org/10.21468/SciPostPhysLectNotes.7}{SciPost
  Phys. Lect. NotesPage~7}~(2019).

\bibitem{LaGaDyn}
J. C. Halimeh \textit{et al.} (in preparation).

\bibitem{MPSKit}
Maarten Van~Damme, Jutho Haegeman, Gertian Roose, and Markus Hauru.
\newblock ``{MPSKit.jl}''.
\newblock \url{https://github.com/maartenvd/MPSKit.jl}~(2020).

\bibitem{banuls2013mass}
M.~C. Ba{\~n}uls, K.~Cichy, J.~I. Cirac, and K.~Jansen.
\newblock ``The mass spectrum of the schwinger model with matrix product
  states''.
\newblock \href{https://dx.doi.org/10.1007/JHEP11(2013)158}{Journal of High
  Energy Physics {\bf 2013}, 158}~(2013).

\bibitem{banuls2016chiral}
Mari~Carmen Ba\~nuls, Krzysztof Cichy, Karl Jansen, and Hana Saito.
\newblock ``Chiral condensate in the schwinger model with matrix product
  operators''.
\newblock \href{https://dx.doi.org/10.1103/PhysRevD.93.094512}{Phys. Rev. D
  {\bf 93}, 094512}~(2016).

\bibitem{vumps}
V.~Zauner-Stauber, L.~Vanderstraeten, M.~T. Fishman, F.~Verstraete, and
  J.~Haegeman.
\newblock ``Variational optimization algorithms for uniform matrix product
  states''.
\newblock \href{https://dx.doi.org/10.1103/PhysRevB.97.045145}{Phys. Rev. B
  {\bf 97}, 045145}~(2018).

\bibitem{McCulloch2008}
I.~P. {McCulloch}.
\newblock ``{Infinite size density matrix renormalization group,
  revisited}''~(2008).
\newblock  \href{http://arxiv.org/abs/0804.2509}{arXiv:0804.2509}.

\end{thebibliography}
\end{document}